\documentclass[acmsmall, screen, nonacm]{acmart}

\usepackage{longtable}
\usepackage{multirow}
\usepackage{makecell}
\usepackage{siunitx}
\usepackage{caption}
\usepackage{subcaption}
\usepackage{wrapfig}
\usepackage{threeparttable}
\usepackage{float}
\usepackage{enumitem}
\ifodd 1
    
    \newcommand{\com}[1]{{\textbf{\color{red}(COMMENT: #1)}}}
    \newcommand{\resp}[1]{{\color{cyan}\textbf{Response}:#1}}
\else
    
    \newcommand{\com}[1]{}
    \newcommand{\resp}[1]{}
\fi

\authorsaddresses{}
\setcopyright{none}
\renewcommand\footnotetextcopyrightpermission[1]{}

\begin{document}

\title{Estimating the Social Cost of Corporate Data Breaches}

\author{Lina Alkarmi}
\affiliation{%
  \institution{University of Michigan}
  \city{Ann Arbor, MI}
  \country{USA}
}
\email{lalkarmi@umich.edu}

\author{Armin Sarabi}
\affiliation{%
  \institution{University of Michigan}
  \country{USA}
}
\email{arsarabi@umich.edu}

\author{Mingyan Liu}
\affiliation{%
  \institution{University of Michigan}
  \country{USA}
}
\email{mingyan@umich.edu}

\authorsaddresses{%
  Corresponding author: Lina Alkarmi (\href{mailto:lalkarmi@umich.edu}{lalkarmi@umich.edu}). 
  Other authors: Armin Sarabi (\href{mailto:arsarabi@umich.edu}{arsarabi@umich.edu}) and Mingyan Liu (\href{mailto:mingyan@umich.edu}{mingyan@umich.edu}).
}

\date{December 2025}

\begin{abstract}
While the size of a data breach is typically measured by the number of (consumer, customer, or user) records exposed or compromised, its economic impact is generally measured from the point of view of the corporation suffering the data breach: cost in crisis management, legal fees, civil penalty, drop in stock price, and so on. This ignores the externalized costs shifted onto individuals whose records were exposed and who may as a result subsequently fall victim to credit fraud, identity theft and other economic crimes.  This study examines whether it is possible to estimate the {\em true} cost, or the {\em social} cost of a data breach, measured by the impact on its victims and their out of pocket costs. To accomplish this we establish two building blocks: (1) the estimation of the average direct financial losses of an identity theft (IDT) victim, including the opportunity cost of lost time, and healthcare expenditures associated with physical and emotional distress associated with identity theft; and (2) the estimation of increases in incidents of IDT that can be attributed to a major breach event. To establish (1), we perform a comprehensive 13-year longitudinal analysis of identity theft by pooling all six waves of the Identity Theft Supplement (ITS) to the National Crime Victimization Survey (2008-2021). For (2) we pair the ITS data with the  breach chronology data from the Privacy Rights Clearinghouse (PRC), augmented by a number of auxiliary datasets; this allows us to perform hypothesis testing to verify whether following a major breach event there is a statistically significant increase in IDTs, and subsequently estimate how much of that increase can be attributed to said breach. This ``breach-to-victim'' conversion, combined with (1), then yields an estimate of the social cost of a given data breach.  Our findings show that the average social cost per victim has declined significantly since 2016, likely driven by the adoption of EMV chip technology and improved fraud detection. Furthermore, we find that there is indeed a statistically significant increase in the number of IDTs following a mega-breach event when accounting for a discovery lag of 1-2 months post-breach. Applying our model to real-world cases allows us to estimate an upper and lower bound social cost of specific mega-breach events. We find that for the 2009 Heartland and 2013 Target breaches, even the conservative lower bound social cost estimate exceeded settlements by factors of 5 and 16, respectively. In contrast, the 2017 Equifax breach resulted in a lower bound estimate of \$263.8 million, falling well within its \$700 million settlement cap. While the Equifax upper bound estimate of \$1.72 billion in social cost more than doubles this settlement, the narrowing gap between institutional liability and an incident's social cost provides empirical evidence of a market saturation effect that reduces the marginal damage of individual compromised records over time.
\end{abstract}

\maketitle
\section{Introduction}
Over recent years, identity theft (IDT) has evolved from a localized crime of physical opportunity to a highly organized and technology-driven one. Where offenders once relied on stolen mail or discarded documents, the modern threat environment is characterized by the digitization of financial and medical records, which has created more opportunities for exploitation. Today, large-scale data breaches occur with such frequency that they have become a predictable feature of the digital economy \cite{Verizon_DBIR_2025}, feeding a dark-web marketplace where personal identifiers are traded as low-cost commodities \cite{Holt2013}. This availability of data means that even a small corporate data compromise can expose millions of individuals' data. Despite this growing prevalence, current reports on the economic impact of data breaches remain skewed by their focus on the corporation. When a breach occurs, industry reports typically quantify the impact in terms of corporate expenses such as regulatory fines, legal settlements, and forensic investigations \cite{IBM_DBR_2025}. However, these figures completely ignore the externalized (social) costs shifted onto the victims. 

This study stems from the observation that while a company may report a fixed dollar amount per compromised record, that figure does not account for the social harm that victims face. As criminal tactics become more and more sophisticated, the burden on a victim has shifted from temporary financial inconvenience to a long struggle for digital recovery. In some cases, victims suffer serious physical and mental ailments as a result of the crime \cite{Golladay2017Consequences}.
Admittedly, there are other forms of externalized cost of data breaches beyond costs associated with financial crimes suffered by consumers: for instance, data breaches may lead a firm to increase spending in cybersecurity, which raises the cost of products and services it offers, which is then at least partially transferred to its customers. In the present study we will limit ourselves to the cost of IDTs, and our goal is to develop a method that can help us estimate the totality of this cost incurred by a major data breach event. 

Toward that end, we will attempt to broaden the typical definition of the cost of an IDT, which is often narrowly measured by a victim's immediate out-of-pocket (OOP) loss, a measure that ignores or underestimates other costs that victims often incur. Quantifying a broader array of direct as well as indirect costs associated with IDT, which includes the hours victims spend on the phone with banks and government agencies, the professional legal help they may need to hire, and the physical and emotional distress that manifests as medical bills, is critical for several reasons. First, a comprehensive cost model allows policymakers to accurately weigh the benefits of security mandates against the burden taken on by the public. Second, understanding these costs helps victim service organizations and insurance providers better allocate resources for the intangible harms, such as mental health support, that traditional financial compensation ignores. By focusing on the victim's perspective, our work provides a more complete insight into the total burden of these breach events. 
It should be noted that while we substantially broaden the array of downstream costs associated with an IDT in our study, there are clearly additional externalized and downstream harm to both individuals and societies that our model does not capture. For this reason, while we will continue to use the term ``social cost'' throughout this paper for lack of a better term, we acknowledge that this term is used in a narrow sense constrained by what we can extract from available data.

Establishing this ``social cost'' requires linking two disconnected types of data: corporate data breaches and the private experiences of IDT victims. In this paper, we bridge this gap by conducting a longitudinal analysis on trends in IDT. We develop a model to quantify the social cost of IDT, and analyze the number of IDT victims compared to the number of data records breached over time. We use over a decade of national survey data to map the life cycle of IDT, from how it is stolen, how it is discovered, and most importantly, what it truly costs the American public in both time and well-being. By pairing this victim data with the chronology of major data breaches, we test the question: \textbf{Does the frequency of mega-breaches drive a measurable increase in IDT, and if so, does the corporate penalty match the social cost?} To address this, we propose two estimation methods: an empirical lower bound of a mega-breach's short-term social cost (a lower bound), and a time-decaying projection of its long-term impact (an upper bound).
Our contributions are as follows:

\begin{enumerate}
    \item \textbf{Longitudinal data integration}: To our knowledge, this is the first study to harmonize and pool all six waves of the ITS (2008-2021) to provide a comprehensive 13-year analysis of identity theft trends.
    \item \textbf{Social cost modeling:} Unlike previous studies that focus primarily on direct financial loss, our model incorporates social costs, including the opportunity cost of lost time and monetized healthcare expenses related to mental and physical distress.
    \item \textbf{Data breach to identity theft victim analysis:} By comparing self-reported survey data with external breach chronology data, we conduct a Wilcoxon signed-rank test to verify whether there is an increase in identity theft cases following a mega-breach event, and calculate a breach to identity theft victim conversion rate to quantify the risk of successful victimization following a breach event. 
    \item \textbf{Corporate liability case studies:} We apply our social cost model and breach-to-victim conversion in three case studies: the 2009 Heartland Payment Systems breach, the 2013 Target breach, and the 2017 Equifax breach. We calculate upper and lower bound estimates for the social cost of these events, and compare that to their settlements.
\end{enumerate}

The remainder of this paper is organized as follows. Section \ref{sec:background} describes related work and the motivation for our study. In Section \ref{sec:data_and_sample}, we describe our data sources and pre-processing procedures.  Section \ref{sec:comprehensive_cost_of_IDT} details our social cost model along with our calculations and results from the ITS dataset. Section \ref{sec:breach_to_victim} compares the PRC data breach chronology data with the  victim data from the ITS survey while also detailing our model for estimating the breach-to-victim conversion rate. Section \ref{sec:breach_to_victim} also tests the relationship between mega-breaches and the number of victims, and calculates the social cost of our three case study breaches. Finally, Section \ref{sec:discussion} discusses limitations and extensions of this study, and \ref{sec:conclusion} concludes the paper. Further technical details regarding the specific data cleaning protocols, social cost calculations, and extended results are provided in the Appendix.

\section{Background and Literature Review}
\label{sec:background}

Consistent with the Bureau of Justice Statistics (BJS), we define IDT as the attempted or successful misuse of an existing account, fraudulent opening of a new account, or the misuse of personal information for other fraudulent purposes \cite{HarrellThompson2024, Harrell2024Stats}. To study the social cost of  and trends of theft over the years, we used data from the \textit{National Crime Victimization Survey (NCVS): Identity Theft Supplement (ITS)} \cite{ICPSR26362, ICPSR34735, ICPSR36044, ICPSR36829, ICPSR37923, ICPSR38501}. The ITS is a large scale survey that collects data about individuals' personal experiences with IDT directly from a nationally representative sample of U.S. households. This survey is sponsored by the U.S. Bureau of Justice Statistics (BJS), and is administered periodically as a supplement to the main NCVS survey. The ITS gathers detailed information from IDT victims, including the types of personal information compromised, how they discovered the theft, personal and financial consequences, and the actions taken in response to the theft. In this study, we use data from six waves of the ITS, conducted in 2008, 2012, 2014, 2016, 2018, and 2021 \cite{ICPSR26362, ICPSR34735, ICPSR36044, ICPSR36829, ICPSR37923, ICPSR38501}. Complimenting this, we use data breach chronology data from Privacy Rights Clearinghouse (PRC) to obtain the number of breaches and exposed records over time, and compare with the number of victims from the ITS dataset \cite{PRC_Chronology}.

Previous research using the ITS dataset has consistently identified a target suitability profile for certain types of IDT. \citet{Nevin2025} and \citet{Copes2010} found that credit card victimization risk is generally higher among individuals who are older, white, and possess higher income and education; groups likely to possess higher credit limits and more complex financial activities. While these demographics are frequently targets of existing account fraud, research shows that lower income, younger, and minority groups are disproportionately victimized by more damaging forms of IDT such as existing bank account fraud and new account fraud. \citet{Reynolds2021Differential} and \citet{DeLiema2021} further explored these disparities, demonstrating that socio-economic status significantly impacts the likelihood and severity of out-of-pocket (OOP) losses. Beyond demographic analysis, studies such as \cite{Hu2021Forecasting} have utilized machine learning on the ITS data to forecast victimization and evaluate preventative actions. Notably, most existing studies only pool a subset of the ITS data, such as three or four waves. To the best of our knowledge, our study is the first to perform analysis of all six waves, which provides a more complete picture of trends. 

Our work also expands the literature on the total cost of IDT from a victim's point of view. For instance, \citet{Miller2021} uses the 2016 ITS survey as the source of its cost estimate of IDT, summing the victims' OOP losses and lost time costs, yet did not consider medical or mental health costs. \citet{Anderson2013Measuring} provides a measure for decomposing cybercrime impact into direct, indirect, and defense costs, but they also note that their model ``generally disregard[s] distress,'' because it is difficult to measure \cite{Anderson2013Measuring}. Similarly, \cite{Romanosky2016Examining} offers an empirical estimation of the costs associated with data breaches and security incidents from a corporate perspective, but notes that their calculations do not include intangible costs such as distress and lost time \cite{Romanosky2016Examining}. By contrast, our study considers a more comprehensive measure of the cost of IDT that includes intangible harms such as monetized health expenses.

Finally, the relationship between data breaches and individual victimization is an integral part of this cost estimate that connects corporate responsibility with (socialized) individual/consumer harm; interestingly, this appears to be a relatively under-explored area. BJS research indicates that victims of IDT are twice as likely as non-victims to have been notified that their data has been exposed in a breach in the past year \cite{Harrell2024Stats}. \citet{Bisogni2020More} use 13 years (2005-2017) of United States incident data and Bayesian modeling to demonstrate that while data breaches and IDT are correlated, their relationship is largely influenced by state population size. However, beyond these general correlations, explicit breach-to-victim calculations that quantify the probability of a leaked record resulting in a successful crime are hard to come by. This stems from the inherent difficulty of these calculations. Corporate breach notifications rarely track downstream victimization, and anonymous victimization surveys lack the data to trace fraud back to a specific exposure event. Consequently, the relationship between a corporate data breach and a consumer's loss remains obscured by the messiness of attribution.

For our purpose of establishing the link between data breaches and IDT victimization, we use PRC's data breach chronology and supplement it with data from corporate disclosures and federal breach portals. By adopting and building on the framework developed in \citet{Graves2018Should} for estimating national record exposure using linear regression to extrapolate from state-level findings when national data is missing or undisclosed, we calculate a conversion rate that measures how effectively criminals convert compromised data into successful IDT.

\section{Datasets Used in the Study}

\label{sec:data_and_sample}

\subsection{Identity Theft Data}
\begin{figure}[htbp]
    \centering
    \includegraphics[width=\textwidth, trim={1cm 6cm 1cm 6cm}, clip]{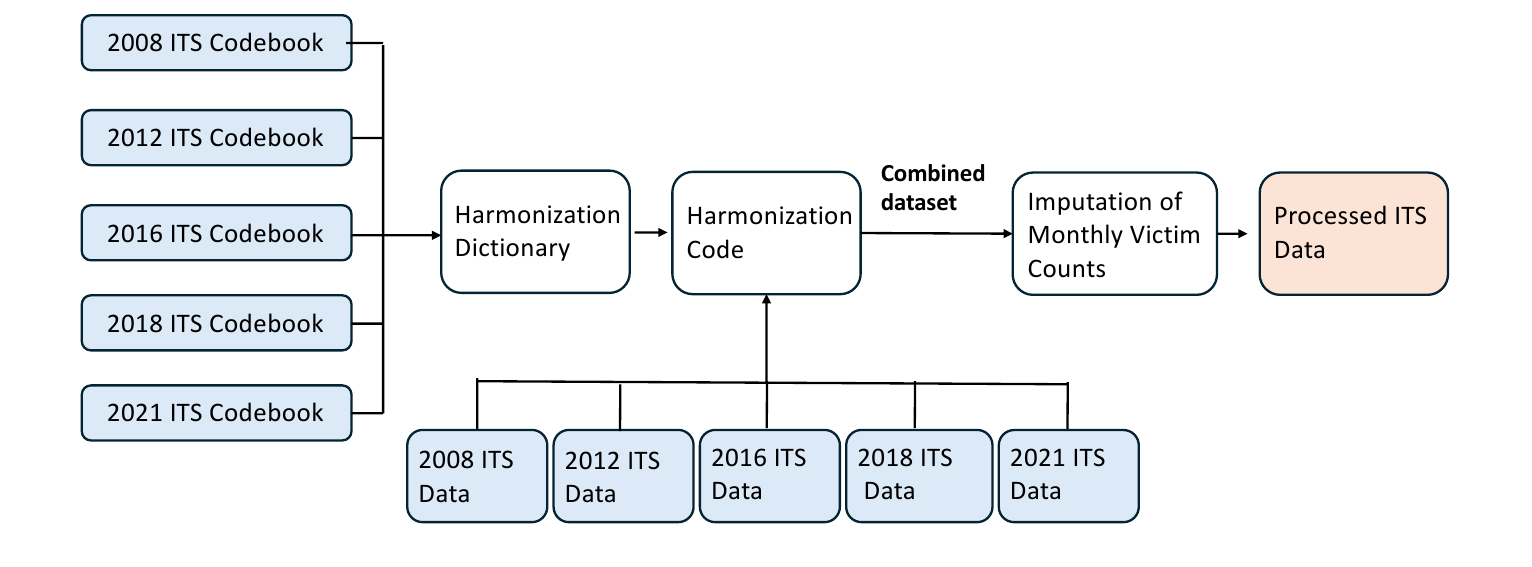}
    \vspace{-10pt}
    \caption{Workflow for the longitudinal harmonization and processing of the ITS datasets (2008-2021).}
    \label{fig:ITS_processing_flow}
\end{figure}

As mentioned previously, in this work we use data from six waves of the ITS, conducted in 2008, 2012, 2014, 2016, 2018, and 2021 \cite{ICPSR26362, ICPSR34735, ICPSR36044, ICPSR36829, ICPSR37923, ICPSR38501}. Each year's survey yields a distinct dataset which contains data about each person interviewed. In each dataset, a row corresponds to a single respondent, and the columns contain their responses to survey questions. Categorical responses within these columns are stored as numeric codes rather than text, requiring the use of the accompanying codebook to map these values to their labels (e.g., mapping ``1'' to ``Yes''). Additionally, each respondent in the survey has a corresponding weight that indicates how many people in the broader U.S. population this person represents. Combining these datasets allows for a longitudinal analysis of trends in IDT victimization and its consequences; however, it comes with some challenges with survey inconsistency over the years as we detail below. The overall workflow for preparing the IDT data is illustrated in Figure \ref{fig:ITS_processing_flow}.

\subsubsection{Combining the ITS Waves (Harmonization)}
A primary challenge in using these multiple waves is the evolution of the survey over time, which resulted in significant inconsistencies in variable names, value codes, and questions asked over the years, including evolving categories for demographic variables such as household income, and changing variable names and codes for core incident characteristics. Another example of survey inconsistency is that in those administered from 2008 through 2018, the initial screening questions were designed by BJS to identify a broad pool of victims which included individuals who experienced \textit{attempted} IDT (e.g., their credit card was stolen but never used), as well as \textit{successful} IDT (e.g., fraudulent charges were made). Follow-up questions in the survey were then used to differentiate between these two outcomes. In contrast, the 2021 ITS survey was redesigned by BJS to more directly identify victims, and the survey questions focused on whether a respondent's personal information was actually misused for fraudulent purposes. Thus, the 2021 dataset inherently represents a population of successful victims, excluding those who only experienced an attempt.

To address this, we harmonized the data by creating a mapping, or harmonization dictionary, which contained a definitive set of rules for recoding and standardizing variables to ensure that they are conceptually consistent and comparable across all six survey waves. This dictionary was created after careful review and comparison of the codebook associated with each year of the ITS data. For full details of the variable harmonization, recoding schemes, and filtering logic, see Appendix \ref{app: dataset_harmonization}. After harmonization, we constructed the final analysis sample by filtering out reported incidents that were ``attempted only,'' resulting in a dataset composed exclusively of victims of successful IDT. Since the survey was administered in waves taken approximately every two years, the harmonized dataset only includes victims in every other year. This means that there are ``blackout'' periods in the survey data, which we then mitigated with imputation detailed  in \ref{sec:ITS_impute}.

The application of this harmonization and filtering resulted in a dataset containing a total unweighted sample of 41,091 victims of successful IDT. These victims are distributed across the six survey waves as follows: 2,815 from 2008, 4,401 from 2012, 4,660 from 2014, 10,227 from 2016, 9,928 from 2018, and 9,060 from 2021. Note that while we used the six survey waves, they are pooled cross-sectional snapshots, not a longitudinal panel of the same individuals. Therefore, we are tracking trends in the population, not the same group of people over 13 years.

To enable generalization of the data findings to the U.S. population, the ITS dataset includes a final survey weight for each respondent. The use of this weight transforms the sample data into nationally representative samples. The weight is calculated by first starting with a base weight that accounts for each household's unequal probability of being selected for the survey. This base weight is then adjusted to account for households that were selected but did not respond, and then calibrated to align the sample's demographic composition with U.S. population totals \cite{ICPSR26362, ICPSR34735, ICPSR36044, ICPSR36829, ICPSR37923, ICPSR38501}. The resulting final weight is given in the dataset and represents a specific number of people in the U.S. population. For example, if a single victim in the dataset has weight 15,000, then their experience is statistically counted as representing 15,000 victims nationwide. {\em Unless otherwise specified, all analyses presented in the rest of this paper utilize data that has been weighted using these final survey weights}. 

\subsubsection{Imputation of Monthly Victim Counts} 
\label{sec:ITS_impute}
Because the ITS survey is administered as a supplement to the NCVS survey with specific reference periods, monthly victim counts are not continuous across the entire 13-year study period (i.e., some months have zero data). To construct a continuous timeline of IDT victimization for comparison against monthly breach data, during the months falling between survey waves where no direct victimization data was collected, we applied a log-linear interpolation. This method assumes a constant rate of change between the observed data points of the adjacent survey waves, allowing us to estimate the victim count during non-survey months and smooth the transitions between the distinct data collection periods.

\subsection{Data Breach Chronology Data}
\label{sec:PRC_processing}

\begin{figure}[htbp]
    \centering
    \includegraphics[width=\textwidth, trim={1cm 8.7cm 1cm 6cm}, clip]{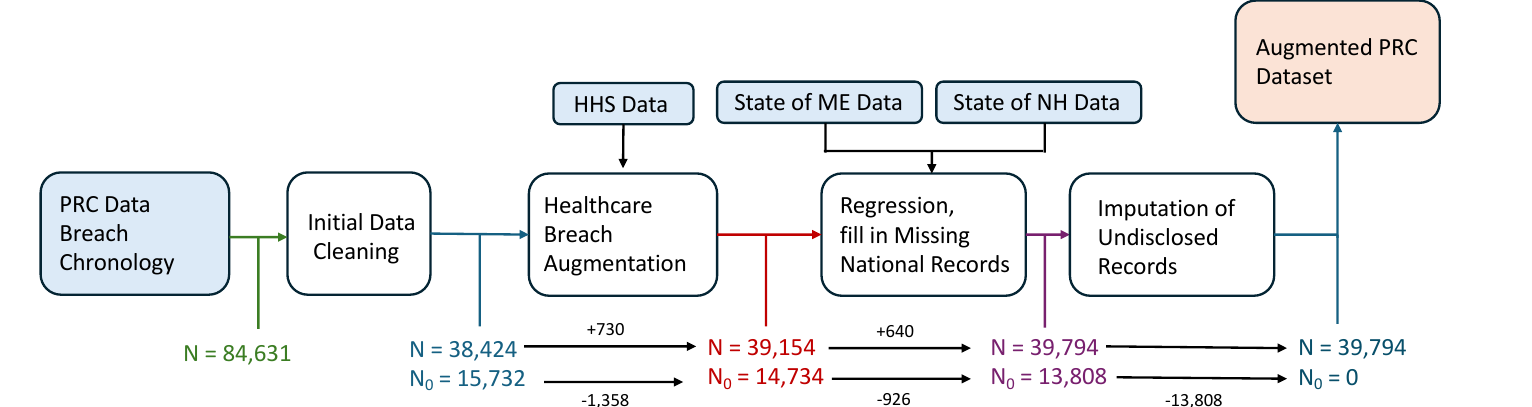}
    \vspace{-10pt}
    \caption{Workflow for the processing and augmentation of the PRC data breach chronology.}
    \label{fig:PRC_processing_flow}
\end{figure}

On the exposed data records front, we utilized the Privacy Rights Clearinghouse (PRC) Data Breach Chronology database (Version 2.0) \cite{PRC_Chronology}.  This database catalogs reported data breaches in the United States, providing details like organization profiles, breach types, timelines, and the resulting impact. This raw data was augmented following a multi-stage process to enhance its completeness and accuracy, following similar methods used in \cite{Graves2018Should}. Throughout this section, we use $N$ to denote the total number of data breach incidents and $N_0$ to represent the subset of incidents with undisclosed or missing record counts. Our augmentation process, illustrated in Figure \ref{fig:PRC_processing_flow}, involved the following steps:

\begin{enumerate}
    \item Initial data cleaning following PRC guidelines
    \item Enrichment via federal healthcare disclosures
    \item Estimating national record counts for incidents where only state-level record counts were disclosed
    \item Imputation of record counts for remaining breaches with undisclosed record counts
\end{enumerate}

\subsubsection{Initial Data Cleaning}

To begin, we queried the database for all recorded incidents occurring between 2008 and 2021, specifically extracting the organization name, reported date, and total number of records exposed for each incident. This initial query yielded 84,631 reports. To ensure the integrity of our incident count, we utilized PRC assigned group IDs to consolidate the multiple (state-level) reports stemming from the same underlying event into single observations. This process removed 46,207 duplicate reports, resulting in $N = 38,424$ unique breach events. Of these unique events, $N_0 = 15,732$ initially reported a value of zero for the number of records exposed.

\subsubsection{Integrating Supplemental Healthcare Breach Disclosures}
Following the methodology established by Graves et al. in 2018 \cite{Graves2018Should}, we supplemented the PRC dataset to address missing record counts and incorporate missing events. To improve the coverage of healthcare-related incidents, we cross-referenced the PRC dataset with breach reports from the U.S. Department of Health and Human Services (HHS) \cite{HHS_BreachPortal}. These additions were done in three distinct stages:

\begin{enumerate}[label=(\alph*)]
    \item We first identified and integrated 730 HHS-reported breaches that were entirely absent from the original PRC database, leading to $N = 39,154$. 
    \item For incidents present in both datasets where magnitude values differed, we retained the higher reported value to ensure the capture of the maximum potential impact. There were 7,372 such instances.
    \item Finally, for incidents present in both datasets where PRC originally reported zero or undisclosed record counts, we utilized the HHS data to populate these missing values. Through this process, we successfully recovered 1,358 previously missing magnitude values, thus reducing the data sparsity for healthcare related breaches, and leaving us with $N_0 = 14,734$. 
\end{enumerate}

\subsubsection{Regression Modeling Using State-Level Records}
To further recover missing record counts, we used a linear regression model to estimate national exposure from state-level data. Following the approach of \cite{Graves2018Should}, we gathered breach filings from state Attorneys General in Maine and New Hampshire. We selected these states because they consistently report the number of affected residents within their jurisdictions and provide comprehensive data coverage for our study period (2008-2021) \cite{NH_DOJ_Breach, Maine_AG_Breach}. We pooled these two state-level sources and deduplicated the records, prioritizing the highest reported victim count for each incident \cite{NH_DOJ_Breach, Maine_AG_Breach}. To minimize the inclusion of purely local incidents, we filtered out organizations containing keywords indicative of limited geographic scope (e.g. ``Town of'', ``plumbing'', ``electric'', ``dealership'', ``restaurant''). Following this filtration, we pooled the data from both Maine and New Hampshire. In the rare instances where a single breach event was reported in both states (there were only 4 such cases), we retained the observation with the higher reported victim count to avoid double-counting while capturing the upper-bound of the state-level impact. Through this process, we integrated breaches previously absent from the PRC database resulting in a pool of 2,298 unique national-level incidents identified from state-level filings. Of these 2,298, 1,372 were entirely absent from the original PRC database, and 640 were missing from the PRC database augmented with healthcare disclosures. Consequently, we added these 640 new incidents to our augmented dataset, resulting in $N = 39,794$. The remaining 926 incidents were present in the PRC but lacked an associated record count. To estimate national impact from these state-level numbers, we used Ordinary Least Squares (OLS) regression to determine the national weight of a single state-level victim. In this model, the independent variable $X$ represents the number of residents impacted within a specific state jurisdiction, while the dependent variable $Y$ represents the total national record exposure for the same incident, obtained from the PRC data. Before running the model, we filtered the data for high-confidence identity matches and verified that the national total number of people affected was at least five times greater than or equal to the state sample of affected residents. This yielded a final sample of 803 unique matched pairs, or incidents where both state-level and national-level data were available. This regression allowed us to recover estimates for these 926 breach events, resulting in $N_0 = 13,808$, and accounting for approximately 121 million records per year.

\subsubsection{Imputation of Undisclosed Records}
\label{sec:imputation_of_undisclosed}

Even after the augmentation by two state-level sources, we still faced a large volume of 13,808 incidents that lacked an associated number of disclosed records. Again following the methodology of \cite{Graves2018Should}, we sought to find an annual baseline for undisclosed records ($n_u$) through the following three-stage process: 

\begin{enumerate}[label=(\alph*)]
    \item \textit{Weight Calculation:} We categorized each breach in the PRC database by its breach type, and then calculated a typical weight for each category using the known, non-outlier data from the PRC database. These categories from the PRC database were as follows: physical payment card compromises, external cyber attacks, internal threats from authorized users, physical document theft or loss, portable device breaches, stationary device breaches, unintended disclosures, and unknown \cite{PRC_Chronology}. The median value of exposed records for each breach type was used as this weight. Due to the skew of the data, the median was chosen over the mean. 
    
    \item \textit{Baseline Estimation $n_u$:} We define the total undisclosed volume for a specific breach category $j$ as the product of the number of undisclosed incidents $l_j$ and the typical weight (median size) of known breaches in that category $W_j$. The total annual baseline is calculated as follows, where $T$ is the total study duration in years ($T=14$), and $k$ represents the eight PRC breach categories.: $n_u = \frac{1}{T} \sum_{j=1}^k (l_j \times W_j)$. This process allowed us to sum the categorized estimates and led to a finding of an annual $n_u = 1,531,784$ records as a conservative baseline that represents the number of undisclosed exposed records each year.

    \item \textit{Proportional Distribution $R_i$: }We distributed that volume across the zero-count incidents proportionally by year. Specifically, let $I_{u,y}$ be the set of incidents with undisclosed record counts in year $y$, and let $|I_{u,y}|$ be the total number of incidents in such a year. For any incident $i \in I_{u,y}$, the imputed number of records $R_i$ is defined as: $R_i = \frac{n_u}{|I_{u,y}|}$.
\end{enumerate}

The above model could in principle be further refined by separately estimating the baseline for each incident type; see more discussion on this in Section \ref{sec:discussion}.
  In the early years from 2008 to 2011, the number of breaches with undisclosed count $|I_{u,y}|$ was relatively low, averaging around 300 incidents per year. This number steadily increased over the years, reaching 1,109 in 2017. In the most recent years of our study (2018 to 2021), the number of breaches with undisclosed count remained high, consistently exceeding 950 incidents annually and peaking at 1,110 in 2021. The above methodology ensures that the total volume of imputed records in any given year remains constant and equals to $n_u$, thereby preventing the over-inflation of data while still accounting for the cumulative impact of undisclosed breaches. This resulted in a more complete data breach chronology dataset that we used for Section \ref{sec:breach_to_victim}. 

To verify our choice of $n_u$, we compared our estimation against the statistical modeling of the PRC dataset performed by \cite{Edwards2016Hype}. \citet{Edwards2016Hype} identified median breach sizes of 383 records for negligent breaches (when records are exposed accidentally) and 3,141 records for malicious breaches (when records are targeted). Applying these medians to our own dataset provides a quantitative verification of our baseline. Our PRC query identified 13,808 incidents with undisclosed record counts over the 13-year study period following enrichment, averaging approximately 1,062 such incidents per year. If we assume these annual incidents with undisclosed record counts were entirely negligent, the estimated volume would be 406,746 records ($1,062 \times 383$). If they were entirely malicious, the volume would be 3,335,742 records ($1,062 \times 3,141$). Our estimate of $n_u = 1,531,784$ is within this range, approximately 4 times higher than a purely negligent scenario, but significantly lower than a purely malicious one. This alignment suggests that our constant is a middle ground estimate consistent with independent modeling of typical breach magnitudes \cite{Edwards2016Hype}. We recognize that a static annual baseline $n_u$ is a simplification of a likely fluctuating reality. However, this approach provides a stable and conservative lower bound of the undisclosed breach market. Furthermore, because the frequency of incidents with undisclosed record counts $|I_{u,y}|$ increased significantly in later years, a static $n_u$ ensures that the imputed record count per incident remains modest.

\section{The Cost of Identity Theft Calculated from ITS Data}
\label{sec:comprehensive_cost_of_IDT}

\subsection{Modeling the Cost of Identity Theft}

To gain a clear picture of the real world impact of IDT, we developed a model to quantify its total cost. Our model calculates a comprehensive, per-victim cost for each survey year by monetizing the consequences of victimization across three categories: (1) direct financial and professional costs, (2) the opportunity cost of lost time, and (3) healthcare costs related to the incident. To ensure a valid cost comparison, all monetary values were adjusted to constant 2021 dollars using the annual average Consumer Price Index for All Urban Consumers (CPI-U) from the U.S. Bureau of Labor Statistics \cite{BLS_CPI, USInflationCalc_CPI}. For more information about the inflation adjustments, see Appendix \ref{app:inflation}.

\subsubsection{Category 1: Direct Financial and Professional Costs}
This category addresses the most immediate and explicit monetary losses that victims face. The primary component of this cost is the  direct/OOP loss that victims personally sustained and were not able to recover (i.e., stolen money not recovered). This excludes any fraudulent charges that were successfully disputed or covered by a financial institution, representing only the final realized financial harm to the victim. In addition to direct/OOP losses, this category also accounts for the cost of hiring professional legal services. While the ITS survey indicates whether a victim hired a lawyer or not, it does not record the amount paid. To monetize this expense, we assign an inflation adjusted cost of \$445 to each victim who reported hiring legal help. This fixed cost estimate is based on an average attorney hourly rate of around \$330, as reported in the 2023 Clio Legal Trends Report, and assumes approximately 1.5 hours of an attorney's time \cite{USNews_Lawyer}.

\subsubsection{Category 2: The Opportunity Cost of Lost Time}
Beyond direct financial loss, a significant cost comes from the time that victims sacrifice to resolve problems that arise from their IDT. This opportunity cost was quantified using the ``Hours Spent Resolving'' variable from the ITS dataset. To assign a monetary value to this lost time, the weighted average hours spent by victims each year was multiplied by the inflation adjusted average hourly wage for that corresponding year. The wage data was sourced from the U.S. Bureau of Labor Statistics' Current Employment Statistics Survey, which tracks average hourly earnings of all private nonfarm employees \cite{FRED_Wage}. This data was used because it represents the vast majority of the U.S. workforce and serves as a strong indicator of national wage trends. For more information regarding this data, see Appendix \ref{app:hourly_wages}.

\subsubsection{Category 3: Health and Well-being Costs}
While the significant emotional and psychological harm that victims suffer is difficult to quantify directly, our model addresses a measurable dimension of this distress by estimating the expenses related to seeking professional care. Using responses from the ITS survey, we identified victims who reported visiting a medical professional, received counseling, or took medication due to distress related to the incident. Inflation adjusted cost estimates were then assigned based on national averages. Each medical visit was valued at \$133 \cite{DebtOrg_Doctor}, each therapy session at \$89 \cite{GoodRx_Therapy}, and a course of medication at \$54 \cite{CBO_Prescription}. For more information on these estimates, see Appendix \ref{app:fixed_service_costs}. While for some victims this expense may have been covered under their healthcare insurance plan, we include it here as a real cost induced by IDTs.

\subsubsection{Final Calculation}
For each survey year, the {\em total cost per victim} was calculated by summing the monetized components: (1) average OOP loss and average legal cost, (2) average lost time cost, and (3) average healthcare cost. Finally, the {\em total national cost} was derived by multiplying this per-victim figure by the total weighted number of victims that year. 

It is important to note how the per victim averages for professional services were calculated. The total estimated cost for a service in a given year (e.g., the total amount spent on lawyers by all victims) was divided by the entire {\em unweighted} victim population for that year. This means that the average cost is spread across all victims, including the vast majority who did not incur that specific expense. For instance, if only one out of one hundred victims hires a lawyer for \$445, the average legal cost across all one hundred victims is just \$4.45. This explains why the average legal and healthcare costs appear low in the final analysis. While the financial burden for the individuals who require these services is substantial, they represent a small fraction of the total victim population, and the final average correctly reflects this distribution. Note that the OOP loss data is positively skewed, meaning that although most victims experienced zero OOP loss, the mean is pulled higher by a small number of higher value outlier cases. For calculating total national burden, using the average OOP loss is appropriate; however, it is important to keep in mind that this average cost may not reflect the cost for an average victim.

\subsection{Results from the ITS Data}
Below we briefly present some basic information on the ITS data, and then detail the result of the model presented in the previous section and associated observations over time.

The IDT victim population is presented in Table \ref{tab:victimization_year}, which displays the trend of victimization over time by counting the number of victims in each of the six survey waves. These numbers originate directly from the harmonized ITS dataset previously detailed in Section \ref{sec:data_and_sample}. For each survey wave, the total estimated number of victims (N) was found by summing the final survey weights of all respondents that were confirmed to be successful victims. Each year's weighted \% was then calculated to show its share relative to the total number of victims across all six surveys combined. From Table \ref{tab:victimization_year}, it is clear that the number of victims is generally increasing over time (we refer an interested reader to  Appendices \ref{app:longitudinal} and \ref{app:demographics} for a more detailed trend and demographic analysis).

\begin{table}[ht!]
  \centering
  \begin{minipage}{0.48\textwidth}
    \centering
    \captionof{table}{Victimization by year, from harmonized ITS dataset.}
    \label{tab:victimization_year}
    \begin{tabular}{lrr}
      \toprule
      \textbf{Year} & \textbf{N} & \textbf{\% of dataset} \\
      \midrule
      2008 & 11,684,672 & 9.83 \\
      2012 & 16,580,475 & 13.95 \\
      2014 & 17,576,206 & 14.79 \\
      2016 & 25,563,022 & 21.51 \\
      2018 & 23,536,881 & 19.80 \\
      2021 & 23,928,598 & 20.13 \\
      \bottomrule
    \end{tabular}
  \end{minipage}
    \hfill 
  \begin{minipage}{0.48\textwidth}
    \centering
    \includegraphics[width=\linewidth]{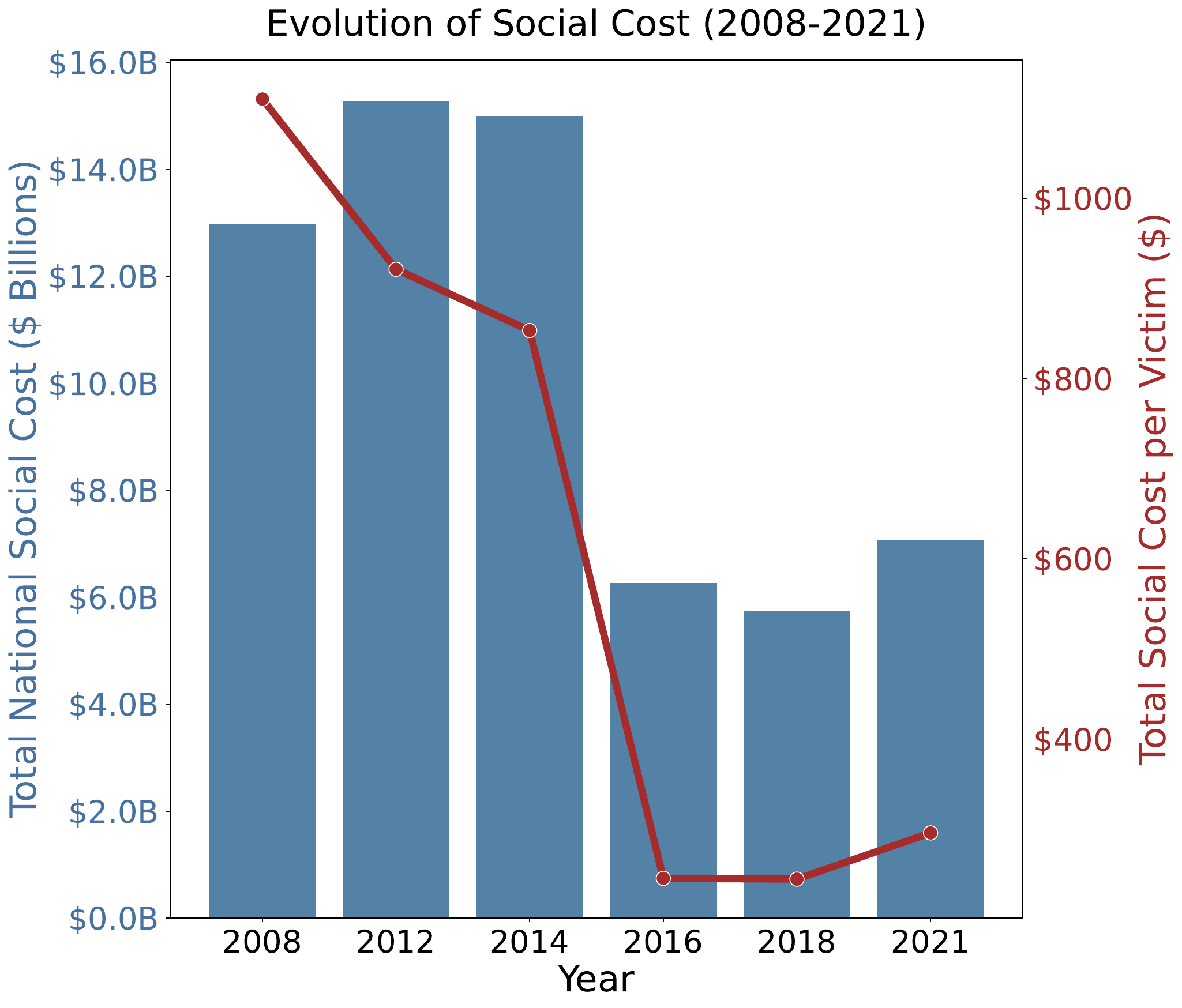}
    \captionof{figure}{Evolution of social costs associated with incidents over the study period.}
    \label{fig:social_cost_evolution}
  \end{minipage}

\end{table}

Next we present the results calculated using our Social Cost model, which monetizes direct financial losses, the opportunity cost of lost time, and healthcare expenses related to physical and emotional distress. Table \ref{tab:social_costs_by_year} displays these calculations. Note that as stated earlier, all monetary values in this table, including average OOP loss, have been adjusted to 2021's inflation. Table \ref{tab:social_costs_by_year} and Figure \ref{fig:social_cost_evolution} reveal two phases of social cost per victim. Between 2008 and 2014, the Total Social Cost per Victim remained high, peaking at \$1,110.31 in 2008. This period was characterized by substantial unrecoverable OOP losses. We observe a dramatic decline in the per-victim cost starting in 2016. The Total Social Cost per Victim fell to approximately \$245 in 2016 and 2018, primarily driven by a sharp reduction in OOP loss (discussed in detail in \ref{sec:trends_in}), and better resolution systems put into place. Despite this per-victim decline, the Total National Social Cost remains a multi-billion dollar problem, totaling over \$7 billion in 2021. This sustained national cost is due to the sheer volume of victims, which reached 23.9 million in the most recent survey wave.

\begin{table}[ht]
\centering
\caption{Social Costs of IDT by Year}
\label{tab:social_costs_by_year}

\small 
\begin{tabular}{lrrrrrrr}
\toprule
\thead{Year} & 
\thead{Total \\ Weighted \\ Victims} & 
\thead{Avg. Out-of- \\ Pocket Loss (\$)} & 
\thead{Avg. Legal \\ Cost (\$)} & 
\thead{Avg. Lost \\ Time Cost (\$)} & 
\thead{Avg. \\ Healthcare \\ Cost (\$)} & 
\thead{Total Social \\ Cost per \\ Victim (\$)} & 
\thead{Total \\ National \\ Social Cost (\$)} \\
\midrule
\textbf{2008} & 11,684,672 & 747.89 & 5.65 & 355.64 & 1.13 & 1110.31 & 12,973,628,242 \\
\textbf{2012} & 16,580,475 & 679.81 & 3.70 & 236.92 & 0.95 & 921.38 & 15,276,880,586 \\
\textbf{2014} & 17,576,206 & 663.01 & 3.31 & 186.41 & 0.68 & 853.41 & 14,999,689,893 \\
\textbf{2016} & 25,563,022 & 116.43 & 3.21 & 124.97 & 0.66 & 245.27 & 6,269,966,976 \\
\textbf{2018} & 23,536,881 & 117.96 & 1.43 & 124.36 & 0.64 & 244.40 & 5,752,387,938 \\
\textbf{2021} & 23,928,598 & 155.52 & 1.83 & 137.58 & 0.80 & 295.73 & 7,076,429,708 \\
\bottomrule
\end{tabular}
\end{table}

\subsubsection{Trends in Social Cost and Identity Theft}
\label{sec:trends_in}

To better understand the factors affecting the social cost measure, we examine several trends observed from the harmonized ITS data. Below we provide a brief summary of the most important trends, with a more detailed analysis presented in Appendix \ref{app:longitudinal}. 

\begin{figure}[htbp]
    \centering
    \begin{subfigure}[b]{0.48\textwidth}
        \centering
        \includegraphics[width=\textwidth]{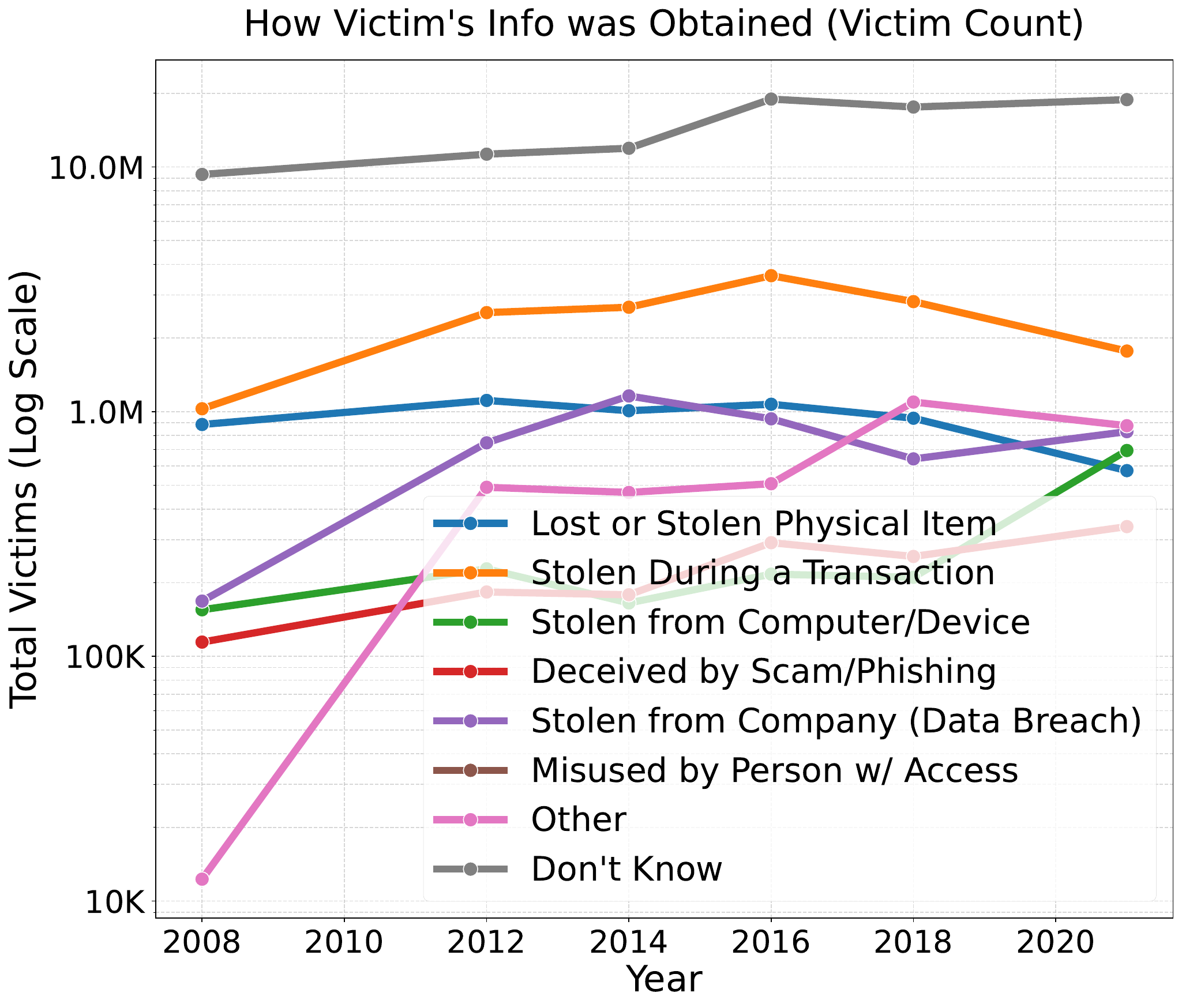}
        \caption{Number of victims categorized by how their information was stolen.}
        \label{fig:theft_method_long_a}
    \end{subfigure}
    \hfill 
    \begin{subfigure}[b]{0.48\textwidth}
        \centering
        \includegraphics[width=\textwidth]{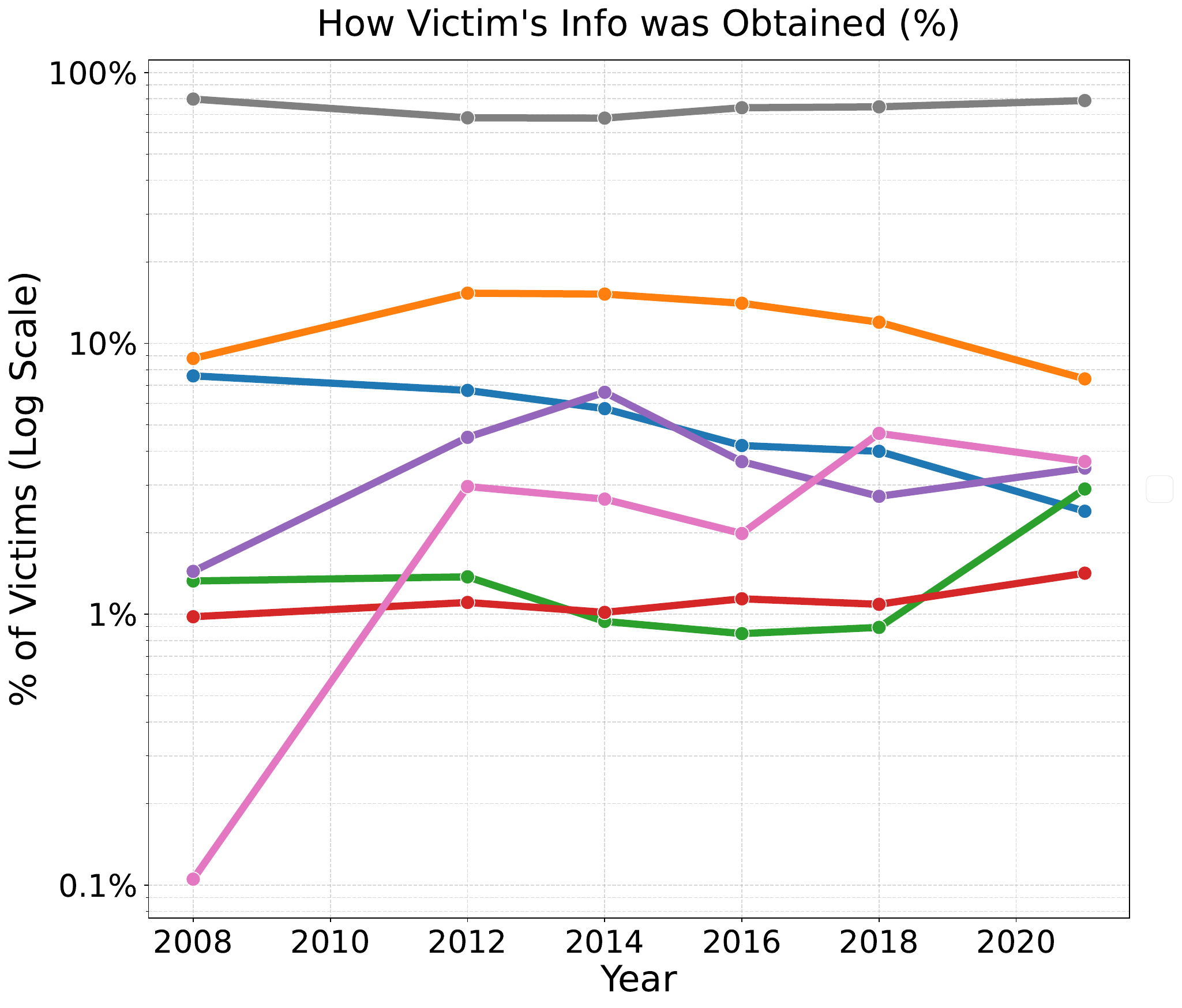}
        \caption{Percentage of victims categorized by how their information was stolen.}
        \label{fig:theft_method_long_b}
    \end{subfigure}
    
    \caption{Analysis of theft methods over the longitudinal study period.}
    \label{fig:theft_method_long}
\end{figure}

Figure \ref{fig:theft_method_long} displays how victims reported their information was obtained by offenders, which adds more context to the nature of the crime they suffered. A noticeable trend is the considerable decline in information being compromised during a transaction. This method, although still the most common means of identified theft, peaked in 2012 and 2014, and then fell sharply after, accounting for less than 8\% in 2021. Note that in October 2015, the major credit card networks (Europay, MasterCard, and Visa) implemented the EMV liability shift, a policy change that transferred financial responsibility for in-person counterfeit card fraud from the card issuer to whichever party in the transaction (either the merchant or the issuer) had not adopted the more secure chip card technology \cite{CRS_EMV_2016}. This was not a government mandated law but rather a policy change by the major credit card networks to financially incentivize the adoption of chip technology. The policy forced a nationwide payment infrastructure upgrade in an effort to counter the vulnerability of the easily cloned magnetic strip, which had been a leading carrier of counterfeit card fraud. The EMV liability shift was a major turning point in the fight against in-person payment fraud. The reduction in information being stolen during a transaction in Figure \ref{fig:theft_method_long} aligns with the time frame of the October 2015 liability shift for EMV chip card adoption in the United States. The efficacy of this transition was reported by the Federal Reserve, which reported that in-person card fraud, including counterfeit cards, declined from \$3.68 billion in 2015 to \$2.91 billion in 2016 \cite{fed2018fraud}. Similarly, Visa also reported that by October 2016, counterfeit fraud dollars at chip-enabled merchants had dropped by 43\% compared to a year earlier \cite{visa2016infographic}, and later data showed this decline reached 66\% by June 2017 \cite{visa2017report}.

\begin{wrapfigure}{r}{0.5\textwidth}
    \centering
    \includegraphics[width=0.48\textwidth]{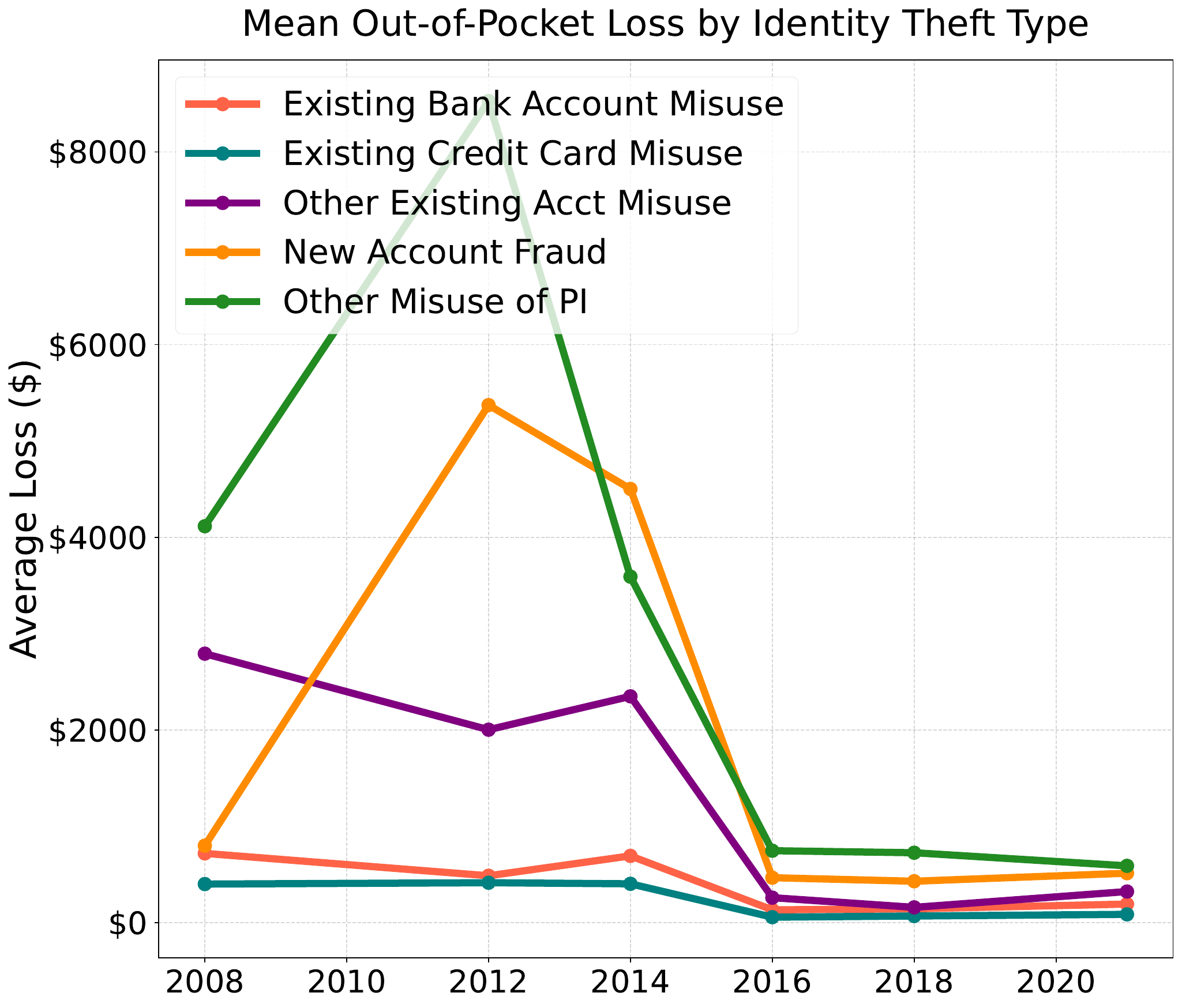}
    \caption{Mean Out-of-Pocket Loss by IDT Type (2008–2021).}
    \label{fig:oop_loss_comparison}
\end{wrapfigure}

Furthermore, a financial trend is evident in Figure \ref{fig:oop_loss_comparison}, which compares the average OOP losses adjusted to 2021 dollars, specifically for victims of different IDT types. It is worth noting that our OOP loss is less than direct loss values noted by BJS official reports \cite{HarrellThompson2024}, because we define OOP loss as the total amount of money stolen that was not recovered or reimbursed, so we include only costs that were not recoverable to the victim, rather than simply considering total amount stolen. Figure \ref{fig:oop_loss_comparison} shows that all categories show a drop and convergence between 2014 and 2016. This is the main driver behind the large drop in social cost observed in Table \ref{tab:social_costs_by_year}. We present two possible explanations for this trend. First, for the existing bank account misuse and existing credit card misuse OOP loss, we believe that the EMV transition played a dominant role as discussed above, because it reduced the most common type of high value, card-present fraud, and for the fraud that did occur, liability was clearer and banks had implemented faster detection systems, leading to more \$0 liability outcomes for consumers \cite{aba2018report}. Victims still reported ``credit card misuse'' because criminals pivoted to online, card-not-present fraud \cite{fed2018fraud}. This remote type of fraud is often more easily and quickly reimbursed by banks:  unlike in-person counterfeit fraud, where a dispute could be complex, as in a remote fraud case the victim is still in physical possession of their card, and bank systems can often use location data or IP addresses to quickly confirm that the transaction was fraudulent. This shift to a more easily reimbursable type of fraud meant the number of victims in our survey experiencing a direct, unrecoverable OOP loss fell dramatically, pulling the average loss for the entire group down. The EMV mandate applied to both credit and debit cards, the latter of which are the primary driver of bank account fraud losses according to a report by the American Bankers Association \cite{aba2018report}. While EMV chips directly impacted counterfeit card fraud, the concurrent drop for bank accounts represents this debit card protection as well as other broader improvements in banks' fraud detection systems \cite{aba2018report}.

However, the fact that this drop is observed over all IDT categories, including those unrelated to the EMV mandate (although not nearly as dramatic), such as ``Other Misuse of PI'' and ``New Account Fraud,'' suggests that the structural changes to the ITS survey during 2016 may also have played a role. As mentioned earlier, in 2016 the survey underwent a sample redesign, where household sample sizes were increased by 41\%, and weights were recalibrated to reflect the 2010 Decennial Census \cite{Harrell2019Victims}. BJS mentions that these changes require caution when comparing 2016 estimates to previous waves \cite{Harrell2019Victims}. This survey shift likely explains the volatility seen in the ``Other Misuse of PI'' (which includes employment fraud or government benefit fraud), ``Other Existing Account Misuse'' (which includes utilities and telephone account misuse), and ``New Account Fraud'' categories, which were characterized by lower incidence rates but potentially very high value outlier losses. The surge in mean loss for these types in 2012 and 2014 is likely due to a few extreme outlier cases that affect the smaller sample mean. The stabilization after 2016 suggests that the improved, larger scale sampling better diluted the impact of high loss outliers that previously skewed averages.

\section{From Data Breach to Identity Theft}
\label{sec:breach_to_victim}

In what follows, we will (1) perform a statistical test to see in what sense a major breach event may be directly correlated with a subsequence increase in reported IDT incidents, and (2) present a model that estimates a ``breach-to-victim'' conversion rate. These are then used to perform a lower-bound and an upper-bound estimate on the social cost of given major breach events, respectively.

\subsection{Testing Whether There is an Increase in IDT Following a Mega-Breach}
\label{sec:wilcoxon_methods}

To analyze the relationship between large scale data breaches and subsequent IDT victimization, we conduct a Wilcoxon signed-rank test \cite{conover1999practical}. Here we define a ``mega-breach'' as a breach that exposes at least 10 million records. For this experiment, we utilize both the augmented PRC data as well as the processed ITS data. The former is used exclusively to identify the time stamps of mega-breaches, while the later serves as the metric to measure the actual change in victimization levels. Essentially, the PRC data marks the events in time, allowing us to align the ITS victim reports into ``pre-breach'' and ``post-breach'' windows for statistical comparison. 

A Wilcoxon signed-rank test was selected to test whether or not the months following a mega-breach exhibit a statistically significant increase in the number of reported IDTs. This test was chosen due to outliers in the data that violate the normality assumptions required for t-tests. To satisfy the Wilcoxon requirement for independent pairs of observations, breaches occurring within 3 months of one another were treated as a single compound event. In such instances, the event month $T_0$ is defined as the month of the initial breach in the cluster. This consolidation resulted in 19 distinct mega-breach events identified in the augmented PRC data.

We used a fixed six month window ($T_0-6$ to $T_0-1$) to measure the baseline median victimization level prior to each mega-breach. Because IDT is rarely discovered at the exact moment of a breach, we conducted a longitudinal sweep across varying ``discovery lags'' ($i$), which we define as the time delay (in months) between data exposure (marked by PRC) and IDT discovery (measured by ITS). Each choice of the $i$ value then defines a ``post-breach'' window of 6 months. For example, if a mega-breach event occurred in May, and if we consider a 1-month discovery lag, then any IDT occurring between June and November of that year would fall into the post-breach window. 

We consider a range of possible discovery lags, $i \in \{0, \dots, 8\}$, and for each value $i$ we compared the pre-breach baseline to the six-month post-breach observation window starting at $T_0 + i$. For each discovery lag value, we performed the Wilcoxon signed-rank test for the following hypotheses:

\begin{enumerate}
    \item $H_0$ (Null Hypothesis): The median rate of IDT discovery in the six-month window following a mega-breach (delayed by lag $i$) is not significantly higher than the median rate in the six-months preceding it.
    \item $H_1$ (Alternate Hypothesis): The median rate of IDT discovery in the six-month window following a mega-breach (delayed by lag $i$) is significantly higher than the median rate in the six-months preceding it.
\end{enumerate}

\begin{figure}[htbp]
    \centering
    \begin{subfigure}[b]{0.48\textwidth}
        \centering
        \includegraphics[width=\textwidth]{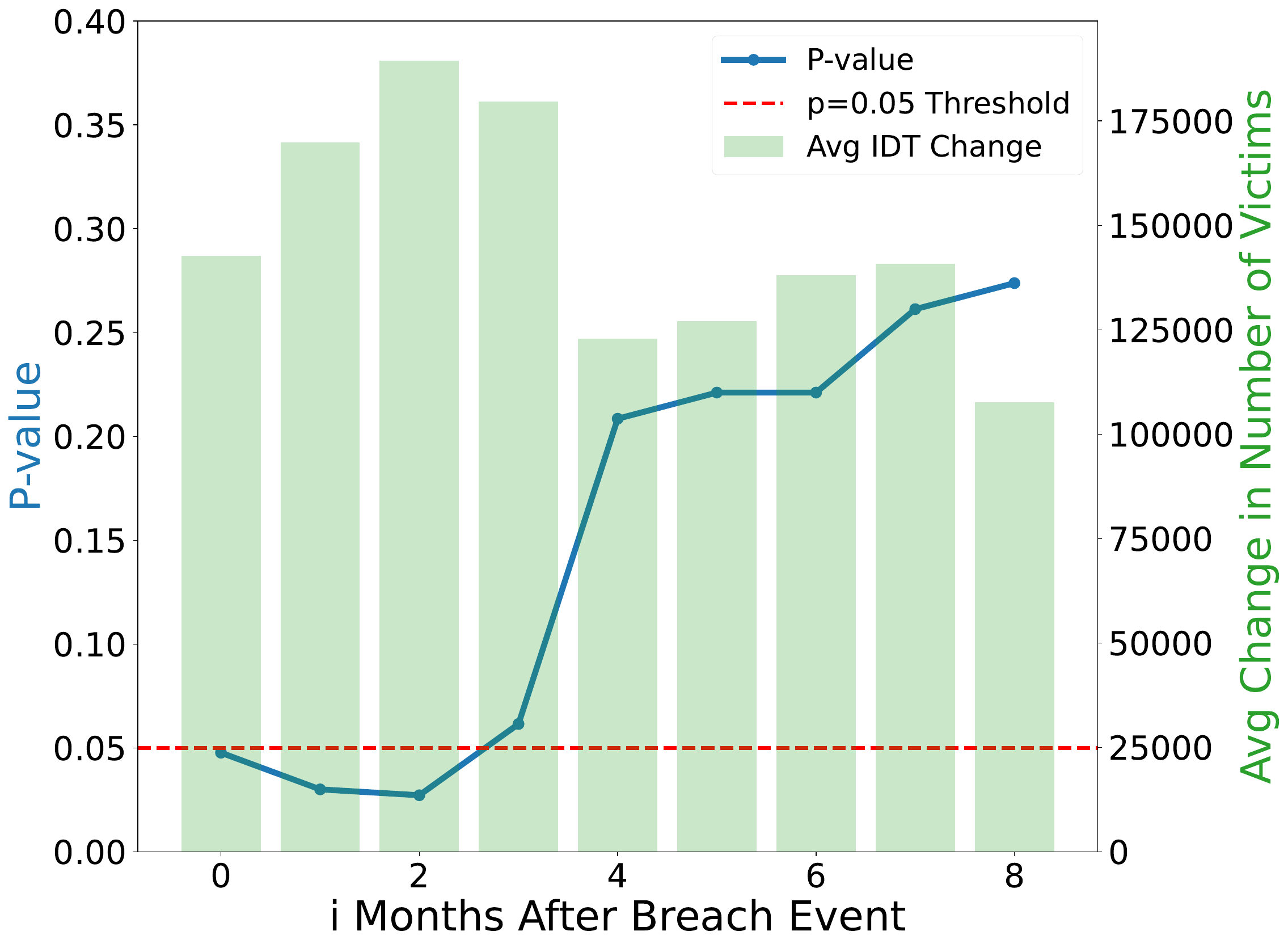}
        \caption{Augmented PRC data.}
        \label{fig:wilcoxon_raw}
    \end{subfigure}
    \hfill
    \begin{subfigure}[b]{0.48\textwidth}
        \centering
        \includegraphics[width=\textwidth]{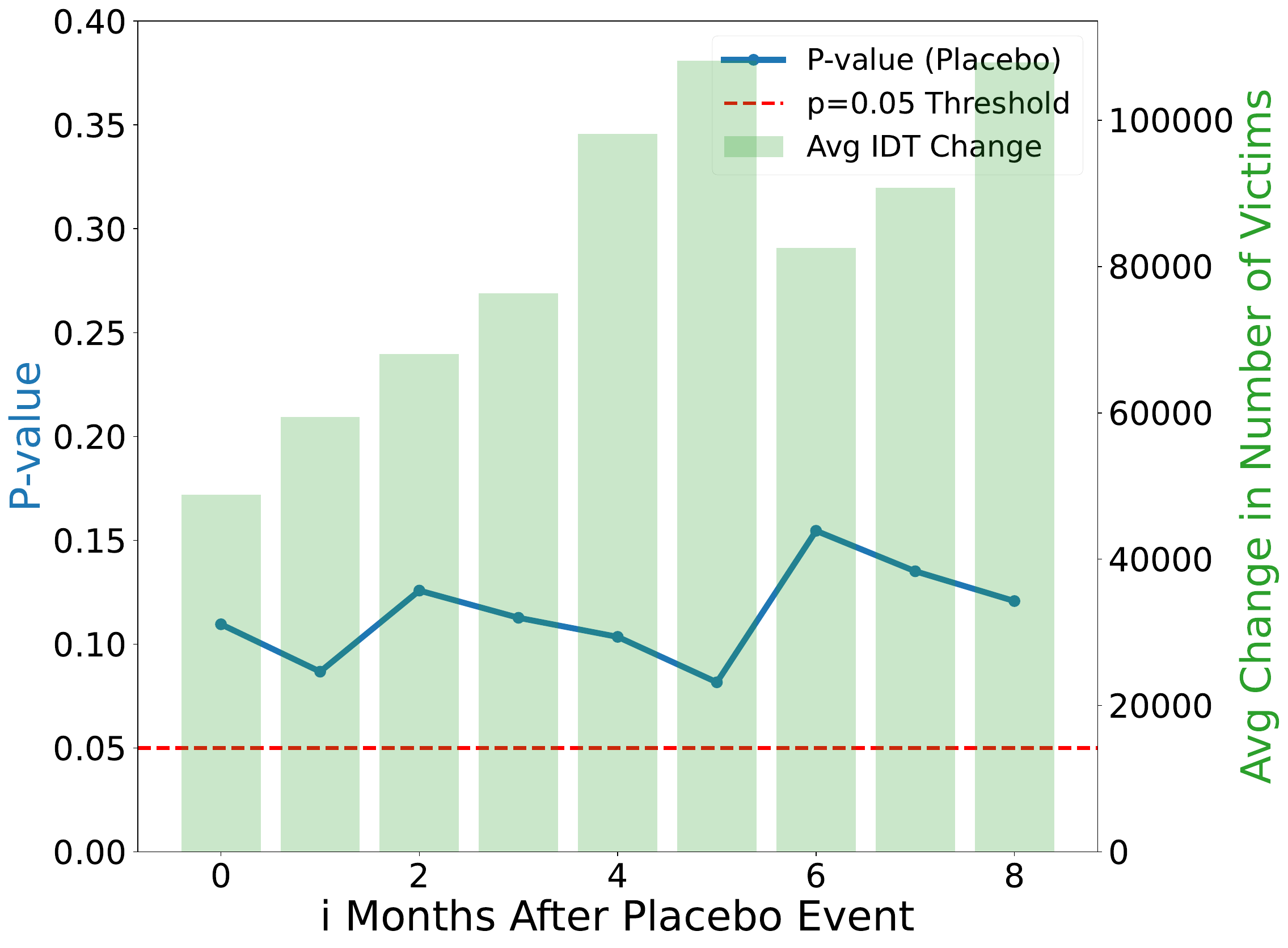}
        \caption{Placebo (No mega-breach event).}
        \label{fig:wilcoxon_placebo}
    \end{subfigure}
    
    \caption{Wilcoxon signed-rank test results: Comparison between augmented PRC data and placebo results.}
    \label{fig:wilcoxon_comparison}
\end{figure}

To validate that the observed spikes are uniquely tied to mega-breaches rather than general trends, we also conducted a placebo test. In this control experiment, we identified all months in our dataset that did not meet the 10 million record mega-breach threshold. We then applied the identical longitudinal sweep methodology to these placebo months, treating them as pseudo-events. Following the same 3-month consolidation rule used for the primary experiment, this identified a set of control points where no major exposure occured. We hypothesized that for these placebo events, the null hypothesis should fail to be rejected across all discovery lags (e.g., we would observe $p > 0.05$).

The results of the longitudinal sweep for both the augmented PRC data and the placebo control are shown in Figure \ref{fig:wilcoxon_comparison}. In these plots, the blue line tracks the $p$-value across varying discovery lags, while the green bars indicate the average change in the estimated number of monthly victims (calculated from the ITS  data), compared to the pre-breach baseline. When examining the results in Figure \ref{fig:wilcoxon_raw}, we observe a clear window of statistical significance: the null hypothesis is rejected ($p<0.05$) for discovery lags of 1 and 2 months. The analysis reveals that the lowest $p$-value occurs for a lag of 2 months. During these periods of statistical significance, the estimated number of IDT victims rises by approximately 175,000 to 190,000 individuals per month compared to the pre-breach baseline. Notably, the $p$-value rises above the 0.05 threshold by the third month, suggesting that the shock of a mega-breach on reported victimization levels dissipates relatively quickly. These findings demonstrate a significant, though time-limited, correlation between massive data exposures and surges in IDT reports. Conversely, the results of the placebo test shown in Figure \ref{fig:wilcoxon_placebo} confirm that when no mega-breach occurs, there is no statistically significant increase in identity theft. Across the entire range of discovery lags ($i \in \{0, \dots, 8 \}$), the $p$-value remains consistently above the 0.05 threshold, often exceeding 0.10.

The result of this Wilcoxon testing will directly inform how we construct a lower-bound on the estimated social cost of a mega-breach in Section \ref{sec:breach_social_cost}, by focusing exclusively on the immediate aftermath of the breach, indicated by the spike in reported IDT as seen above, even though many of the exposed records may lead to IDTs months and years later, which blends into the background and becomes part of the pre-breach ``median'' of the next event. 

\subsection{A Breach-to-Victim Conversion Model}
\label{sec:conversion_rate}

In reality, the impact of a data breach can be long-lasting: exposed records will remain somewhere on the dark web indefinitely waiting for the next exploiter; many unwitting consumers whose records were exposed in breaches never bother to take appropriate actions such as updating their passwords or freezing their credit. In this section we develop a model that attempts to capture this long-term impact, and this model will then directly inform the construction of an upper-bound estimate of the social cost of a mega-breach in Section \ref{sec:breach_social_cost}.

\begin{figure}[htbp]
    \centering
    \begin{minipage}[t]{0.48\textwidth}
        \centering
        \includegraphics[width=\textwidth]{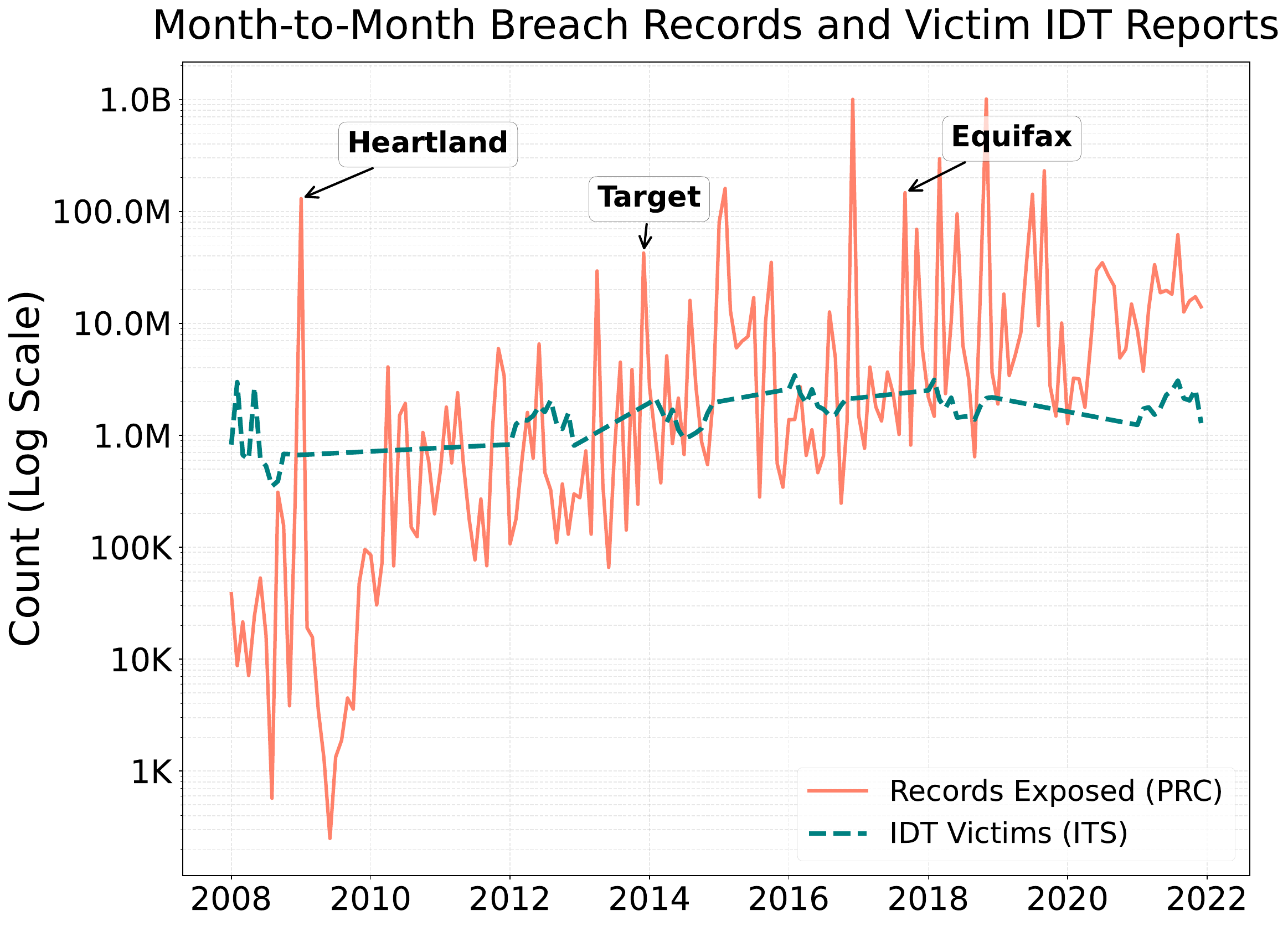}
        \caption{Number of records exposed (PRC) and number of reported IDT victims (ITS) over the study period.}
        \label{fig:section7_fig5}
    \end{minipage}
    \hfill 
    \begin{minipage}[t]{0.47\textwidth}
        \centering
        \includegraphics[width=\textwidth]{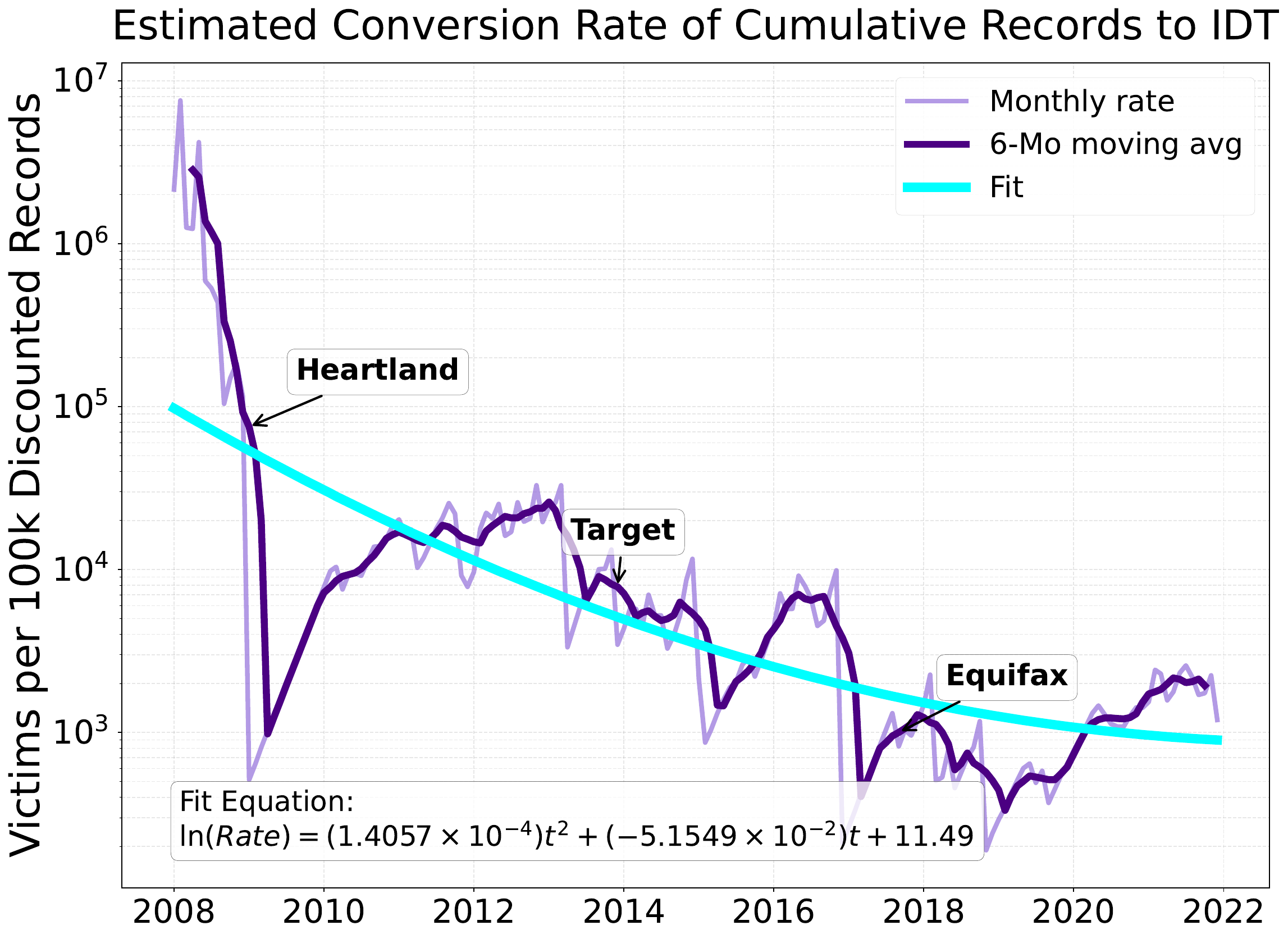}
        \caption{Analysis of conversion rates from data being breached to becoming an IDT victim. }
        \label{fig:conversion_rate}
    \end{minipage}
    
\end{figure}

Figure \ref{fig:section7_fig5} shows the monthly counts for both records exposed (the augmented PRC data, in orange) and reported IDT victims (the processed ITS data, in teal). Unlike cumulative metrics often found in industry reports, these figures represent month-to-month totals, allowing for a direct comparison of how record exposure spikes correlate with surges in reported IDTs over time (detailed in Section \ref{sec:wilcoxon_methods}). We specifically highlight three mega-breaches: Heartland Payment Systems (January 2009 \cite{USCourts_Heartland}), Target (December 2013 \cite{Target_Breach}), and Equifax (September 2017 \cite{House_Equifax}). These three incidents appear as prominent spikes on the record exposure curve in Figure \ref{fig:section7_fig5}. As can be seen, while the volume of exposed records exhibits volatility over the 13-year study period, it is generally increasing over time. 
The number of successful IDT reports follows a more stable trajectory. 

We define an estimated conversion rate of cumulative breached records (up to month $t$) to IDT (in month $t$), denoted by $\mathcal{C}_t$. Here, $t$ represents the number of months since the start of the study period ($t=0$ for January 2008). This method acknowledges that IDT in any given month is rarely the result of a single isolated breach; rather, it is drawn from a longitudinal supply of previously compromised records that remain ``fresh'' or exploitable on the criminal market. We assume that compromised/stolen data has an infinite ``shelf-life'' but with diminishing utility over time as victims change passwords, credit cards expire, or financial institutions improve fraud detection. To model this, we define the total available pool of exposed data at time $t$ as a discounted cumulative sum, $D_t$, calculated recursively as follows, where $M_k$ represents the monthly record exposure for month $k$ (from the augmented PRC dataset) and $\alpha$ is a discount factor representing the decreasing utility of stolen data over time: 
\begin{equation}D_t = \sum_{k=0}^{t} \alpha^{t-k} M_k~. 
\end{equation}
In our experiments we will set $\alpha = 0.8$, which implies that roughly 20\% of the utility of a leaked record is lost each month. It should be noted that this particular choice of the discount factor value, while arbitrary, is not critical: this modeling step is followed by a regression (discussed shortly below), so any change in this choice will largely be compensated by the fit equation. This is because the log-quadratic regression essentially treats $\alpha$ as a scaling parameter; while different discount factors may shift the absolute magnitude of the available data pool, they do not fundamentally alter the longitudinal trend of the conversion rate.

The conversion rate is then expressed as the number of reported victims (from the ITS data) at time $t$ for every 100,000 discounted cumulative records available on the market by month $t$. 

\begin{equation}\mathcal{C}_t = \frac{\text{(Number of IDT Victims)}_t}{D_t} \times 100,000~. 
\end{equation}

The results of applying this model are shown in Figure \ref{fig:conversion_rate}, where we calculated this conversion rate at monthly intervals (light purple) and applied a six-month moving average (dark purple) to identify long-term trends. Again, we explicitly highlight the Heartland, Target, and Equifax breaches as consistent reference points to allow for a direct comparison between raw record exposure in Figure \ref{fig:section7_fig5} and conversion rate in Figure \ref{fig:conversion_rate}. The localized peaks in the conversion rate primarily coincide with the ITS survey waves. This suggests that the volatility is largely driven by the periodic transition from interpolated victim estimates back to raw weighted survey data. The conversion rate in Figure \ref{fig:conversion_rate} exhibits a generally decreasing trend; as the cumulative pool of stolen data $D_t$ grows by orders of magnitude, the marginal ``yield'' or criminal value of a single record decreases. This is also reflected in Figure \ref{fig:section7_fig5}, where the volume of records exposed in later years is significantly higher than in initial years; however, because the number of successful IDTs did not grow at a matching rate, the resulting conversion rate dropped.

To model this trend for our purpose, we performed a time-based log-quadratic regression on the six-month moving average. This model estimates the expected conversion rate as a function of time $t$, where $t$ is again defined as the number of months since January 2008. The resulting fit equation is:
\begin{equation}
\label{eqn:fit}
    \ln(\mathcal{C}_t) = (1.41 \times10^{-4})t^2 + (-5.15 \times 10^{-2})t + 11.49~. 
\end{equation}

To retrieve the (instantaneous) conversion rate $\mathcal{C}_t$ for a breach, we first calculate the $t$ value (months after January 2008) for the breach given its date. Plugging this $t$ value into Equation \ref{eqn:fit} and taking the exponential (to undo the natural logarithm) yields the conversion rate for that specific breach. For example, the conversion rates at the times of the Heartland  ($t=12$), Target ($t=71$), and Equifax ($t=116$) breaches were 75,205, 7,841, and 1,001 IDT victims per 100,000 discounted exposed records, respectively.

In this model, the negative linear coefficient ($-5.15 \times 10^{-2}$) represents the overall downward trend on conversion rate over time, likely due to improved fraud detection systems. The quadratic coefficient ($1.41 \times10^{-4}$) accounts for the slight flattening of the curve in recent years, reflecting a stabilizing risk. Equation \ref{eqn:fit} allows us to estimate the expected victim yield for specific breach events based on their historical timing. We will use this model in Section \ref{sec:breach_social_cost} to provide upper-bound estimates on the social costs of specific breach events. 

\subsection{The Social Cost of Mega-breaches}
\label{sec:breach_social_cost}
To apply our social cost model to real-world mega-breach events, we performed an analysis of the three landmark security events highlighted earlier: the 2009 Heartland Payment Systems breach, the 2013 Target breach, and the 2017 Equifax breach. We chose these three incidents because they were large breaches that compromised significant portions of the U.S. population (around 130 million for Heartland \cite{USCourts_Heartland}, around 40 million for Target \cite{Target_Breach}, and around 147 million for Equifax \cite{House_Equifax}), and because they have well documented corporate settlement figures that allow direct comparison with our social cost estimates. For each of these case studies, we establish both a lower bound and an upper bound estimate. 

\begin{enumerate}[label=(\alph*)]
\item \textbf{Lower Bound (Empirically Measured Short-Term Impact):} This estimate utilizes the Wilcoxon-signed rank analysis presented earlier to measure the statistically significant increase in IDT victims discovered in the 6-month window (with a discovery lag of $i=2$) immediately following the breach. This essentially means if a breach occurred at time $t$, we measure the average increase in IDT victims in months $t+2$ to $t+8$.
By multiplying the increase in number of measured identity victims post-breach with the social cost per victim at the time of the breach, we calculate an estimate of the breach's social cost. This is a lower-bound estimate because it is based on the short-term impact of the breach empirically measured by the increase in the number of victims immediately following the breach. 

\item \textbf{Upper Bound (Projected Long-Term Impact)}: This estimate utilizes our log-quadratic fit (Equation \ref{eqn:fit}) and the discount factor $\alpha = 0.8$ to project the total victim yield for a specific breach as well as the life-time cost incurred by the breach. To maintain consistency with our empirical findings, the projection begins after a discovery lag of two months. For a breach of size $B$ that occurred at time $t$, the remaining ``fresh'' records from the breach is estimated to be $B \cdot \alpha^{k-t}$ from month $k=t+2$ through the end of the study period; applying to this the monthly conversion rate $\mathcal{C}_k$ we obtain the estimated number of victims in month $k$ attributed to the breach; summing up these estimates then gives us the projected total number of victims over the long term. By weighting the monthly victim yield by an interpolated monthly social cost of IDT $S_k$, we obtain the estimated social cost in month $k$ attributed to the breach. Summing these values across the study period then provides the total projected social cost of the incident over the long-term. 
This serves as an upper bound because it accounts for the victimization that extends beyond the immediate post-breach shock, assuming the conversion of stolen records encompasses the full scale of the exposure over time.
\end{enumerate}

\subsubsection{The Lower Bound Estimate}
For our lower bound estimate, we calculated the change in the number of IDT victims statistically attributable to each breach. This was derived from the Wilcoxon signed-rank test as described in Section \ref{sec:wilcoxon_methods},  using a discovery lag of 2 months. We compared the median monthly victimization count in the six months prior to the breach against the median monthly victimization count in the six months immediately following the breach with a two-month delay in between. The difference between these two periods represents the estimated increase in victim count that coincides with the breach window. 
For each breach, we identified the statistically significant victim count increase following the breach. We then translated these victim counts into the  cost estimate by multiplying the social cost data from Table \ref{tab:social_costs_by_year}:  
$$\text{Total Social Cost} = \text{Social Cost per Victim} \times \text{Number of Victims}$$

Using the social cost per victim from the years closest to the incidents, we noted \$1,110.31 for the 2008 period for the Heartland breach, \$853.41 for the 2014 period for the Target breach, and \$244.40 for the 2018 period for the Equifax breach. 
More specifically, 
\begin{enumerate}
\item For the Heartland breach, our analysis identified a statistically significant median increase of 88,956 victims per month in the post-breach window. Multiplying this by the six-month observation period and the 2008 social cost per victim of \$1,110.31 yields a total social cost of approximately \$592.7 million. Compared to its cumulative \$107 million settlement \cite{heartland_amex_2009, heartland_mastercard_2010, heartland_visa_2010, heartland_consumer_2012}, the social cost was over five times higher than the institutional payout.

\item For the Target breach, our analysis identified a statistically significant median increase of 59,714 victims per month, resulting in an estimated 358,284 total victims in the post-breach window compared to the pre-breach baseline. Applying the \$853.41 social cost per victim from the 2014 period results in a total estimated social cost of approximately \$305.8 million.  In contrast, Target's highly publicized multi-state settlement for the breach was only \$18.5 million \cite{TexasAG_TargetSettlement}. This indicates that the realized social harm was about 16 times higher than the institutional payout. 

\item For the Equifax breach, our analysis found a median increase of 179,889 victims per month, leading to an estimated 1,079,334 victims following the breach. Using the \$244.40 social cost per victim from the 2018 period, the estimated total social cost is \$263.8 million. Despite the severity of this exposure, Equifax's global settlement was capped at \$700 million \cite{FTC_EquifaxSettlement}. In this specific instance, the compensation settlement actually exceeded the estimated immediate social costs. This reversal, where the settlement exceeds the estimated social costs, may reflect the saturation effect observed in our earlier analysis. By 2017, many victims were likely already compromised, reducing the marginal impact of a new breach. 
\end{enumerate}

We acknowledge that this methodology relies on the attribution of a change in the number of IDT victims to specific breach events. By isolating the specific window surrounding a breach, we have attempted to provide a  causal estimate. It is also important to note that our current model treats all compromised records as equal units of risk, yet we know that breaches involving sensitive personal information such as SSNs inherently carry a higher potential for long-term damage than those involving changeable credentials like credit card numbers. While our analysis in the Appendix \ref{app:social_cost_breach} briefly touches on the higher costs associated with SSN exposure, the case study calculations above do not apply a specific weight to the breached records that reflects how sensitive the records are. Future refinements of this model should aim to quantify this distinction, as the lower social cost for Equifax may mask a higher cost due to SSN exposure.

\subsubsection{The Upper Bound Estimate}
The upper bound victim yield for a breach occurring at time $t$ is calculated by summing the anticipated victim yield over the remaining months of the study period. As detailed in \ref{sec:conversion_rate}, this approach assumes that every compromised record has a utility that decays as security measures are implemented or the data becomes stale. We use the same discount factor $\alpha = 0.8$ assumed in \ref{sec:conversion_rate} to represent the monthly retention of a record's exploitative value. To maintain consistency with our empirical findings regarding discovery lags in \ref{sec:wilcoxon_methods}, the projection begins accumulating victims at month $t+2$. The total projected victims for a breach of size $B$ occurring at time $t$ is defined as $\mathcal{V}_B$: 

\begin{equation}
    \mathcal{V}_B = \sum_{k=t+2}^{T} (B \cdot \alpha^{k-t}) \cdot \frac{\mathcal{C}_k}{100,000}~, 
\end{equation}
where $\mathcal{C}_k$ is the instantaneous or immediate conversion rate at month $k$, defined from our log-quadratic fit (Equation \ref{eqn:fit}), and $T$ is the terminal month of the study period ($T=167$, December 2021). 
We then multiply this number of projected victims by a time-varying social cost $S_k$ as follows:

\begin{equation}
\label{eq:social_cost_breach_upper}
\text{Total Estimated Social Cost} = \sum_{k=t+2}^{T} \left( B \cdot \alpha^{k-t} \cdot \frac{\mathcal{C}_k}{100,000} \cdot S_k \right)~, 
\end{equation}
where $S_k$ is the social cost per victim in month $k$, derived through linear interpolation of the social cost estimates presented in Table \ref{tab:social_costs_by_year}. Because the ITS only provides specific survey years, we utilized a piecewise linear function to estimate the social cost for intervening months. This approach acknowledges that the factors influencing social cost, such as the efficacy of bank fraud detection and the value of lost time, shift gradually over time rather than in abrupt annual steps. By embedding this time-varying social cost into our summation, we ensure that victims occurring later are valued at the contemporary victim social cost relevant to their time of discovery, rather than a static cost fixed at the date of initial record exposure. 

Applying this model to our three landmark mega-breaches, we obtain the following: 

\begin{enumerate}
\item For the 2009 Heartland Payment Systems breach, the model projects approximately 171.8 million victims resulting from the mega-breach event. 
Using Equation \ref{eq:social_cost_breach_upper}, we obtain an upper bound social cost estimate of the Heartland breach at \$179.1 billion, which is approximately 1,673 times higher than Heartland's cumulative settlement of \$107 million \cite{heartland_amex_2009, heartland_mastercard_2010, heartland_visa_2010, heartland_consumer_2012}.  This reflects the vulnerability of a breached record in early years, where a single record has a high probability of successful conversion into financial crime. 

\item For the 2013 Target breach, we estimate a projected yield of 5.49 million victims and an upper bound social cost of \$4.06 billion. This figure is over 219 times higher than Target's \$18.5 million settlement \cite{TexasAG_TargetSettlement} highlighting how the majority of social costs are externalized onto the public. 

\item 
Finally, for the 2017 Equifax breach, the model projects 6.95 million victims resulting from the breach. Despite the high volume of records exposed in this breach (147 million), the increased market saturation and improved fraud detection in 2017 results in a lower per-record victim yield than in previous years. Our upper bound estimated social cost for the Equifax breach is \$1.72 billion. This projection still more than doubles Equifax's \$700 million settlement cap \cite{FTC_EquifaxSettlement}, suggesting that even in a saturated market, the long-term impact of data exposure can create a multi-billion dollar social burden.

\end{enumerate}

\subsubsection{Summary}
A summary of our upper and lower bound estimates is displayed in Table \ref{tab:case_study_summary}, where we see a profound disparity between corporate liability and realized social harm, even when this harm is rather narrowly defined. In two out of the three case studies (Heartland and Target), even our lower bound social cost estimate (which only captures the short-term impact) exceeded the breach's settlement. The upper bound projections reveal a far more severe social cost, as the projected long-term impact from the Heartland, Target, and Equifax breaches exceeded their respective settlements by factors of approximately 1,673, 219, and 2.

These results also provide empirical evidence of the saturation effect within the stolen data market. While the total number of records exposed in the 2017 Equifax breach was greater than that of the 2009 Heartland breach, the projected social harm for Equifax is two orders of magnitude lower. This decline in marginal yield reflects the diminishing utility of new data as the national supply of compromised identifiers reaches saturation. Nevertheless, the \$1.72 billion upper bound social cost for Equifax still more than doubles the \$700 million settlement cap, suggesting that the long-term exposure of data in even a saturated market imposes a persistent financial burden on the consumer that current regulatory penalties fail to fully address.

\begin{table}[h]
\centering
\caption{Comparison of Corporate Settlements and Estimated Social Costs for Mega-Breach Events}
\label{tab:case_study_summary}
\begin{tabular}{lcccccc}
\hline
\textbf{Case Study} & \textbf{Date} & \textbf{$t$} & \textbf{Records ($M$)} & \textbf{Settlement} & \textbf{Lower Bound} & \textbf{Upper Bound} \\ \hline
Heartland & 2009-01 & 12 & 130.0 M& \$107 M & \$592.7 M & \$179.1 B \\
Target & 2013-12 & 71 & 40.0 M& \$18.5 M & \$305.8 M & \$4.06 B \\
Equifax & 2017-09 & 116 & 147.0 M& \$700 M & \$263.8 M & \$1.72 B \\ \hline
\end{tabular}
\end{table}
\section{Discussion}
\label{sec:discussion}

\subsection{Limitations and Future Work}
\label{sec:limitations}

\paragraph{Limitations of the data sources}
While this study provides a comprehensive 13-year analysis of IDT, several limitations must be acknowledged. First, our IDT data (from the ITS survey) relies on self-reported survey responses, which may be subject to recall bias or a lack of technical knowledge regarding how information was obtained. As noted in Table \ref{tab:theft_summary_avg}, nearly 74\% of victims were unaware of the specific method used to compromise their data, limiting our ability to definitively attribute individual thefts to data breaches. Second, the structural redesign of the NCVS in 2016 introduced sampling inconsistencies as mentioned earlier: the 41\% increase in sample size and the recalibration of weights based on the 2010 Census instead of the 2000 Census could have contributed to the drop in observed OOP losses across all categories of theft. While we harmonized variables to the best of our ability, these methodological survey shifts limit our longitudinal comparisons between pre- and post-2016 survey waves. 
The IDT data also does not contain geographical information to allow us to account for regional price variations; as a result, our social cost model utilized fixed national averages to monetize professional services and healthcare. Perhaps the biggest limitations of the PRC data is the significant volume of  incidents with zero or undisclosed record exposure, which necessitated the augmentation with the HHS reports and state-level filings, a process that is inherently noisy.

\paragraph{Limitations of our methodology}
Our attribution of IDT to specific mega-breaches relies on statistical correlation (Wilcoxon tests) rather than direct causal tracing. While we identify significant increases in reported IDTs following a mega-breach, the number of victims we attribute to major  breach events represent a modeled potential outcome.
Furthermore, our long-term social cost projections are naturally constrained by the time boundaries of the study. Because our calculations end at December 2021, the upper-bound estimates for more recent events, such as the 2017 Equifax breach, only reflect the realized portion of social cost within the study period. The true long-term cost likely extends well beyond this terminal month. As more waves of the ITS survey are released, this work can be extended to include the new data. Furthermore, as mentioned earlier, our social cost model is narrowly defined, constrained by what is captured in the ITS surveys: it focuses exclusively on harms resulting from identity theft and subsequent financial or psychological distress. Consequently, our estimates do not account for other dimensions of externalized harm, such as broader privacy costs, risks to national security, or the downstream impact of data being used for non-IDT cybercriminal activities.

\paragraph{Future directions}
While our log-quadratic conversion model provides a robust fit for the conversion rate, it utilizes a uniform discount factor $\alpha$ to represent the decay of record utility. In reality, the ``shelf-life'' of stolen data likely varies based on the sensitivity of the identifiers involved. For example, Social Security numbers may maintain their exploitative value longer than revocable credentials like credit card numbers. Modeling this serves as a potential future extension of this study. 
Another line of future work involves improving the imputation process for breaches with undisclosed record counts. An alternative to our procedure (detailed in Section \ref{sec:imputation_of_undisclosed}) is to calculate a category-specific annual baseline $n_u^j$ for each of the $k$ breach types. Under this more granular approach, the imputed records for an incident $i$ of type $j$ in year $y$ would be defined as $R_{i_j}=n_u^j/|I_{u,y}^j|$. While this would account for the different typical magnitudes associated with various breach types (e.g. malicious attacks versus accidental disclosures), we maintained the global $n_u$ approach to provide a stable, conservative baseline for the overall number of undisclosed records. More effort can be made to identify and integrate more diverse datasets beyond the ITS and PRC datasets to better quantify the number of exposed records and number of IDTs. Finally, with the rise of generative AI in phishing and impersonation scams, research is needed to quantify the costs associated with deepfake-related IDT.

\subsection{A Dynamic Saturation Model to Estimate the Number of Unique Individuals Exposed in Breaches}
\label{sec:estimating_number_records}

The analysis in Section \ref{sec:breach_social_cost} shows a potential saturation of the stolen records market, whereby even as many more records are compromised, the number of new incidents of IDT grows much slower. Below we attempt to explicitly model this saturation through a de-duplication of the record exposure.

The objective of this model is to transform the monthly raw number of exposed records into an estimated number of {\em unique individuals whose data was compromised}. Specifically, this model incorporates the U.S. population for ages 16+ (to match the ITS survey, which was administered to those 16+) and calculates a dynamic saturation index $\mu_t$ that tracks the portion of the population already compromised, beginning at $t=0$ and asymptotically approaching 1 as the stolen data market reaches full saturation. This ensures that as the cumulative volume of data exposure increases, the model identifies the diminishing probability that a reported record belongs to an individual whose data had not previously been compromised. 

Our model iterates using the following variables for the $t$-th period, which could be a month or a year:

\begin{enumerate}
    \item $N_t$ (U.S. Population 16+): The potential victim pool at time $t$ based on annual population averages from the U.S. Bureau of Labor Statistics \cite{FRED_Population}.
    
    \item $r_t$ (Raw Number of Records Exposed): We take the total volume of records exposed in period $t$, $r_t$, directly from the augmented PRC data, and scale it to $\theta r_t$ with $\theta = 1.75$, to account for under-reporting as described in \cite{Graves2018Should}.  Here, we use ``records'' and not ``individuals,'' because the input to the model is number of records, and the output is unique number of individuals. 
       
    \item $c_t$ (Estimated Number of Newly Exposed Unique Individuals): The estimated number of previously uncompromised individuals exposed during period $t$. 
        
    \item $C_t = \sum_{i=1}^{t} c_i$ (Estimated Number of Cumulative Exposed Unique Individuals): The total volume of estimated unique individuals at the end of period $t$. 
    
    \item $\mu_t$ (Saturation Index): The ratio of the number of cumulative unique individuals at the end of period $t$ over the potential victim pool. 
    
    \item $\gamma_t$ (Dynamic Identity Density): The probability that a record in a new breach represents a newly unique individual, where $\gamma_t = \gamma_0 \times (1-\mu_{t-1})$. We set the base density as $\gamma_0 = 0.8$.
\end{enumerate}

We initialize the model at $t=0$ (corresponding to the start of 2008), where $C_0 = 0$ and $\mu_0 = 0$. Our model updates $\mu_t$ recursively. We define the update as follows, where $Y$ is a scaling factor: 

\begin{equation}
\label{eq:sat_model}
    1 - \mu_t = (1 - \mu_{t-1}) \cdot \exp\left( - \frac{\theta r_t \cdot \gamma_0 }{N_t \cdot Y } \right)~. 
\end{equation}

The role of $Y$ as a scaling factor is to define the total capacity of the identity market, effectively representing how many distinct ``identity units'' (such as different accounts) exist per person within the population. When $Y=1$, the model assumes that each individual in the population $N_t$ 
possesses exactly one ``identity unit'' that can be compromised. When $Y>1$, we scale the potential record pool of our model to $N_t \cdot Y$. This follows the assumption that one individual may have multiple distinct accounts (e.g. financial, social media, medical).

Rearranging the terms in Equation \ref{eq:sat_model} allows us to solve for the new saturation level $\mu_t$ based on the previous year's level $\mu_{t-1}$ and the influx of newly breached records $r_t$:

\begin{equation}
\mu_t = 1 - \left[ (1 - \mu_{t-1}) \cdot \exp\left( -\frac{\theta r_t \cdot \gamma_0}{N_t \cdot Y } \right) \right]
\end{equation}

This formulation ensures that even as $r_t$ grows arbitrarily large, $\mu_t$ asymptotically approaches 1 but never exceeds it. Once the new saturation level $\mu_t$ is determined, we calculate the number of newly unique individuals $c_t$ exposed during the period by taking the difference in the cumulative pool: $$c_t = (\mu_t - \mu_{t-1}) \cdot (N_t \cdot Y)$$ This value of $c_t$ represents the net number of new individuals that were compromised during this period. 

In Appendix \ref{app:sat_model}, we recreate Figures \ref{fig:section7_fig5}, \ref{fig:conversion_rate}, and \ref{fig:wilcoxon_raw} using the output of the above saturation model (estimated number of unique individuals compromised) rather than the number of records exposed. The implementation of the dynamic saturation model provides several key advantages for understanding the long-term relationship between data breaches and victimization by accounting for the diminishing marginal impact of redundant record exposure. As illustrated in the updated overlay analysis (Figure \ref{fig:saturation_fig5}), the estimated number of unique compromised individuals aligns much more closely with the trajectory of reported successful identity theft victims throughout the study period than the raw counts alone in Figure \ref{fig:section7_fig5}. Thus, this modeling approach can be used to address the data sparsity issues present in the early years of the augmented PRC dataset. 
Secondly, when calculating the month-to-month breach-to-victim conversion rate using the number of unique individuals rather than time-discounted cumulative records (Figure \ref{fig:saturation_conversion_rate}), the resulting victim conversion rate remains relatively stable, fluctuating around a 1:1 ratio. 
Finally, despite the de-duplication of the record pool, the statistical relationship between mega-breaches and IDT remains robust; the Wilcoxon signed-rank test results for the saturation model (Figure \ref{fig:wilcoxon_sat}) preserve the significance observed in the raw data, with $p$-values remaining below 0.05 for discovery lags of one and two months. Ultimately, the dynamic saturation model offers an alternative framework for measuring the ``yield'' of a breach in an increasingly crowded data market, and future work can include refining this model to account for the varying sensitivity of specific data types, such as SSNs, to better quantify long-term social harm.

\section{Conclusion}
\label{sec:conclusion}
This paper presents a novel longitudinal framework for quantifying the social cost of corporate data breaches by bridging the gap between institutional data exposure and the externalized costs faced by IDT victims.
To accomplish this, we performed a comprehensive 13-year harmonization of all six waves of the Identity Theft Supplement (2008-2021) to the National Crime Victimization Survey, pooling a sample of over 41,000 IDT victims. Our primary contribution is the development of a multi-dimensional social cost model that captures the full social cost of IDT, and the application of this model to quantify the social costs of mega-breach corporate events. Our model monetizes direct financial OOP losses, the opportunity cost of lost time, lawyer costs, and healthcare expenditures associated with the physical and emotional distress of IDT victimization. We document a significant shift in IDT patterns over the study period, noting a transition from traditional financial account fraud toward the misuse of digital credentials. We also find that the social cost of IDT is decreasing over the study period (2008-2021). 

By pairing this victim data with a multi-stage augmented chronology of corporate data breaches (primarily from the PRC data breach chronology), we performed hypothesis testing to verify the link between massive data exposure and subsequent victimization. Our Wilcoxon signed-rank analysis confirms a statistically significant increase in IDT discovery when accounting for a discovery lag of 1-2 months. Furthermore, our longitudinal ``breach-to-victim'' conversion model reveals that while the volume of exposed records is increasing, the marginal utility of a single compromised record has decreased as the data market becomes more saturated. 

The application of our model to landmark security incidents reveals a failure in current regulatory penalties to capture the full scope of damage externalized to consumers. For the 2009 Heartland and 2013 Target breaches, the social cost (even when measured through a conservative short-term lens) exceeded institutional settlements by factors of 5 and 16, respectively. While the costs calculated for the 2017 Equifax breach suggest a narrowing gap between corporate settlements and the short-term social cost, our long-term projected social cost indicates that the social cost of such massive exposure could remain in the billions, with an estimated upper-bound cost of \$1.72 billion that more than doubles the corporate settlement. Ultimately, these results suggest that as the data market reaches saturation, the individual risk per record may decline but the total social burden remains a systemic crisis.

\bibliographystyle{plainnat}
\bibliography{references}

\clearpage
\appendix

\section{Trends in Identity Theft from 2008-2021}
\label{app:longitudinal}

\begin{figure}[htbp]
    \centering
    \begin{subfigure}[b]{0.48\textwidth}
        \centering
        \includegraphics[width=\textwidth]{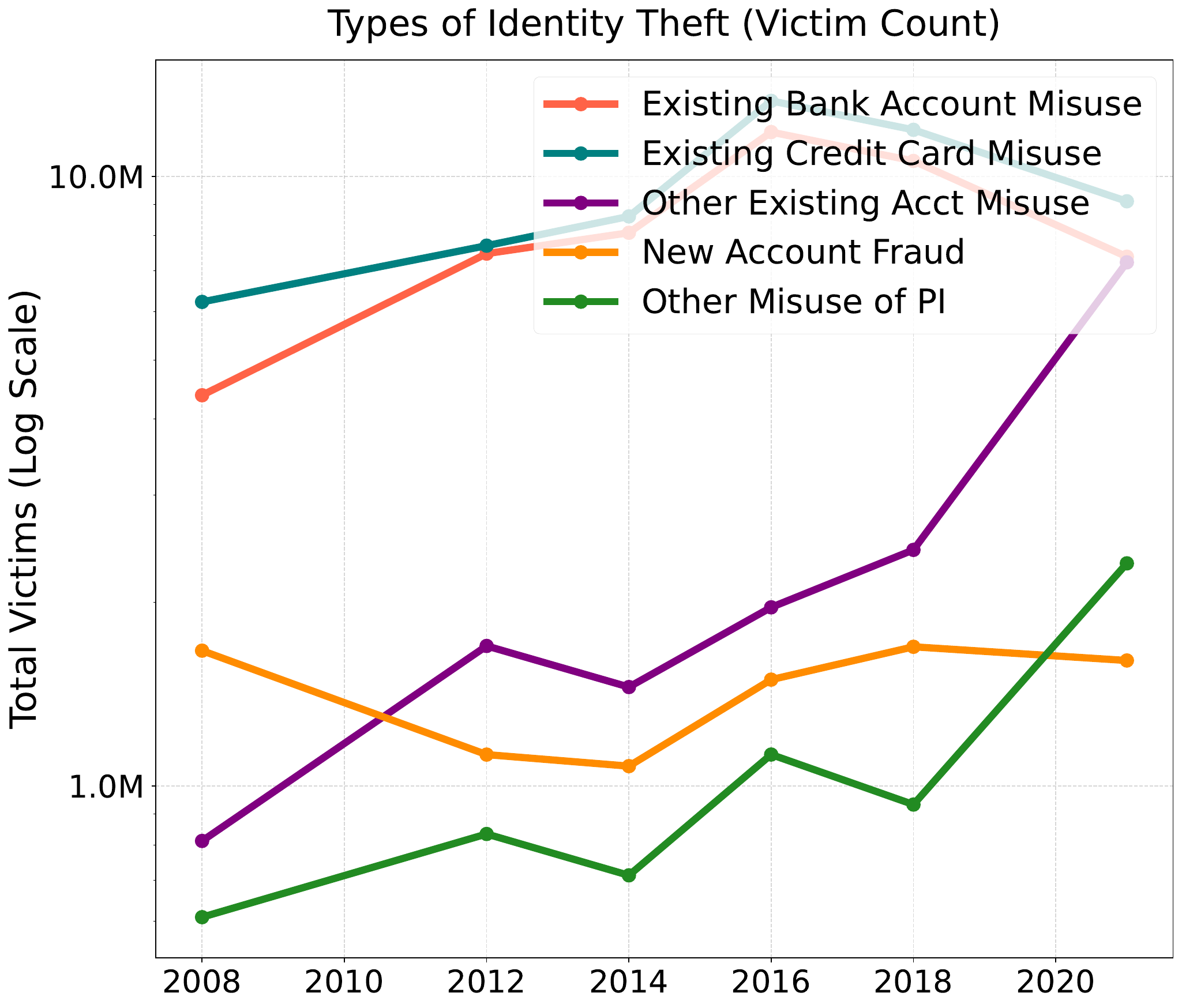}
        \caption{Number of victims categorized by the type of identity theft they suffered from.}
        \label{fig:theft_type_long_a}
    \end{subfigure}
    \hfill
    \begin{subfigure}[b]{0.48\textwidth}
        \centering
        \includegraphics[width=\textwidth]{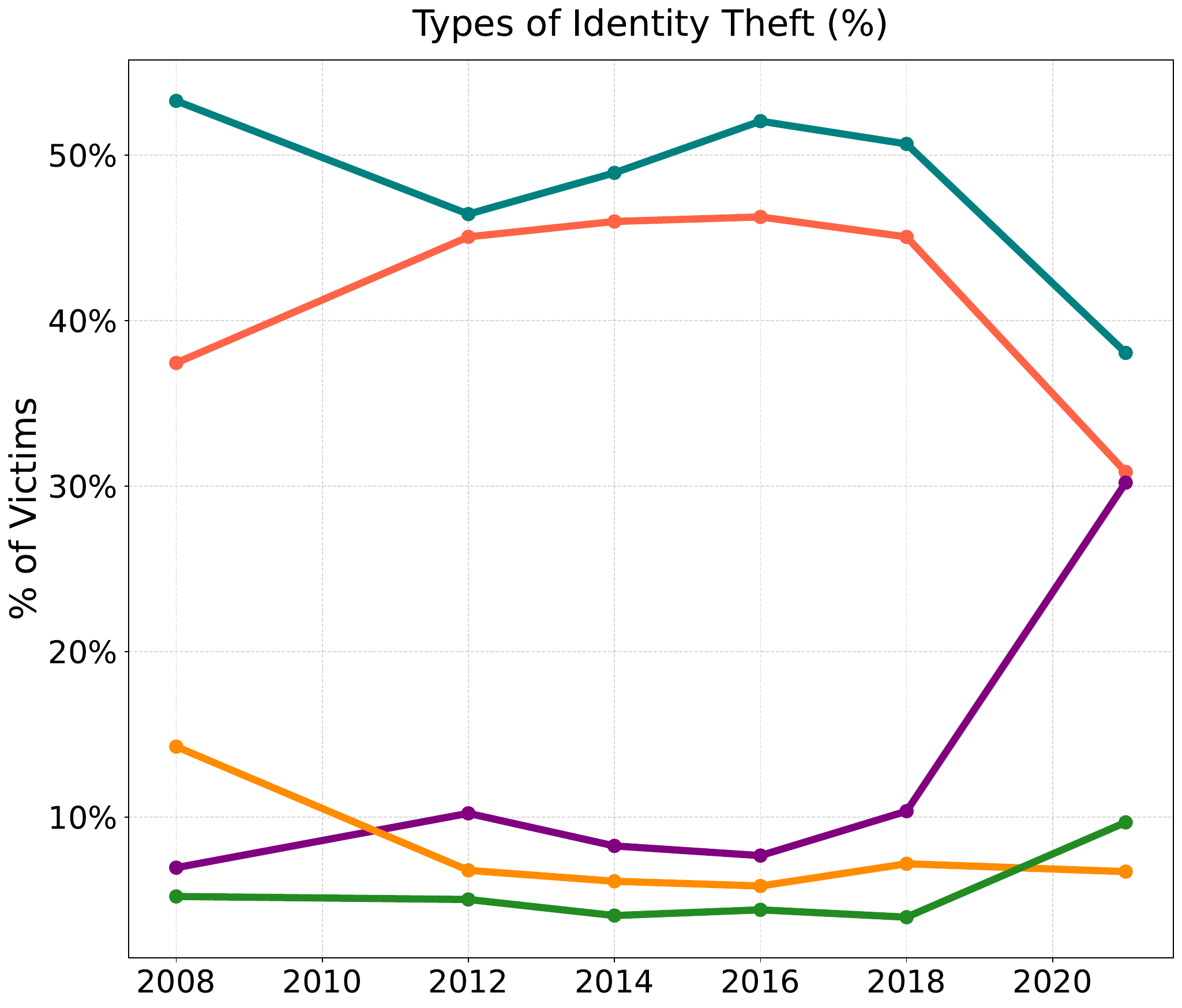}
        \caption{Percentage of victims categorized by the type of identity theft they suffered from.}
        \label{fig:theft_type_long_b}
    \end{subfigure}
    
    \caption{Analysis of theft types over the longitudinal study period.}
    \label{fig:theft_type_long}
\end{figure}

\begin{figure}[htbp]
    \centering
    \begin{subfigure}[b]{0.48\textwidth}
        \centering
        \includegraphics[width=\textwidth]{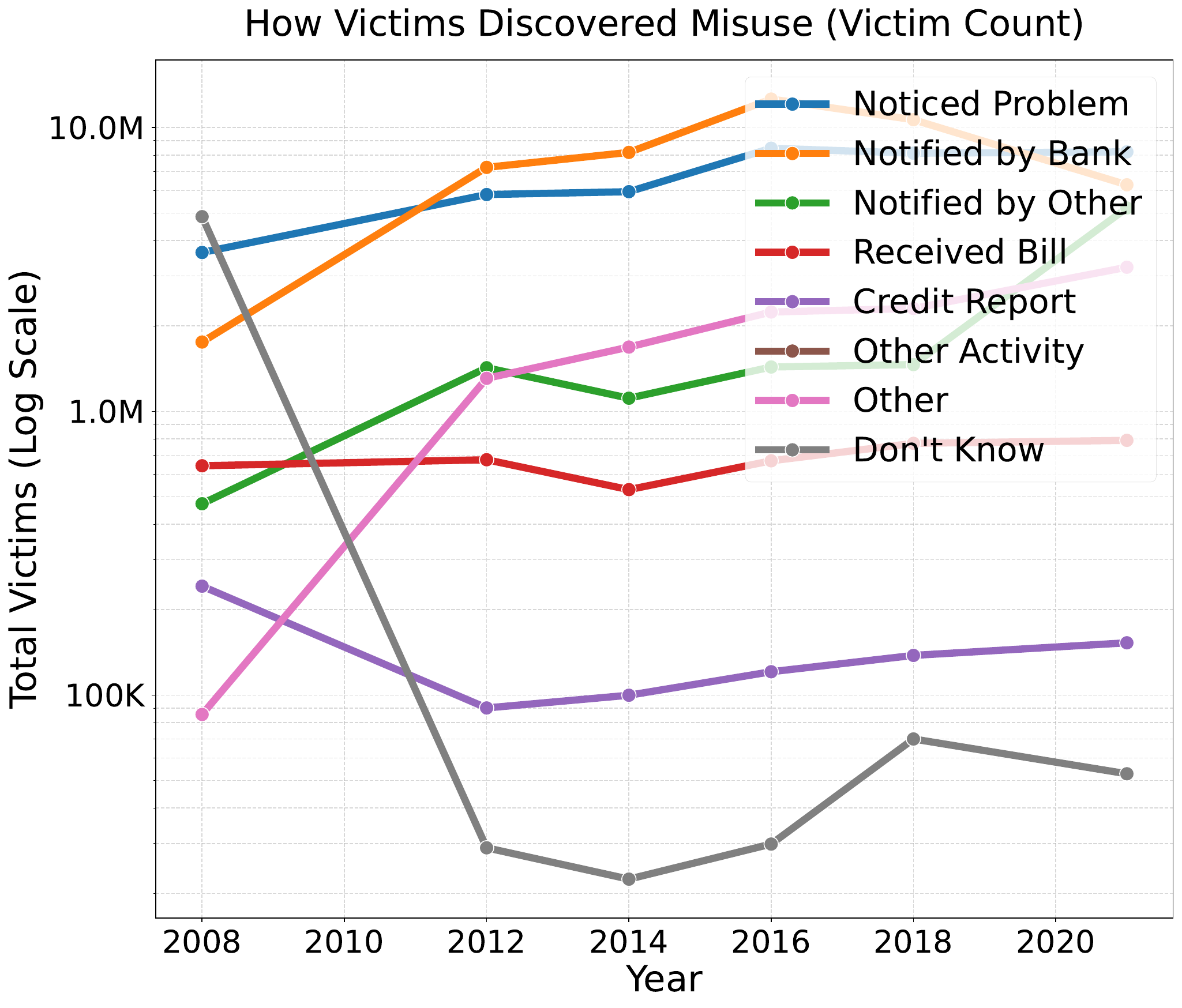}
        \caption{Number of victims categorized by their discovery method.}
        \label{fig:discovery_method_long_a}
    \end{subfigure}
    \hfill
    \begin{subfigure}[b]{0.48\textwidth}
        \centering
        \includegraphics[width=\textwidth]{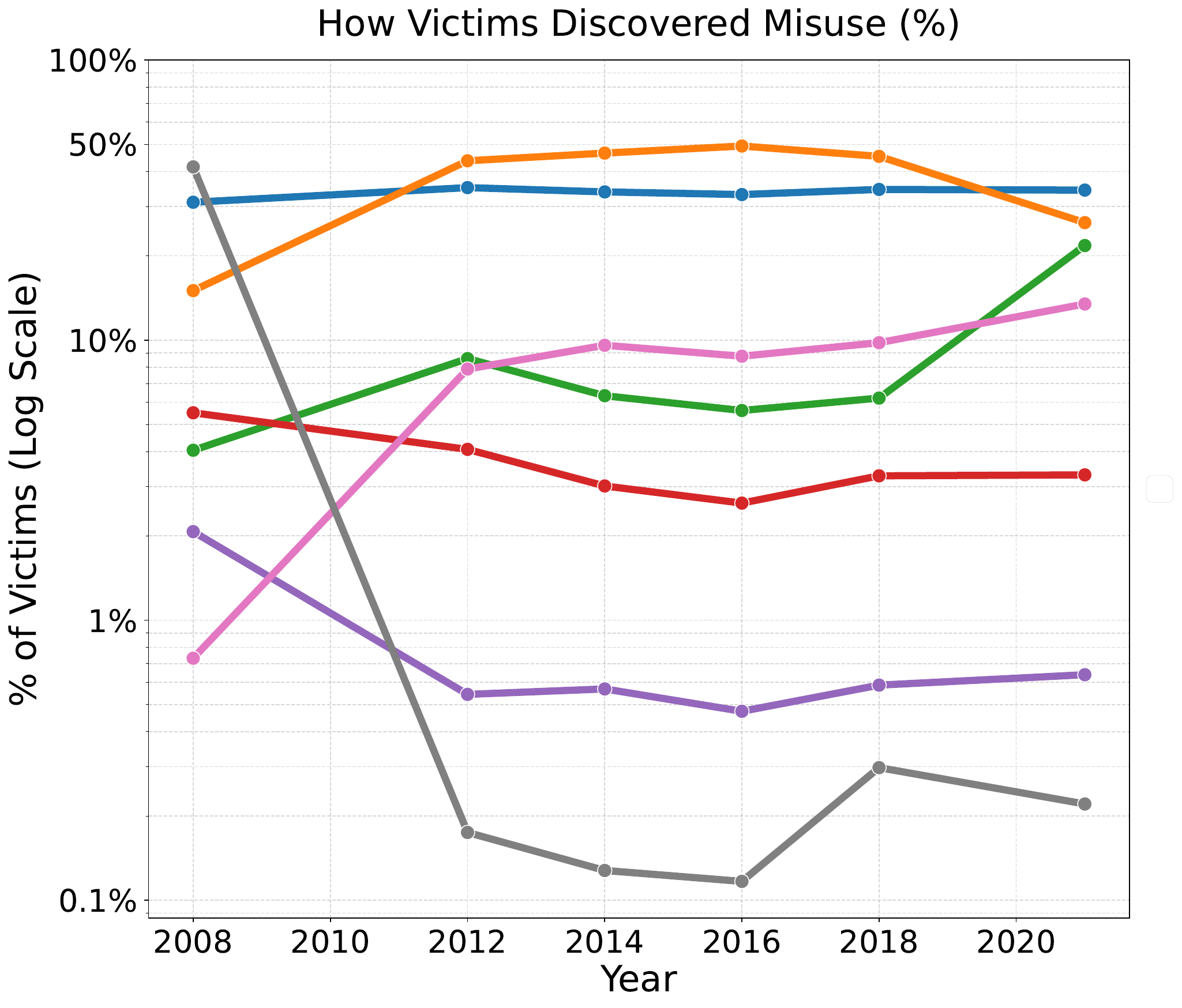}
        \caption{Percentage of victims categorized by their discovery method.}
        \label{fig:discovery_method_long_b}
    \end{subfigure}
    
    \caption{Longitudinal trends in how victims discovered the theft incidents.}
    \label{fig:discovery_method_long}
\end{figure}

Beyond the average characteristics over the entire study period of 2008 to 2021, examining the trends year by year reveals shifts in the type of theft, as well as how victims experience them. Figures  \ref{fig:theft_type_long} and \ref{fig:discovery_method_long}  illustrate these patterns, showing how the crime itself, the way victims discover it, and its financial impact have changed from 2008 to 2021. 

First, considering the types of identity theft experienced, Figure \ref{fig:theft_type_long} shows a shift from traditional financial account misuse to other forms, which is particularly evident in the most recent survey year, 2021. As shown, existing bank account or credit card misuse remain the most common types of identity theft experienced. Misuse of existing credit cards, historically the most prevalent type (affecting over 50\% of victims in some years), saw a substantial decline between 2018 and 2021, dropping to roughly 38\%. Existing bank account misuse followed a similar pattern, also decreasing significantly in 2021. In contrast, the category of "Other Existing Account Misuse," which includes misuse of online accounts like email or social media, experienced a dramatic surge in 2021, rising from around 10\% in 2018 to over 30\% of victims in 2021. This indicates a significant change in criminal activity towards compromising online credentials. This could also be due to the rapid rise of social media over the last few years. New account fraud and other misuses of personal information remained relatively stable at lower prevalence levels throughout the period.

Next, Figure \ref{fig:discovery_method_long} illustrates how victims became aware of the misuse, and its trends mirror the shift in crime types. "Notified by Bank," which rose significantly between 2008 and 2016 to become the most common discovery method (peaking at nearly 50\%), saw a sharp reversal, dropping considerably by 2021 to around 26\%. On the other hand, "Notified by Other Person/Organization," which began as a less common method, rose dramatically in 2021, increasing from roughly 6\% in 2018 to over 21\%. This aligns with the rise of social media and email account misuse as seen in Figure \ref{fig:theft_type_long}, suggesting victims are increasingly learning about compromises through their social networks or other non-financial organizations.

    

\section{ITS Identity Theft Data Overview}
\label{app:demographics}

\subsection{Incident Characteristics}
\label{sec:incident_characteristics}

Next, summaries of the incidents affecting these victims were examined. The averages in Table \ref{tab:theft_summary_avg} represent the mean of the weighted survey values across the six survey years. It is important to note that for the "Type of Identity Theft" analysis, the percentages are calculated as a share of the total victim population for a given year. Because a single individual can experience multiple forms of identity theft (e.g., both credit card misuse and new account fraud), these categories are not mutually exclusive, and their percentages are not expected to sum to 100\%. Similarly, the percentages for the "How Misuse Was Discovered" and "How Information Was Obtained" sections do not total 100\%. This is because many respondents did not know the answer, and due to the skip logic of the survey. The questionnaire uses skip patterns, where a respondent's answer to one question determines whether they are asked subsequent, more detailed questions. For example, the question detailing the specific method of theft was only asked if a victim indicated that they knew how their information was obtained. If they answered "no," they were skipped past the follow up, and thus are not represented in any of those categories.

\begin{table}[h!]
  \centering
  \caption{Average Summary of Identity Theft Incidents (2008-2021).}
  \label{tab:theft_summary_avg}
  \begin{tabular}{lrr}
    \toprule
    \textbf{Category} & \textbf{N} & \textbf{\% of victims} \\
    \midrule
    \multicolumn{3}{l}{\textbf{Type of Identity Theft\tnote{*}}} \\
    \quad Existing Bank Account Misuse & 8,290,072 & 41.77 \\
    \quad Existing Credit Card Misuse & 9,476,109 & 48.23 \\
    \quad Other Existing Account Misuse & 2,598,402 & 12.28 \\
    \quad New Account Fraud & 1,443,137 & 7.82 \\
    \quad Other Misuse of Personal Information & 1,088,330 & 5.39 \\
    \quad Multiple Types & 2,734,282 & 13.76 \\
    \midrule
    \multicolumn{3}{l}{\textbf{How Misuse Was Discovered\tnote{**}}} \\
    \quad Noticed Problem with Account & 6,698,163 & 33.65 \\
    \quad Notified by Financial Institution & 7,790,053 & 37.69 \\
    \quad Notified by Other Person/Organization & 1,852,203 & 8.76 \\
    \quad Received Bill/Item Not Ordered & 679,982 & 3.63 \\
    \quad Checked Credit Report & 140,646 & 0.81 \\
    \quad Other/Unspecified & 1,807,364 & 8.37 \\
    \quad Unanswered/Not Applicable & 1,518,891\tnote{***} & 7.09 \\
    \midrule
    \multicolumn{3}{l}{\textbf{How Information Was Obtained\tnote{**}}} \\
    \quad Lost or Stolen Physical Item & 932,295 & 5.10 \\
    \quad Stolen During Transaction & 2,403,504 & 12.12 \\
    \quad Hacking/Computer Theft & 278,239 & 1.38 \\
    \quad Deceived by Scam/Phishing & 227,154 & 1.12 \\
    \quad Data Breach (Company/Employer) & 746,451 & 3.73 \\
    \quad Other & 574,991 & 2.67 \\
    \quad Unanswered/Not Applicable & 16,108,164\tnote{***} & 73.88 \\
    \bottomrule
  \end{tabular}
\end{table}
An analysis of Table \ref{tab:theft_summary_avg} verifies the trends discussed in \ref{app:longitudinal}. Table \ref{tab:theft_summary_avg} displays that on average, the misuse of existing accounts is the most common form of identity theft. Existing credit card misuse affected the largest share of victims (48.23\%), followed by the fraudulent use of existing bank accounts (41.77\%). These two categories alone demonstrate that the overwhelming majority of incidents involve the compromise of accounts that are already held by the victim. The creation of new fraudulent accounts and the misuse of personal information for other purposes, such as filing a fraudulent tax return, are notably less frequent, impacting 7.82\% and 5.39\% of victims, respectively. Furthermore, a significant portion of the victim population, 13.76\% on average, experienced more than one type of identity theft.

The most common methods by which victims discovered the fraud were being notified by a financial institution/bank (37.69\%), or personally noticing a problem with their account, such as not being able to make purchases anymore (33.65\%), which is a reflection of the prevalence and efficacy of fraud detection algorithms used by financial institutions. On the other hand, the proactive measure of checking credit reports was far less common in discovering misuse, with only 0.81\% of victims reporting that they discovered their incident in this way. Finally, Table \ref{tab:theft_summary_avg} shows that in terms of how the victims' personal information was obtained, no single method is dominant. The most frequently cited pathway was information being stolen during a transaction, either online or in-person, accounting for 12.12\% of incidents. Interestingly, while digital threats are a major public concern, direct hacking or computer theft was reported in only 1.38\% of cases, and data breaches from companies or employers accounted for just 3.73\% on average. It is also noteworthy that traditional, non-digital methods remain relevant. For example, 5.10\% of incidents resulted from a lost or stolen physical item, such as a wallet or mail. However, all these values only reflect the victims who knew how their information was obtained. The vast majority of victims did not know, therefore these statistics may not be an accurate reflection of how the information of most victims was obtained.

\subsection{Summary of Data Demographics}

Recall that Table \ref{tab:victimization_year} showed that the number of identity theft victims has been generally increasing with each survey wave. While identity theft affects all segments of the population, comparison of the data sample's victimization with U.S. Census data reveals some disparities in victimization risk, detailed in Table \ref{tab:demographics_percent} in Appendix \ref{app:demographics_census}. Specifically, victims aged 30-64 are overrepresented among victims. Additionally, white individuals are overrepresented among victims, while  individuals of Black, Asian, American Indian or Alaskan Native descent, or of two or more races, are all underrepresented among victims. On the impact of education level, individuals with a Bachelor's degree or a Graduate/Professional degree are overrepresented among victims. Similar to education, with respect to income, households earning \$75,000 and over are the most victimized group, being slightly overrepresented compared to their share of the population. Households in the lower income brackets (\$25,000 or less) are underrepresented among victims. 

To further investigate why certain groups are disproportionately represented, we conducted a series of deeper analyses, including weighted proportions tests and logistic regressions, extending the work of \cite{Reynolds2021Differential, DeLiema2021} to all six waves of the ITS survey. Our results are detailed in \ref{app:demographics_census} and in \ref{app:underrep}, and are summarized as follows. First, we found that victims aged 30-64 are ``digitally active but not digitally native,'' meaning that they are significantly more vulnerable to digital theft than the 16-29 ``digital native'' group. Next, we found disparities in the specific types of fraud experienced across different demographics. Our analysis in \ref{app:underrep} reveals that the over-representation of White and highly educated victims is primarily driven by existing credit card misuse, aligning with the results of \cite{Nevin2025, Copes2010}. It was also found that Black individuals and lower-SES groups are disproportionately victimized by more invasive forms of identity theft, such as existing bank account and new account fraud. For example, we found that Black victims report existing bank account misuse at a rate higher than the White victim population, a finding that aligns with prior research on target suitability and financial accessibility \cite{Copes2010}. Finally, the results of our regression analysis in \ref{app:underrep} verified the trends described in \cite{Reynolds2021Differential}, where higher levels of age, education, and income increase the odds of identity theft victimization. Thus, the trends observed in previous studies using a fraction of ITS surveys continue to be observed when all six waves are pooled.


\section{Detailed Data Demographics and Census Comparison}
\label{app:demographics_census}

Here we display a set of descriptive statistics for the combined dataset,  shown in Table \ref{tab:demographics_percent}, which provides an averaged demographic profile of victims and juxtaposes it against data for the general U.S. population from the Census Bureau. The Victim \% column reflects the characteristics of the sample of 41,091 victims from the harmonized data. To achieve this, the survey weights for victims within a specific category (for example, those aged 30-49 or holding a Bachelor's degree) were summed and then divided by the total weighted victim population. The result is an averaged demographic of a typical identity theft victim between 2008 and 2021. 

For the Census \% column, data was aggregated from several U.S. Census Bureau surveys corresponding to the ITS years (2008, 2012, 2014, 2016, 2018, and 2021). Race data was sourced from the American Community Survey (ACS) 1-Year Data Profile DP05 \cite{Census_ACS_DP05}, while household income data came from the Current Population Survey's (CPS) Table H-17 \cite{Census_CPS_H17}. The remaining categories of age, sex, and educational attainment were derived from ACS tables on U.S. age and sex composition \cite{Census_AgeSex}. The final percentages represent the average demographic makeup of the U.S. population across these years. Note that in this process, age related categories were aligned to ensure a valid comparison, because the ITS sample includes only individuals aged 16 or older. Thus, because ACS groups individuals aged 15-19, a proportional allocation of 80\% or four fifths of this group was used to estimate the 16-19 population. A similar adjustment was made for the educational attainment data where a 15-17 age bracket appeared, and two thirds of that population was taken to estimate the count for ages 16-17. 

\begin{longtable}{lrr}
\caption{Comparison of victim demographics to Census data (\%), averaged over 2008-2021.} \label{tab:demographics_percent} \\
\toprule
\textbf{Category} & \textbf{Victim \%} & \textbf{Census \%} \\
\midrule
\endfirsthead
\multicolumn{3}{c}%
{{\tablename\ \thetable{} -- continued from previous page}} \\
\toprule
\textbf{Category} & \textbf{Victim \%} & \textbf{Census \%} \\
\midrule
\endhead
\midrule
\multicolumn{3}{r}{{Continued on next page}} \\
\endfoot
\bottomrule
\endlastfoot
\multicolumn{3}{c}{\textbf{Age Group}} \\
\midrule
16-29 & 18.03 & 24.35 \\
30-49 & 38.25 & 33.32 \\
50-64 & 28.14 & 24.44 \\
65+ & 15.58 & 17.89 \\
\midrule
\multicolumn{3}{c}{\textbf{Victim Sex}} \\
\midrule
Male & 47.43 & 47.73 \\
Female & 52.57 & 52.27 \\
\midrule
\multicolumn{3}{c}{\textbf{Victim Race/Ethnicity}} \\
\midrule
White alone & 81.96 & 75.3 \\
Black or African American alone & 10.35 & 13.2 \\
Asian alone & 4.96 & 5.5 \\
American Indian/Alaska Native alone & 0.62 & 0.9 \\
Native Hawaiian/Pacific Islander alone & 0.29 & 0.2 \\
Two or More Races & 1.83 & 4.9 \\
\midrule
\multicolumn{3}{c}{\textbf{Victim Educational Attainment}} \\
\midrule
Less than High School & 5.60 & 15.17 \\
High School Graduate (or equivalent) & 18.12 & 28.38 \\
Some College/Associate Degree & 30.16 & 27.18 \\
Bachelor's Degree & 28.10 & 18.87 \\
Graduate/Professional Degree & 18.02 & 10.38 \\
\midrule
\multicolumn{3}{c}{\textbf{Victim Household Income}} \\
\midrule
Under \$15,000 & 5.78 & 8.32 \\
\$15,000 to \$24,999 & 5.46 & 7.95 \\
\$25,000 to \$49,999 & 19.92 & 19.12 \\
\$50,000 to \$74,999 & 18.26 & 15.78 \\
\$75,000 and over & 50.58 & 48.90 \\
\end{longtable}

Table \ref{tab:demographics_percent} provides several insights about the differences between the demographic profile of a victim versus the entire population. For example, adults aged 30-49 and 50-64, are overrepresented among victims compared to their share of the general population. Younger individuals aged 16-29 and seniors aged 65+ are underrepresented. This could suggest that identity theft disproportionately affects individuals who are in their prime working and earning years. On the other hand, victimization rates between sexes are remarkably aligned with the national population, suggesting that sex is not a significant differentiating factor in identity theft victimization. 

Table \ref{tab:demographics_percent} also highlights disparities in victimization rates across racial lines. White individuals are overrepresented among victims, making up 82\% of the victim population but only 75\% of the general population. In contrast, individuals of Black, Asian, American Indian or Alaskan Native descent, or of two or more races, are all underrepresented among victims. It is also noticeable that individuals with a Bachelor's degree or a Graduate/Professional degree are overrepresented among victims. For example, those with a graduate degree make up over 18\% of victims but only about 10\% of the census population. Those with a High school education or less are significantly underrepresented among victims. The impact of income is also shown, where households earning \$75,000 and over are slightly overrepresented, and  households in the lower income brackets (\$25,000 or less) are underrepresented. These disparities are further explored in \ref{app:underrep}.

\section{Analysis of Victimization Factors}
\label{app:underrep}

\subsection{Why are middle aged individuals (ages 30-64)  overrepresented among victims?}
The data indicates that individuals in their prime earning and working years are victimized at a higher rate than their younger and older counterparts. For instance, the 30-64 age bracket contains over 66\% of the victim population while representing less than 58\% of the general population. This overrepresentation may be due to a combination of their financial activity, the value they represent to criminals, and their patterns of digital purchasing. 

A testable hypothesis is that this demographic is "digitally active but not digitally native," meaning that vulnerability is a function of both digital nativity and digital activity. This suggests that while this middle aged group is highly active online for banking, work, and commerce, they may be more susceptible to sophisticated online scams (e.g. phishing) than younger, more "digitally native" generations who grew up with the technology. To investigate this hypothesis, a weighted proportions test was conducted using the harmonized dataset. This analysis required first isolating the age, theft method, and weight variables. Victims were categorized into the same age groups as in Table \ref{tab:demographics_percent}. To isolate the method of compromise, the theft method variable was filtered to only include unambiguous cases. This means that any theft method that is not clearly digital or physical was filtered out. This step was essential because some of the survey's categories are too broad for this analysis. For example, the category "Stolen during a transaction" is ambiguous, as it does not distinguish between an online transaction/digital theft and in in-person/physical theft. This category, along with the "Other" response was therefore excluded. 

This filtering process resulted in two mutually exclusive categories for analysis. The "Digital" group included victims whose information was obtained via "Stolen from a Computer/Device (Hacking)," "Deceived by a scam (e.g. Phishing)," and "Stolen from a Company/Employer (Data Breach)." The "Physical" group included victims whose information came from a "Lost or Stolen Physical Item (Wallet, Mail, etc.)." After filtering the dataset to these cases, a weighted proportions test, using the ITS weight variable, was run to compare the two age groups. Within each resulting cell (e.g., "16-29" and "Digital"), the final survey weights of all respondents were summed. This yielded a table of weighted totals, representing the estimated national count for each combination. Finally, the proportion of digital versus physical theft within each age group was calculated by dividing the weighted sum for each theft method by the total weighted sum for that age group. This process produces nationally representative proportions, allowing for accurate comparisons of victimization patterns across demographics.

\begin{table}[htbp]
  \centering 
  \caption{Weighted Proportion of Theft Method by Age Group.}
  \label{tab:age_theft_method} 
  \begin{tabular}{@{}lcc@{}} 
    \toprule
    \textbf{Age Group} & \textbf{Digital Theft (\%)} & \textbf{Physical Theft (\%)} \\
    \midrule
    16--29    & 44.7               & 55.3                \\
    30--49    & 58.5               & 41.5                \\
    50--64    & 63.4               & 36.6                \\
    65+       & 63.1               & 36.9                \\
    \bottomrule
  \end{tabular}
  \par 
  \vspace{0.5ex} 
  \small
 
\end{table}

The results of this analysis, shown in Table \ref{tab:age_theft_method}, support the hypothesis that middle aged individuals are digitally active but not digitally native." The proportion of victimization from digital methods is lowest for the 16-29 age group (44.7\%), which was the only age group more likely to be victimized by physical means (55.3\%). This suggests that being a digital native can lead to a degree of resilience to online scams. For all other age groups, the vulnerability to digital theft is high and similar. The 30-49 age group (58.5\%), 50-64 age group (63.4\%), and the 65+ age group (63.1\%) all show a vulnerability to digital compromise. This confirms that the non-digital native groups are more susceptible to these digital methods of theft. This finding, however, does not explain why the 65+ age group has a lower overall victimization rate. We theorize this is due to the second factor, the level of digital activity. The 30-64 age groups likely represent a unique target for criminals because they combine high vulnerability (not digitally native) with high activity (a larger digital footprint from banking, commerce, and work). The 65+ age group shares the same vulnerability, but likely has a smaller average digital footprint, presenting fewer targets for compromise. While this theory provides an explanation for the pattern seen in Table \ref{tab:demographics_percent}, it is a limitation of this study that the ITS dataset does not contain variables to directly measure digital activity. Therefore, we present this as our interpretation of our findings that warrants further research with more specialized data.

\subsection{Why are White individuals overrepresented among victims?}

While victimization occurs among all racial groups, the data shows that White individuals are victims at a rate higher than their share of the U.S. population (82\% of victims vs. 75\% of the census population). It is hypothesized that this disparity is a reflection of underlying socioeconomic variables. Overrepresentation of White individuals among victims may be due to them being overrepresented in the higher income and higher education brackets, which are also noted in the data as being groups with higher rates of victimization.

To explore this relationship more deeply, a series of weighted logistic regression models were developed using the full dataset including all survey respondents (not solely victims). The logistic regression models calculate Odds Ratios (OR), which quantify how the odds of being a victim change in relation to each predictor, relative to a baseline group. An OR that is greater than 1 indicates increased odds compared to the baseline, and vice versa. All models incorporated the final survey weight provided in the dataset to ensure findings generalize to the U.S. population, not just the sample characteristics. 

The primary predictor of interest was a binary indicator identifying respondents who reported their race solely as White versus all other respondents combined (who were the reference group). Additional control variables included categorical representations of the highest level of education completed and income, mirroring the groups in Table \ref{tab:demographics_percent}. Respondent age was also included as a continuous control variable in the third model. To integrate the categorical predictors for education and income into the regression, they were represented by sets of binary indicators. For each variable, one category was designated as a reference category ('Less than high school' for education, and 'Under than \$15,000' for income). These categories were chosen as they represent the lowest levels on their scales, providing a baseline against which to measure the effects of higher attainment or income.  Separate binary indicators then represented the remaining categories. This method allows the model coefficients to reflect the change in odds associated with belonging to a specific category relative to the baseline. Respondents missing any key variables were excluded, resulting in a sample of 544,837 individuals for these models.

Each model produces coefficients and corresponding p-values for each predictor. The p-value indicates statistical significance and indicates the probability of observing an association as strong as (or stronger than) the one found in the sample, if no true association existed in the population. A small p-value suggests that the observed relationship is unlikely due to random chance and is considered statistically significant. The modeling results are shown in Tables \ref{tab:logit_results}, \ref{tab:logit_age_results} and \ref{tab:race_profile_comparison}. In Table \ref{tab:logit_results}, the initial model ("Race Only"), containing only the White race indicator, confirmed that identifying as White was associated with significantly higher odds of victimization (OR = 1.320, p $<$ 0.001), compared to non-White respondents.

\begin{table}[htbp]
  \centering
  \begin{threeparttable}
  \caption{Weighted Logistic Regression Predicting Likelihood of Identity Theft Victimization.}
  \label{tab:logit_results}
  \small
  \sisetup{
    table-format=1.3,
    table-align-text-post=false,
    table-number-alignment=center,
    input-symbols = <, 
    round-mode=places,
    round-precision=3 
  }
  
  \setlength{\tabcolsep}{3pt} %
  \begin{tabular}{@{} l
                     S[table-format=1.3] S[table-format=<1.3]
                     S[table-format=1.3] S[table-format=<1.3]
                     S[table-format=1.3] S[table-format=<1.3]
                     S[table-format=1.3] S[table-format=<1.3]
                     @{}}
    \toprule
    & \multicolumn{2}{c}{\textbf{Race Only}} & \multicolumn{2}{c}{\textbf{+ SES Controls}} & \multicolumn{2}{c}{\textbf{+ SES + Age}} & \multicolumn{2}{c}{\textbf{+ Interaction}} \\
    \cmidrule(lr){2-3} \cmidrule(lr){4-5} \cmidrule(lr){6-7} \cmidrule(lr){8-9}
    Predictor & {OR} & {p} & {OR} & {p} & {OR} & {p} & {OR} & {p} \\
    \midrule
    \textbf{Race} & & & & & & & & \\
    \quad White (Ref: Non-White) & 1.320 & <0.001 & 1.268 & <0.001 & 1.257 & <0.001 & 1.048 & 0.427 \\
    \addlinespace
    \textbf{Education} (Ref: $<$ HS) & & & & & & & & \\
    \quad HS Graduate & & & 1.954 & <0.001 & 1.928 & <0.001 & 1.630 & <0.001 \\
    \quad Some College/Assoc. & & & 3.173 & <0.001 & 3.144 & <0.001 & 2.930 & <0.001 \\
    \quad Bachelor's & & & 3.852 & <0.001 & 3.809 & <0.001 & 3.085 & <0.001 \\
    \quad Graduate/Professional & & & 4.743 & <0.001 & 4.654 & <0.001 & 3.812 & <0.001 \\
    \addlinespace
    \textbf{Income} (Ref: $<$ \$15k) & & & & & & & & \\
    \quad \$15k -- \$24,999 & & & 1.002 & 0.963 & 0.995 & 0.867 & 1.004 & 0.895 \\
    \quad \$25k -- \$49,999 & & & 1.024 & 0.349 & 1.021 & 0.420 & 1.028 & 0.289 \\
    \quad \$50k -- \$74,999 & & & 1.188 & <0.001 & 1.187 & <0.001 & 1.192 & <0.001 \\
    \quad \$75k and over & & & 1.366 & <0.001 & 1.369 & <0.001 & 1.371 & <0.001 \\
    \addlinespace
    \textbf{Age} & & & & & & & & \\
    \quad Age & & & & & 1.002 & <0.001 & & \\
    \addlinespace
    \textbf{Interaction: White*Edu} & & & & & & & & \\
    \quad White * HS Grad & & & & & & & 1.245 & 0.002 \\
    \quad White * Some College & & & & & & & 1.105 & 0.122 \\
    \quad White * Bachelor's & & & & & & & 1.305 & <0.001 \\
    \quad White * Graduate & & & & & & & 1.301 & <0.001 \\
    \midrule
    N (Respondents, thousands) & \multicolumn{2}{c}{544,837} & \multicolumn{2}{c}{544,837} & \multicolumn{2}{c}{544,837} & \multicolumn{2}{c}{544,837} \\
    \bottomrule
  \end{tabular}
    \begin{tablenotes}[para,flushleft] 
      \small 
      \item \textit{Note:} Table displays odds ratios (OR) and p-values (p) from weighted logistic regressions. SES = Socioeconomic Status. Reference category for Education is Less than High School ($<$ HS). $p$-values less than 0.001 are shown as "$<$0.001".
    \end{tablenotes}
  \end{threeparttable}
\end{table}

The second model ("+SES controls") added indicators for education and income levels. Adding these control variables allows the model to statistically isolate the specific association between the primary predictor (the White race indicator) and the outcome (victimization), while holding the influence of these other factors constant. The odds ratio for the White race indicator decreased in this model (OR = 1.268), which is consistent with the hypothesis that socioeconomic factors contribute to the disparity. This reduction confirms that education and income differences account for some, but not all, of the initial race based difference.

The third model, ("+SES, + Age") incorporated the respondent's age. The odds ratio for the White race indicator reduced slightly again (OR = 1.257), but stayed highly significant (p$<$0.001). This finding strengthens the conclusion that race maintains an independent association with victimization risk, beyond what can be explained by education, income, and age differences. Age itself also emerged as having a small, statistically significant positive relationship with victimization odds (OR=1.002, p$<$0.001) in this model. 

The final model, ("+Interaction (Race*Edu)") tested whether the relationship between education and victimization risk differs systematically for White versus non-White individuals. This involved adding interaction terms to the model. An interaction term tests if the effect of one predictor (like having a Bachelor's degree) depends on, or is modified by, the level of another predictor (like being White). Creating these interaction terms involves mathematically combining the indicators for the two variables in question. In our analysis code, this was done by multiplying the binary indicator for being White (0 or 1) by the binary indicator for each specific education level (0 or 1). The resulting term (e.g., "White * Bachelor's") will only equal 1 if a person is both White and has a Bachelor's degree; otherwise, it is 0. By including these combined terms in the regression, the model can estimate this joint effect separately from the individual effects of race and education alone.

This fourth model yielded the most interesting findings. The main effect associated with the White race indicator (representing the effect of being White compared to non-White specifically among those in the reference education group, 'Less than High School') became non significant (OR = 1.048, p = 0.427). However, several interaction terms were statistically significant (e.g., White * HS Grad: OR = 1.245, p = 0.002; White * Bachelor's: OR = 1.305, p $<$ 0.001). These results refute the initial hypothesis that race serves only as a proxy for socioeconomic status. The analysis reveals a significant interaction, which is the way educational attainment relates to the odds of identity theft victimization is different for White individuals compared to non-White individuals. The positive interaction odds ratios suggest that higher levels of education correspond to a greater increase in victimization risk for White respondents than for non-White respondents, relative to those with less than a high school education. This relationship requires deeper examination than originally anticipated.

Regarding the observation that most odds ratios in Table \ref{tab:logit_results} are above 1, this simply reflects the patterns in the data relative to the chosen reference groups. For education and income, the higher categories consistently show increased odds of victimization compared to the lowest categories ('Less than High School' and 'Under \$15,000'). Similarly, the initial models showed White individuals had higher odds than non-White individuals. Odds ratios close to 1 (like those for the lower middle income brackets, which are not statistically significant) indicate little difference in odds compared to the reference group. An odds ratio significantly less than 1 would indicate lower odds of victimization, but such a pattern was not observed for most primary predictors in these models. A similar analysis was done for Tables \ref{tab:logit_age_results} and \ref{tab:race_profile_comparison}, where Table \ref{tab:race_profile_comparison} performs a comparative profile analysis that measures the percentage of each reported category with respect to race.

\begin{table}[htbp]
  \centering
  \begin{threeparttable}
  \caption{Weighted Logistic Regression Predicting Likelihood of Identity Theft Victimization by Age Group.}
  \label{tab:logit_age_results}
  \small
  \sisetup{
    table-format=1.3,
    table-align-text-post=false,
    table-number-alignment=center,
    input-symbols = <, 
    round-mode=places,
    round-precision=3 
  }
  \setlength{\tabcolsep}{5pt}
  \begin{tabular}{@{} l 
                     S[table-format=1.3] S[table-format=<1.3] 
                     S[table-format=1.3] S[table-format=<1.3] 
                     S[table-format=1.3] S[table-format=<1.3] 
                     @{}}
    \toprule
    & \multicolumn{2}{c}{\textbf{Age Only}} & \multicolumn{2}{c}{\textbf{+ SES}} & \multicolumn{2}{c}{\textbf{+ SES/Race}} \\
    \cmidrule(lr){2-3} \cmidrule(lr){4-5} \cmidrule(lr){6-7}
    \textbf{Predictor} & {OR} & {p} & {OR} & {p} & {OR} & {p} \\
    \midrule
    \textbf{Age Group (Ref: 65+)} & & & & & & \\
    \quad 16--29 & 0.722 & <0.001 & 0.817 & <0.001 & 0.836 & <0.001 \\
    \quad 30--49 & 1.312 & <0.001 & 1.149 & <0.001 & 1.176 & <0.001 \\
    \quad 50--64 & 1.319 & <0.001 & 1.210 & <0.001 & 1.223 & <0.001 \\
    \addlinespace
    \textbf{Education (Ref: $<$ HS)} & & & & & & \\
    \quad HS Graduate & {} & {} & 1.845 & <0.001 & 1.842 & <0.001 \\
    \quad Some College/Assoc. & {} & {} & 3.016 & <0.001 & 3.009 & <0.001 \\
    \quad Bachelor's & {} & {} & 3.585 & <0.001 & 3.629 & <0.001 \\
    \quad Graduate/Professional & {} & {} & 4.322 & <0.001 & 4.429 & <0.001 \\
    \addlinespace
    \textbf{Income (Ref: $<$ \$15k)} & & & & & & \\
    \quad \$15k -- \$24,999 & {} & {} & 0.997 & 0.916 & 0.986 & 0.653 \\
    \quad \$25k -- \$49,999 & {} & {} & 1.022 & 0.389 & 1.005 & 0.848 \\
    \quad \$50k -- \$74,999 & {} & {} & 1.180 & <0.001 & 1.154 & <0.001 \\
    \quad \$75k and over & {} & {} & 1.341 & <0.001 & 1.311 & <0.001 \\
    \addlinespace
    \textbf{Race (Ref: Other)} & & & & & & \\
    \quad White & {} & {} & {} & {} & 0.827 & <0.001 \\
    \quad Black & {} & {} & {} & {} & 0.673 & <0.001 \\
    \quad Asian & {} & {} & {} & {} & 0.500 & <0.001 \\
    \midrule
    N (Respondents, thousands) & \multicolumn{2}{c}{544,837} & \multicolumn{2}{c}{544,837} & \multicolumn{2}{c}{544,837} \\
    \bottomrule
  \end{tabular}
    \begin{tablenotes}[para,flushleft] 
      \small 
      \item \textit{Note:} Table displays odds ratios (OR) and p-values (p) from weighted logistic regressions. Model 1 includes only Age Group. Model 2 adds Socioeconomic Status (SES: Income and Education). Model 3 adds Race (White, Black, Asian) to the controls. $p$-values less than 0.001 are shown as "$<$0.001". Models used final survey weights. 
    \end{tablenotes}
  \end{threeparttable}
\end{table}

\begin{table}[htbp]
  \centering
  \begin{threeparttable}
  \caption{Victim Profile Comparison: Theft and Discovery Methods by Race (Weighted \%)}
  \label{tab:race_profile_comparison}
  \small 

  \begin{tabular}{@{} l *{4}{c} *{3}{c} @{}}
  
    \toprule
    \textbf{Category} & \multicolumn{4}{c}{\textbf{Weighted \% of Victims in Group}} & \multicolumn{3}{c}{\textbf{Difference vs. All Other (pp)}} \\
    \cmidrule(lr){2-5} \cmidrule(lr){6-8}
     & {White} & {Black} & {Asian} & {All Other} & {Diff} & {Diff} & {Diff} \\
     & {(\%)} & {(\%)} & {(\%)} & {(\%)} & {(White)} & {(Black)} & {(Asian)} \\
    \midrule
    \multicolumn{8}{@{} l}{\textbf{Type of Identity Theft}} \\
    \midrule
    Existing Bank Account Misuse & 40.89 & 54.03 & 27.01 & 51.89 & -11.00 & +2.14 & -24.87 \\
    Existing Credit Card Misuse & 49.81 & 28.01 & 64.55 & 31.12 & +18.69 & -3.11 & +33.43 \\
    Other Existing Account Misuse & 12.48 & 17.25 & 12.61 & 18.20 & -5.72 & -0.95 & -5.59 \\
    New Account Fraud & 6.52 & 12.77 & 6.31 & 12.30 & -5.78 & +0.46 & -5.99 \\
    Other Misuse of PI & 5.21 & 8.54 & 3.85 & 6.15 & -0.95 & +2.39 & -2.30 \\
    \midrule
    \multicolumn{8}{@{} l}{\textbf{How Misuse Was Discovered}} \\
    \midrule
    Problem with Account & 33.18 & 36.51 & 36.85 & 37.92 & -4.74 & -1.41 & -1.07 \\
    Notified by Bank & 40.86 & 28.65 & 39.04 & 31.24 & +9.62 & -2.59 & +7.80 \\
    Notified by Other  & 9.02 & 12.03 & 7.21 & 13.23 & -4.21 & -1.20 & -6.02 \\
    Checked Credit Report & 0.57 & 1.89 & 0.45 & 1.34 & -0.78 & +0.55 & -0.89 \\
    Other/Unspecified & 8.74 & 12.19 & 9.59 & 9.13 & -0.38 & +3.06 & +0.46 \\
    Don't Know & 0.00 & 0.00 & 0.00 & 0.00 & +0.00 & +0.00 & +0.00 \\
    \midrule
    \multicolumn{8}{@{} l}{\textbf{How Information Was Obtained}} \\
    \midrule
    Lost or Stolen Physical Item & 4.20 & 8.71 & 4.13 & 6.84 & -2.65 & +1.87 & -2.71 \\
    Stolen During Transaction & 12.45 & 9.92 & 10.22 & 13.36 & -0.91 & -3.43 & -3.13 \\
    Hacking/Computer Theft & 1.34 & 1.53 & 1.56 & 2.43 & -1.09 & -0.90 & -0.88 \\
    Data Breach (Company/Employer) & 3.79 & 3.46 & 3.41 & 4.74 & -0.95 & -1.28 & -1.33 \\
    Other & 2.78 & 3.85 & 1.88 & 4.71 & -1.93 & -0.86 & -2.83 \\
    \bottomrule
  \end{tabular}
  \begin{tablenotes}
    \small
    \item \textit{Note}: Table displays weighted percentage profiles of victims by racial group. "All Other Victims" includes American Indian/Alaska Native, Native Hawaiian/Pacific Islander, Two or More Races, and Latino groups, and serves as the reference category. Categories for theft type may sum to $>100\%$ as victims can experience multiple types of fraud.
  \end{tablenotes}
  \end{threeparttable}
\end{table}

\subsection{Why are individuals with higher educational attainment and income overrepresented among victims?}

The data also shows that individuals with higher socioeconomic status (individuals with more education or higher education levels) are overrepresented in the victim sample. For example, individuals with a graduate or professional degree make up 18\% of victims but only 10\% of the general population. In addition, Table \ref{tab:logit_results} revealed a strong association between education and likelihood of victimization. Even after controlling for race, age, and income, as education level increased, the odds ratio of becoming a victim increased as well. Having a graduate or professional degree was especially notable, being associated with an odds ratio of 4.743, meaning those individuals are almost five times as likely to be victims. This suggests that they could be more attractive or accessible targets for criminals.

One prominent hypothesis suggests that these individuals represent more valuable targets due to potentially greater assets, higher credit limits, and larger savings. An alternative but not mutually exclusive hypothesis is that these individuals might have a larger digital and commercial footprint, thereby creating more opportunities for their data to be compromised. 

This analysis focused on testing the "More to steal" hypothesis by examining whether higher socioeconomic status correlates with greater financial losses among those who experienced identity theft. The initial, most intuitive step was to calculate the simple weighted average out of pocket loss for victims in each education and income category. The results of this analysis, presented in Table \ref{tab:weighted_avg_loss}, were counterintuitive. They appeared to contradict the hypothesis, showing that the highest average losses were borne by victims in the lowest socioeconomic groups, such as those with less than a high school education (\$414.96) and those earning under \$15,000 (\$565.86).

\begin{table}[htbp]
  \centering
  \begin{threeparttable}
  \caption{Weighted Average Out-of-Pocket Loss per Victim by Socioeconomic Group}
  \label{tab:weighted_avg_loss}
  \begin{tabular}{@{} l S[table-format=3.2] @{}}
    \toprule
    \textbf{Category} & {\textbf{Weighted Average Loss (\$)}} \\
    \midrule
    \multicolumn{2}{@{}l}{\textbf{Education Level}} \\
    \hspace{1em} Less than High School & 414.96 \\
    \hspace{1em} High School Graduate & 281.56 \\
    \hspace{1em} Some College/Assoc. Degree & 309.63 \\
    \hspace{1em} Bachelor's Degree & 166.51 \\
    \hspace{1em} Graduate/Professional Degree & 213.25 \\
    \midrule
    \multicolumn{2}{@{}l}{\textbf{Income Level}} \\
    \hspace{1em} Under \$15,000 & 565.86 \\
    \hspace{1em} \$15,000 to \$24,999 & 412.49 \\
    \hspace{1em} \$25,000 to \$49,999 & 439.74 \\
    \hspace{1em} \$50,000 to \$74,999 & 250.81 \\
    \hspace{1em} \$75,000 and over & 128.05 \\
    \bottomrule
  \end{tabular}
    \begin{tablenotes}[para,flushleft]
      \small
      \item \textit{Note:} Table displays the simple weighted average out-of-pocket loss for victims.
    \end{tablenotes}
  \end{threeparttable}
\end{table}

To further explore potential reasons why individuals with higher educational attainment are overrepresented, a comparative profile analysis was also conducted. Using the victim-only data, respondents were segmented into three mutually exclusive groups based on their highest education level: Graduate/Professional, Bachelor's Degree, and All Other Victims. The average weighted percentage for each type of theft, discovery method, and theft method was then calculated for each of these three groups. The results are presented in Table \ref{tab:victim_profile_education}. The most significant finding in Table \ref{tab:victim_profile_education} is a dramatic difference in the type of fraud experienced. Both high education groups are far more likely to be victims of Existing Credit Card Misuse. The Bachelor's Degree group experiences this fraud at a rate of 18.39\% higher than the baseline group, while the Graduate/Professional group's rate is 27.53\% higher than the baseline. This finding is mirrored by Existing Bank Account Misuse, where both high education groups are underrepresented (13.43 and 21.93\% lower than baseline, respectively). This strongly suggests that the high victimization odds ratio observed in the regression models is driven mainly by credit card fraud, not attacks on debit cards or bank accounts. This is further explained by the findings on discovery methods. Both the Bachelor's Degree and Graduate/Professional groups were significantly more likely to be notified by a financial institution (7.30 and 9.43\% higher than baseline, respectively). They also were less likely to have noticed a problem with their accounts themselves.

\begin{table}[htbp]
  \centering
  \begin{threeparttable}
  \caption{Victim Profile Comparison by Education Level}
  \label{tab:victim_profile_education}
  
  \sisetup{
    table-align-text-post=false,
    round-mode=places,
    round-precision=2
  }
  
  \begin{tabular}{@{} l
                     S[table-format=2.2]
                     S[table-format=2.2]
                     S[table-format=2.2]
                     S[table-format=+2.2]
                     S[table-format=+2.2] @{}}
    \toprule
    Category & {Grad./Prof.} & {Bach.'s} & {All Other} & {(Grad - Other)} & {(Bach. - Other)} \\
             & {(\%)}           & {(\%)}               & {(\%)}               & {(\%)}            & {(\%)}            \\
    \midrule
    
    \multicolumn{6}{@{}l}{\textbf{Type of Identity Theft}} \\
    \quad Existing Bank Account    & 27.62 & 36.12 & 49.55 & -21.93 & -13.43 \\
    \quad Existing Credit Card     & 65.29 & 56.16 & 37.76 &  27.53 &  18.39 \\
    \quad Other Existing Account   & 12.28 & 12.64 & 13.61 &  -1.34 &  -0.97 \\
    \quad New Account Fraud               & 6.15  & 6.25  & 8.17  &  -2.02 &  -1.92 \\
    \quad Other Misuse  & 5.22  & 4.47  & 6.04  &  -0.82 &  -1.57 \\
    \quad Multiple Types                  & 14.72 & 14.23 & 13.24 &   1.48 &   0.99 \\
    \addlinespace

    \multicolumn{6}{@{}l}{\textbf{How Misuse Was Discovered}} \\
    \quad  Problem with Account     & 29.76 & 32.00 & 36.18 & -6.42 & -4.19 \\
    \quad Notified by Bank & 45.02 & 42.89 & 35.59 &  9.43 &  7.30 \\
    \quad Notified by Other    & 8.95  & 8.74  & 9.74  & -0.79 & -1.00 \\
    \quad Received Bill Not Ordered   & 2.94  & 3.01  & 3.82  & -0.88 & -0.81 \\
    \quad Checked Credit Report            & 0.45  & 0.63  & 0.84  & -0.40 & -0.22 \\
    \quad Other/Unspecified                & 9.10  & 8.99  & 9.15  & -0.05 & -0.16 \\
    \quad Don't Know                     & 0.15  & 0.09  & 0.10  &  0.05 & -0.01 \\
    \addlinespace 

    \multicolumn{6}{@{}l}{\textbf{How Info Was Obtained}} \\
    \quad Lost/Stolen Physical Item   & 3.43  & 3.81  & 5.56  & -2.13 & -1.75 \\
    \quad  During Transaction       & 11.91 & 12.45 & 12.12 & -0.21 &  0.33 \\
    \quad Hacking/Computer Theft         & 1.39  & 1.27  & 1.50  & -0.11 & -0.23 \\
    \quad Scam/Phishing       & 1.08  & 0.93  & 1.29  & -0.21 & -0.36 \\
    \quad Data Breach  & 4.11  & 4.05  & 3.51  &  0.60 &  0.54 \\
    \quad Other                          & 1.91  & 2.20  & 3.59  & -1.68 & -1.39 \\
    
    \bottomrule
  \end{tabular}
  \begin{tablenotes}
    \small
    \item \textit{Note}: Table compares identity theft profiles by educational attainment. "All Other" serves as the reference group.
  \end{tablenotes}
  \end{threeparttable}
\end{table}

This indicates that the overrepresentation of high education individuals as victims could be due to them having a heavier reliance on credit cards, which are a high-volume target for fraud. Their high odds of victimization in the survey are likely a function of their increased exposure to this specific fraud type, and the real-time fraud detection systems that credit card companies use. This effective monitoring means that frauds against this group are more likely to be detected and reported to the victim, thus captured by the survey.

\section{Social Cost by Demographic Group}

\subsection{Demographic Disparities in Social Cost}

\begin{figure}[htbp]
    \centering
    \begin{subfigure}[b]{0.48\textwidth}
        \centering
        \includegraphics[width=\textwidth]{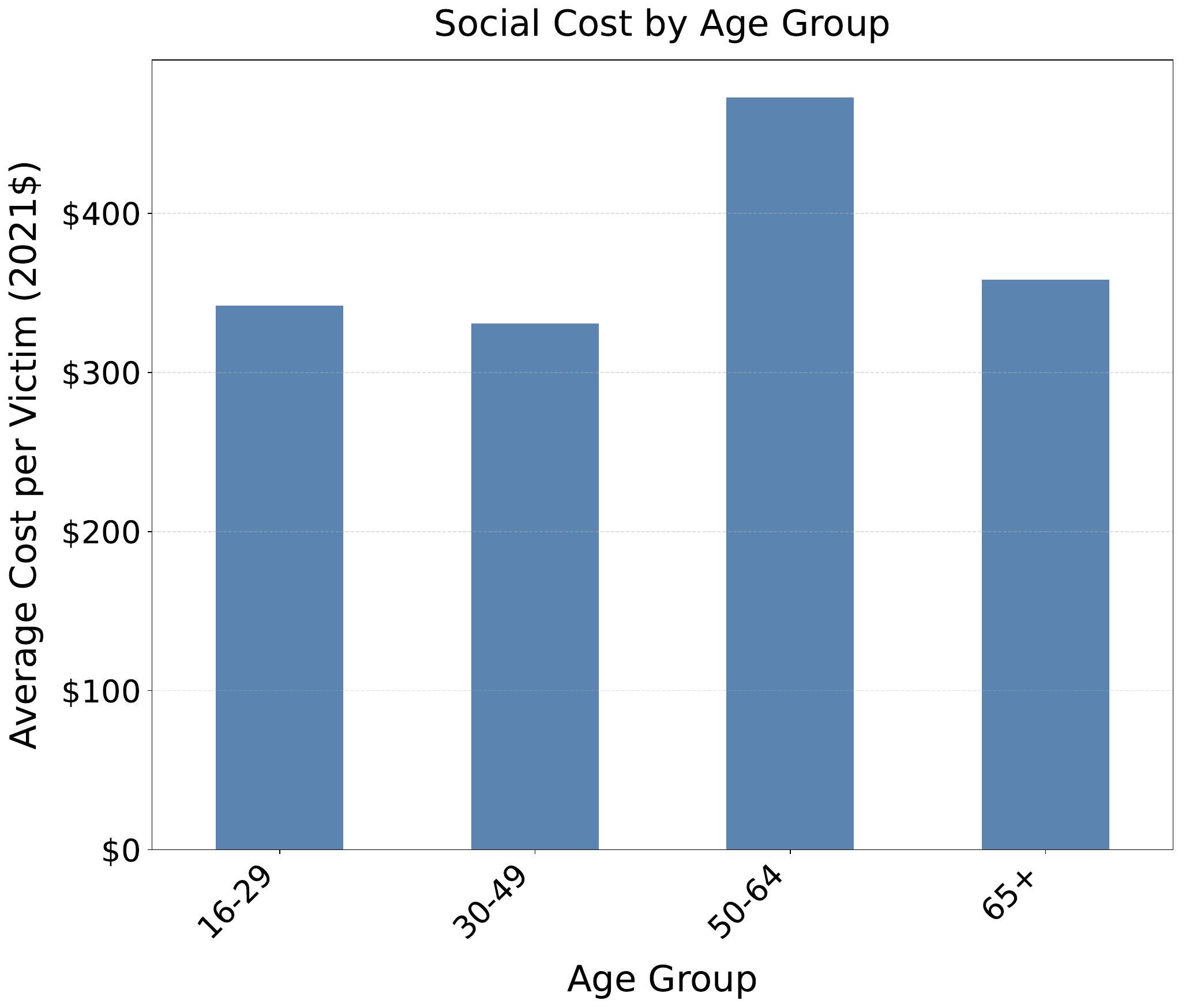}
        \caption{By age group.}
        \label{fig:social_cost_age}
        
    \end{subfigure}
    \hfill
    \begin{subfigure}[b]{0.48\textwidth}
        \centering
        \includegraphics[width=\textwidth]{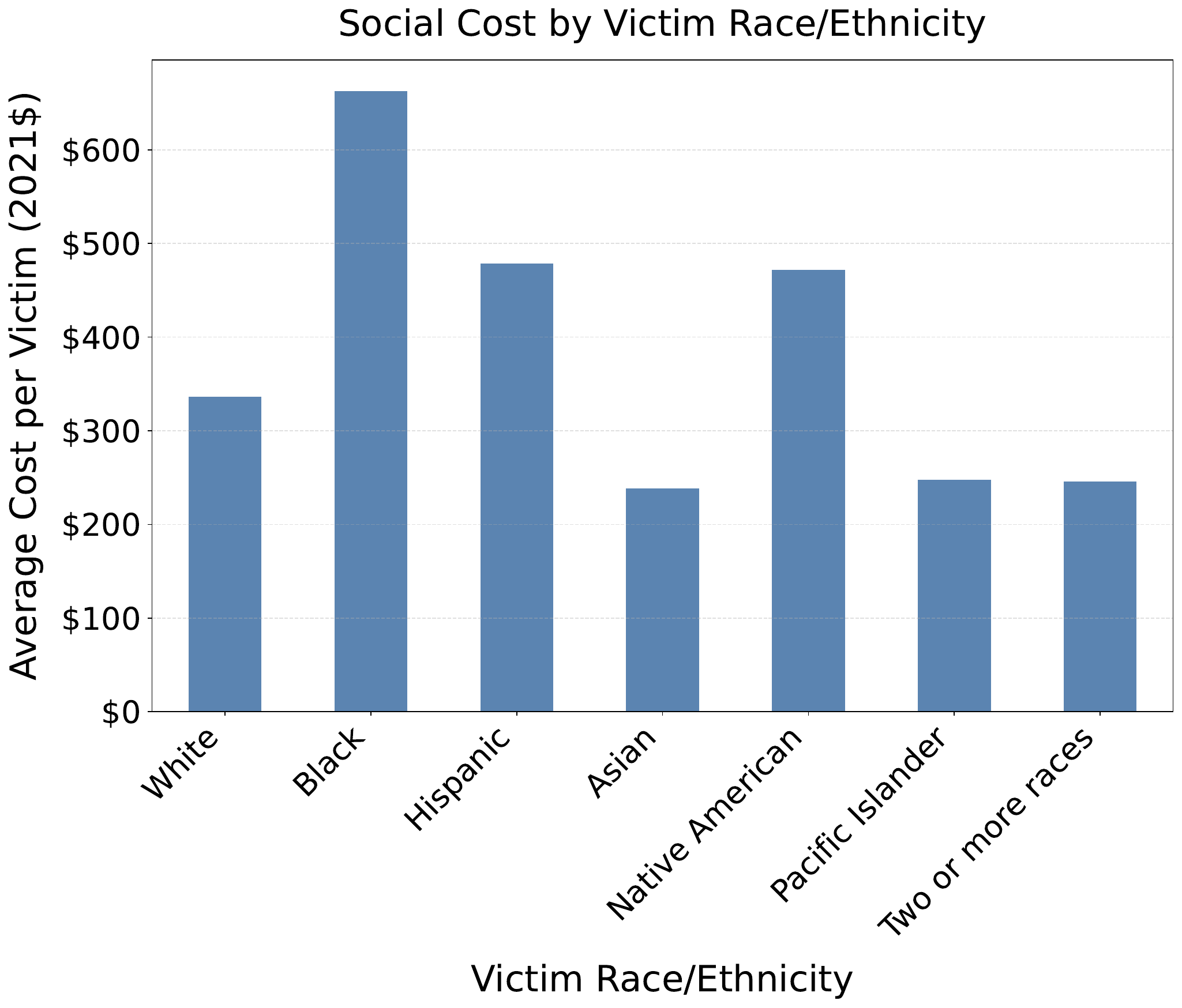}
        \caption{By race.}
        \label{fig:social_cost_race}
    \end{subfigure}
    
    \caption{Average social cost of data breaches by demographic characteristics (Age and Race).}
    \label{fig:social_cost_demographics}
\end{figure}

\begin{figure}[htbp]
    \centering
    \begin{subfigure}[b]{0.48\textwidth}
        \centering
        \includegraphics[width=\textwidth]{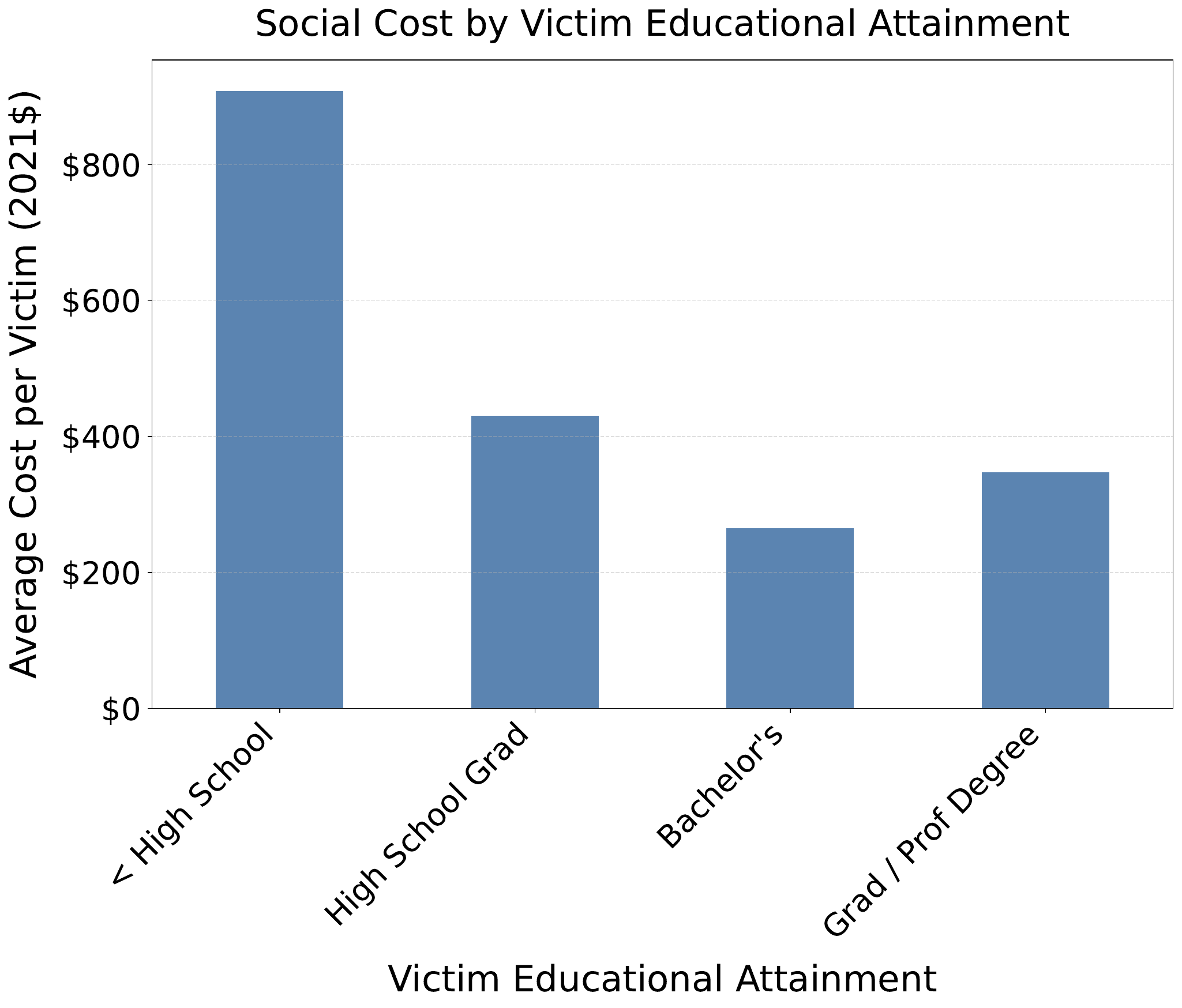}
        \caption{By education.}
        \label{fig:social_cost_education}
    \end{subfigure}
    \hfill 
    \begin{subfigure}[b]{0.48\textwidth}
        \centering
        \includegraphics[width=\textwidth]{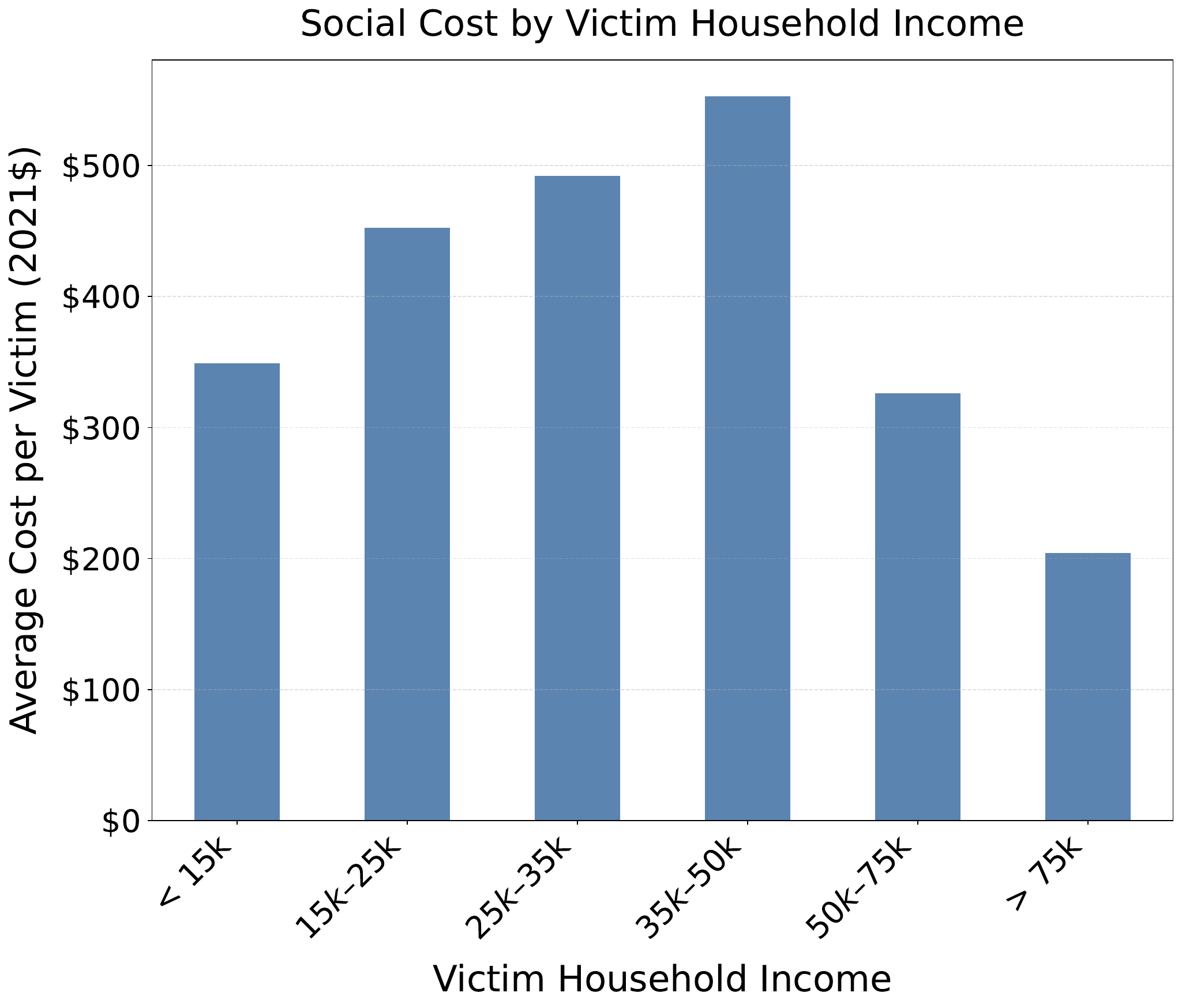}
        \caption{By income.}
        \label{fig:social_cost_income}
    \end{subfigure}
    
    \caption{Average social cost of data breaches segmented by demographic factors.}
    \label{fig:social_cost_combined}
\end{figure}

While our detailed demographic analysis in Appendix \ref{app:demographics} showed that high socio-economic status (SES) individuals have higher odds of victimization, the social cost results shown in Figures \ref{fig:social_cost_age}, \ref{fig:social_cost_race}, \ref{fig:social_cost_education}, and \ref{fig:social_cost_income} demonstrate that the severity of the burden falls disproportionately on vulnerable populations. In terms of age, the highest average social cost affects the 50-64 age group, exceeding \$400 per victim, as shown in Figure \ref{fig:social_cost_age}. This aligns with our hypothesis that this group is ``digitally active but not digitally native,'' discussed in Appendix \ref{app:underrep}. Another interesting finding is that completely opposite of victimization odds, the per-victim social cost is highest for those with the lowest educational attainment and income. Victims with less than a high school education face an average cost of nearly \$850, compared to roughly \$350 for those with graduate degrees, as seen in Figure \ref{fig:social_cost_education}. Similarly, Figure \ref{fig:social_cost_income} shows that victims in the lowest income bracket (under \$15k) bear nearly \$700 in social costs, while the highest earners face only about \$200. Disparities are also evident in terms of race, as seen in Figure \ref{fig:social_cost_race}. Black victims experience the highest average social cost at over \$600 per victim, followed by Hispanic and Native American victims. In contrast, White and Asian victims report significantly lower average social costs, around \$300 and \$250, respectively. These results indicate that the social harm of IDT is far more devastating for low-SES and minority victims.

\label{app:social_cost_groups}

\section{Social Cost for Victims who were Notified of a Data Breach}
\label{app:social_cost_breach}

\subsection{Social Cost for Victims Notified that their Data was Breached}
\label{sec:breach_victim_social_cost}

While the overall social cost of IDT provides an interesting overview, the method by which a criminal obtains personal information can significantly influence the nature and severity of consequences for the victim. To investigate this, we compare victims who were notified that their data was leaked in a data breach to victims who were not notified of such (this was a question in the ITS survey). It is important to note that this distinction is only a proxy. The IDT of victims who were notified that their data was exposed could be attributed to the breach, but could also be just a coincidence (e.g. the data used for IDT was stolen separately). Our comparative analysis, detailed in Table \ref{tab:social_costs_by_type}, reveals an interesting trend, where victims who were notified of a data breach often incur lower per-victim social costs than those who were not. For example, in 2008, breach-notified victims faced a total social cost of \$428.72 per person, while victims not notified of a breach faced \$716.61. After 2016, the difference in social costs faced by breach-notified victims and victims not notified of a breach became much smaller, again likely as a result of the EMV shift as well as structural changes to the ITS survey. In 2021, breach-notified victims faced a total social cost of \$221.61 per person, while non-breach-notified victims faced \$222.64. These results could be due to the fact that breach victims are often alerted early by institutional safeguards, described in Section \ref{sec:incident_characteristics}, allowing for faster remediation. 

When considering the effect of social security number (SSN) compromise on the social cost, we found that SSN compromise was associated with higher social costs. As shown in Table \ref{tab:social_costs_ssn}, victims of an SSN breach consistently faced higher costs across nearly all cost categories compared to victims of a data breach whose SSNs were not leaked. In 2021, the per-victim social cost for an SSN-related breach victim was \$337.24, more than double the \$155.84 faced by victims whose SSN was not exposed. This difference is largely attributable to the lost time cost, since resolving an SSN requires more intensive interaction with government agencies and credit bureaus, leading to higher lost time costs and more emotional distress. Note that in Tables \ref{tab:social_costs_by_type} and \ref{tab:social_costs_ssn}, the breach victim numbers are relatively low compared to the hundreds of millions of users whose records have been exposed. This is because the vast majority of people whose records were exposed did not become IDT victims.

\begin{table}[ht]
\centering
\caption{Social Costs of Identity Theft for Data Breach Victims versus Victims not Notified of a Breach.}
\label{tab:social_costs_by_type}

\sisetup{
    group-separator={,}, 
    group-minimum-digits=4,
    round-mode=places,
    round-precision=2 
}

\footnotesize 

\begin{tabular}{
  l l 
  S[table-format=8.2] 
  S[table-format=3.2] 
  S[table-format=1.2] 
  S[table-format=3.2] 
  S[table-format=1.2] 
  S[table-format=4.2] 
  S[table-format=11.2] 
}
\toprule
{\thead{Year}} & {\thead{Group}} & {\thead{Total \\ Weighted \\ Victims}} & {\thead{Avg. Out-of- \\ Pocket Loss (\$)}} & {\thead{Avg. \\ Legal \\ Cost (\$)}} & {\thead{Avg. Lost \\ Time \\ Cost (\$)}} & {\thead{Avg. \\ Healthcare \\ Cost (\$)}} & {\thead{Total Social \\ Cost per \\ Victim (\$)}} & {\thead{Total \\ National \\ Social Cost (\$)}} \\
\midrule

\multirow{2}{*}{\textbf{2008}} & Breach & 1607267.02 & 151.84 & 6.38 & 268.50 & 2.01 & 428.72 & 689074940.96 \\
& Non-Breach & 9884125.17 & 494.79 & 5.64 & 215.17 & 1.01 & 716.61 & 7083094274.71 \\
\midrule

\multirow{2}{*}{\textbf{2012}} & Breach & 2108805.36 & 62.79 & 1.58 & 162.33 & 1.21 & 227.91 & 480627748.03 \\
& Non-Breach & 14298624.65 & 510.84 & 4.06 & 233.81 & 0.93 & 749.63 & 10718748301.44 \\
\midrule

\multirow{2}{*}{\textbf{2014}} & Breach & 3511656.24 & 953.50 & 2.28 & 206.12 & 0.90 & 1162.80 & 4083351744.90 \\
& Non-Breach & 13941551.04 & 291.84 & 3.60 & 169.56 & 0.63 & 465.63 & 6491616163.16 \\
\midrule

\multirow{2}{*}{\textbf{2016}} & Breach & 5426070.22 & 79.43 & 3.59 & 135.56 & 0.73 & 219.31 & 1190005853.99 \\
& Non-Breach & 19851764.64 & 78.57 & 3.16 & 116.15 & 0.65 & 198.52 & 3941058367.52 \\
\midrule

\multirow{2}{*}{\textbf{2018}} & Breach & 6432045.33 & 69.51 & 0.44 & 127.64 & 0.83 & 198.43 & 1276280960.88 \\
& Non-Breach & 15869463.46 & 85.00 & 1.67 & 118.13 & 0.52 & 205.32 & 3258310823.36 \\
\midrule

\multirow{2}{*}{\textbf{2021}} & Breach & 5747212.66 & 79.33 & 1.82 & 139.16 & 1.30 & 221.61 & 1273655029.05 \\
& Non-Breach & 17868922.65 & 90.67 & 1.78 & 129.54 & 0.66 & 222.64 & 3978280747.92 \\
\bottomrule
\end{tabular}
\end{table}

\begin{table}[ht]
\centering
\caption{Social Costs of Identity Theft for Breach Victims by SSN Breach Status and Year.}
\label{tab:social_costs_ssn}

\sisetup{
    group-separator={,}, 
    group-minimum-digits=4,
    round-mode=places,
    round-precision=2
}

\footnotesize 

\begin{tabular}{
  l l 
  S[table-format=7.2] 
  S[table-format=4.2] 
  S[table-format=1.2] 
  S[table-format=3.2]
  S[table-format=1.2] 
  S[table-format=4.2] 
  S[table-format=10.2] 
}
\toprule
{\thead{Year}} & {\thead{Group}} & {\thead{Total \\ Weighted \\ Victims}} & {\thead{Avg. Out-of- \\ Pocket Loss (\$)}} & {\thead{Avg. \\ Legal \\ Cost (\$)}} & {\thead{Avg. Lost \\ Time \\ Cost (\$)}} & {\thead{Avg. \\ Healthcare \\ Cost (\$)}} & {\thead{Total Social \\ Cost per \\ Victim (\$)}} & {\thead{Total \\ National \\ Social Cost (\$)}} \\
\midrule

\multirow{2}{*}{\textbf{2008}} & SSN Exposed & 770793.01 & 215.16 & 6.55 & 434.98 & 2.48 & 659.18 & 508087860.87 \\
& SSN Not Exposed & 693362.57 & 32.22 & 2.98 & 119.82 & 1.90 & 156.93 & 108810349.32 \\
\midrule

\multirow{2}{*}{\textbf{2012}} & SSN Exposed & 707503.97 & 62.25 & 2.87 & 260.87 & 1.79 & 327.77 & 231897760.95 \\
& SSN Not Exposed & 1299621.39 & 65.07 & 1.01 & 112.66 & 0.65 & 179.40 & 233150967.13 \\
\midrule

\multirow{2}{*}{\textbf{2014}} & SSN Exposed & 548160.54 & 2547.90 & 2.46 & 453.16 & 3.49 & 3007.00 & 1648321393.54 \\
& SSN Not Exposed & 2818797.89 & 690.37 & 2.36 & 164.17 & 0.44 & 857.35 & 2416687892.50 \\
\midrule

\multirow{2}{*}{\textbf{2016}} & SSN Exposed & 2635897.53 & 101.19 & 5.54 & 165.78 & 1.11 & 273.62 & 721247422.80 \\
& SSN Not Exposed & 2400462.74 & 64.46 & 2.04 & 115.06 & 0.44 & 181.99 & 436868321.44 \\
\midrule

\multirow{2}{*}{\textbf{2018}} & SSN Exposed & 1345760.50 & 62.70 & 0.65 & 174.79 & 1.44 & 239.57 & 322404919.80 \\
& SSN Not Exposed & 1425157.50 & 70.06 & 0.00 & 118.03 & 0.37 & 188.46 & 268591628.42 \\
\midrule

\multirow{2}{*}{\textbf{2021}} & SSN Exposed & 1177409.42 & 137.56 & 4.58 & 193.08 & 2.02 & 337.24 & 397069893.06 \\
& SSN Not Exposed & 1422882.78 & 37.73 & 0.99 & 116.04 & 1.09 & 155.84 & 221742562.24 \\
\bottomrule
\end{tabular}
\end{table}

\section{Inflation Adjustments}
\label{app:inflation}
The formula used to convert a nominal monetary value from a given year to its real value in 2021 dollars is as follows:
\begin{equation}
\label{eq:inflation_adjustment}
\text{Adjusted Value (in 2021 \$)} = \text{Nominal Value} \times \left( \frac{\text{CPI}_{2021}}{\text{CPI}_{\text{Original Year}}} \right)
\end{equation}

Table \ref{tab:appendix_cpi} displays all the CPI values that were used for these calculations. This inflation adjustment was applied to any monetary value in this paper. These CPI values were obtained from the U.S. Bureau of Labor Statistics \cite{BLS_CPI, USInflationCalc_CPI}.

\begin{table}[ht]
\centering
\caption{Annual Average Consumer Price Index (CPI)}
\label{tab:appendix_cpi}
\begin{tabular}{lr}
\toprule
\textbf{Year} & \textbf{Annual Average CPI} \\
\midrule
2008 & 215.297 \\
2012 & 233.165 \\
2014 & 236.736 \\
2016 & 240.007 \\
2018 & 251.107 \\
2021 & 270.970 \\
\bottomrule
\end{tabular}
\end{table}

\section{Average Private Nonfarm Hourly Wages}
\label{app:hourly_wages}
To estimate opportunity cost of lost time in the social cost calculations, time lost was multiplied by the respective year's average hourly wage, shown in Table \ref{tab:appendix_wage}. This private nonfarm hourly wage data represents the vast majority of the U.S. workforce, and was obtained from \cite{FRED_Wage}.

\begin{table}[ht]
\centering
\caption{Nominal and Inflation-Adjusted Average Private Hourly Wage}
\label{tab:appendix_wage}
\begin{tabular}{lrr}
\toprule
\textbf{Year} & \multicolumn{2}{c}{\textbf{Average Private Hourly Wage}} \\
\cmidrule(lr){2-3} 
& \textbf{Nominal} & \textbf{Adjusted} \\
\midrule
2008 & 21.19 & 26.67 \\
2012 & 23.26 & 27.03 \\
2014 & 24.23 & 27.73 \\
2016 & 25.37 & 28.64 \\
2018 & 26.72 & 28.83 \\
2021 & 29.92 & 29.92 \\
\bottomrule
\end{tabular}
\end{table}

\section{Fixed Service Costs for Social Cost Calculations}
\label{app:fixed_service_costs}
To estimate the cost of hiring a lawyer, a doctor or therapy appointment, or obtaining medication, we constructed the following Table \ref{tab:appendix_cost_estimates_sourced}, where costs were also adjusted for inflation.

\begin{table}[ht]
\centering
\caption{Cost Estimates and Sources for Social Cost Calculations}
\label{tab:appendix_cost_estimates_sourced}
\small
\begin{tabular}{l l r c r l l}
\toprule
\thead{Item} & 
\thead{Reported \\ Nominal Cost} & 
\thead{Assumed \\ Nominal \\ Cost (\$)} & 
\thead{Source \\ Year} & 
\thead{Adjusted \\ Cost \\ (2021 \$)} & 
\thead{Notes} &
\thead{Source} \\
\midrule
Lawyer & \$327 per hour & 500 & 2023 & 444.65 & Based on approx. 1.5 hours & \cite{USNews_Lawyer}\\
Doctor Visit & \$79--172 & 100 & 2024 & 86.38 & Without insurance & \cite{DebtOrg_Doctor} \\
Therapist Visit & \$100--200 & 100 & 2024 & 86.38 & Without insurance & \cite{GoodRx_Therapy} \\
Medication & \$50 & 50 & 2018 & 53.96 & Generic prescription & \cite{CBO_Prescription} \\
\bottomrule
\end{tabular}
\end{table}

\section{Details of Dataset Filtering}

To create the sample of identity theft victims, the dataset was filtered to include any respondent who answered "Yes" (coded as 1) to any of the following five questions concerning events in the past 12 months: misuse of an existing bank account, misuse of an existing credit card, misuse of another existing account, fraudulent opening of a new account, or the use of personal information for other fraudulent purposes. This is detailed in Table \ref{tab:victim_filtering}.

\begin{table}[ht]
\centering
\caption{Filtering Rules to Identify Identity Theft Victims}
\label{tab:victim_filtering}
\begin{tabular}{p{0.6\textwidth}l}
\toprule
\textbf{Condition Met if Variable...} & \textbf{Equals Value...} \\
\midrule
\multicolumn{2}{l}{\textit{A respondent is a victim if ANY of the following are true:}} \\
\addlinespace 
EXISTING\_BANK\_ACCT\_MISUSE\_12MO   & 1 (Yes) \\
EXISTING\_CC\_MISUSE\_12MO         & 1 (Yes) \\
EXISTING\_OTHER\_ACCT\_MISUSE\_12MO  & 1 (Yes) \\
NEW\_ACCT\_OPENED\_12MO            & 1 (Yes) \\
OTHER\_FRAUDULENT\_PURPOSE\_12MO   & 1 (Yes) \\
\bottomrule
\end{tabular}
\end{table}

After identifying all identity theft victims (both attempted and successful), a second filtering step was applied to remove records where the theft was attempted but not successful. The specific variables and codes used to identify an "attempted only" case vary by survey year due to changes in the survey instrument. For the 2021 survey, no filtering was necessary, as the questionnaire was designed to capture only successful incidents. The precise logic for each year is detailed in Table \ref{tab:successful_theft_filtering}.

\begin{table}[ht]
\centering
\caption{Rules for Identifying and Removing 'Attempted Only' Identity Theft Incidents}
\vspace{0.2cm}
\label{tab:successful_theft_filtering}
\begin{tabular}{lp{0.65\textwidth}}
\toprule
\textbf{Survey Year} & \textbf{Condition to REMOVE a record because theft was not successful} \\
\midrule
\addlinespace 
2008 & A record is removed if \texttt{VS012=2} AND \texttt{VS014=2} AND \texttt{VS016=2} AND \texttt{VS018=2} AND \texttt{VS020=2}. \\
\addlinespace
2012 & A record is removed if the variable \texttt{VS098 = 009} (Not applicable, it was not actually used). \\
\addlinespace
2014 & A record is removed if the variable \texttt{VS093 = 009}  (Not applicable, it was not actually used). \\
\addlinespace
2016 & A record is removed if the variable \texttt{VS093 = 9}  (Not applicable, it was not actually misused). \\
\addlinespace
2018 & A record is removed if the variable \texttt{VS093 = 9}  (Not applicable, it was not actually misused). \\
\addlinespace
2021 & No filtering action is needed. The survey design for this year ensures all identified victims are successful incidents. \\
\bottomrule
\end{tabular}
\end{table}

\section{ITS Victim Timeline}
\label{app:timeline}
To plot the number of reported identity thefts per month from the ITS dataset, the following assumptions were made. The primary variable used for the timeline is the month in which the victim first discovered the identity theft. For survey years where this specific "discovery month" was explicitly recorded (such as the redesigned 2021 ITS), that value was used directly. For survey waves or individual records where a specific discovery month was missing or not collected, the month was estimated using the Interview Quarter variable. In cases where only the interview quarter was available, incidents were mapped to the mid-month of that quarter to provide a representative point estimate (e.g., Quarter 1 $\rightarrow$ February, Quarter 2 $\rightarrow$ May, Quarter 3 $\rightarrow$ August, Quarter 4 $\rightarrow$ November). This ensures the timeline reflects the general seasonal distribution of reports while acknowledging the six-month reference period typically used in the NCVS core interview. Monthly counts were calculated by summing the final ITS weights of all confirmed victims discovered or reported in that month. This allows the timeline to represent national-level estimates of identity theft discovery rather than raw sample counts. 

\section{Dynamic Saturation Model}
\label{app:sat_model}
Here we display the month-to-month overlay and conversion rate results, as well as Wilcoxon signed-rank testing results, for the output of our dynamic saturation model (with $Y=1$)(proposed in Section \ref{sec:estimating_number_records}). Recall that the purpose of this model is to take the augmented PRC data as an input, and output the estimated number of unique individuals whose data was compromised.

\begin{figure}[htbp]
    \centering
    \begin{minipage}[t]{0.32\textwidth}
        \centering
        \includegraphics[width=\textwidth]{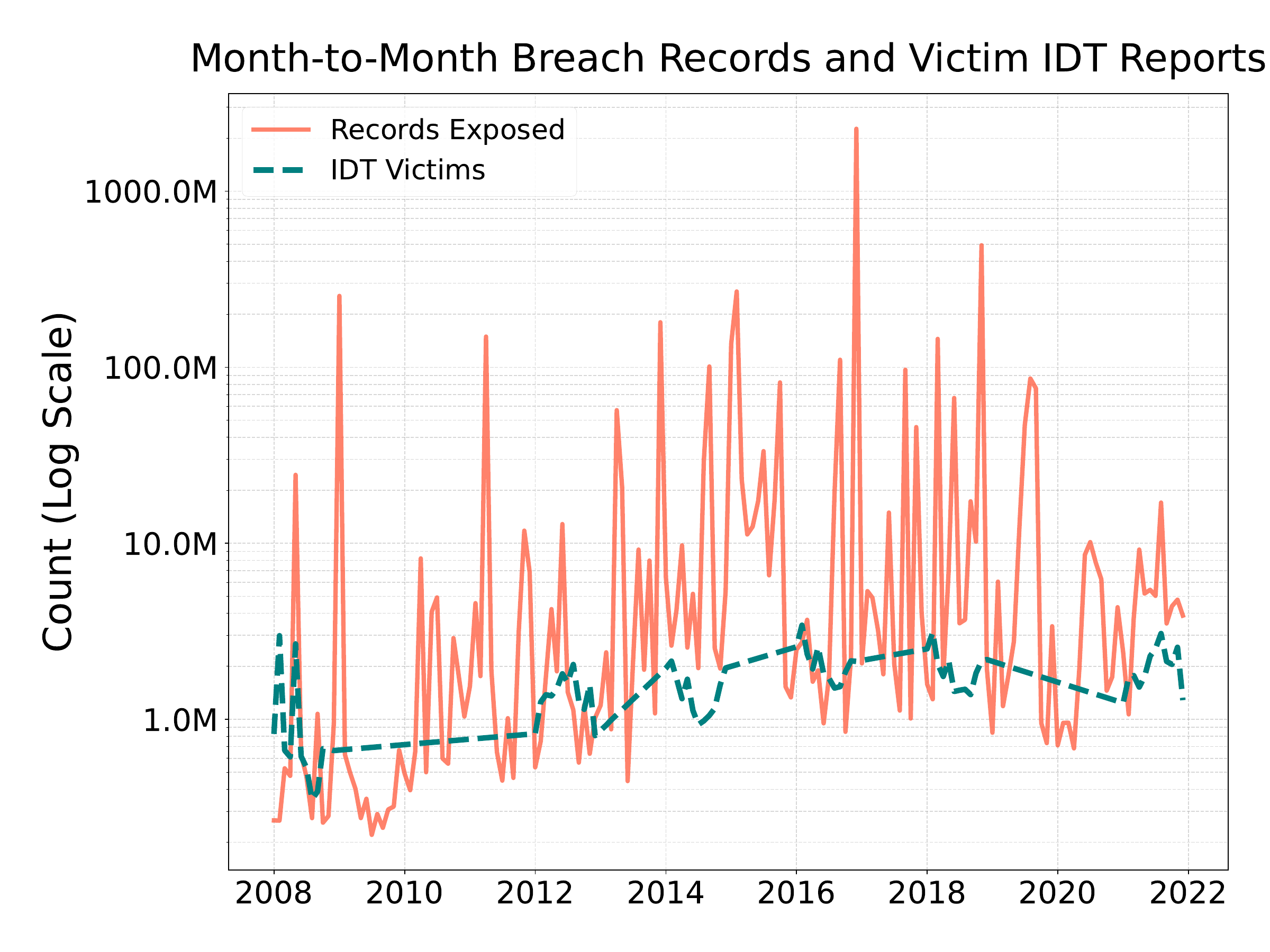}
        \caption{Estimated number of unique individuals compromised (saturation model PRC) and number of reported IDT victims (ITS).}
        \label{fig:saturation_fig5}
    \end{minipage}
    \hfill 
    \begin{minipage}[t]{0.32\textwidth}
        \centering
        \includegraphics[width=\textwidth]{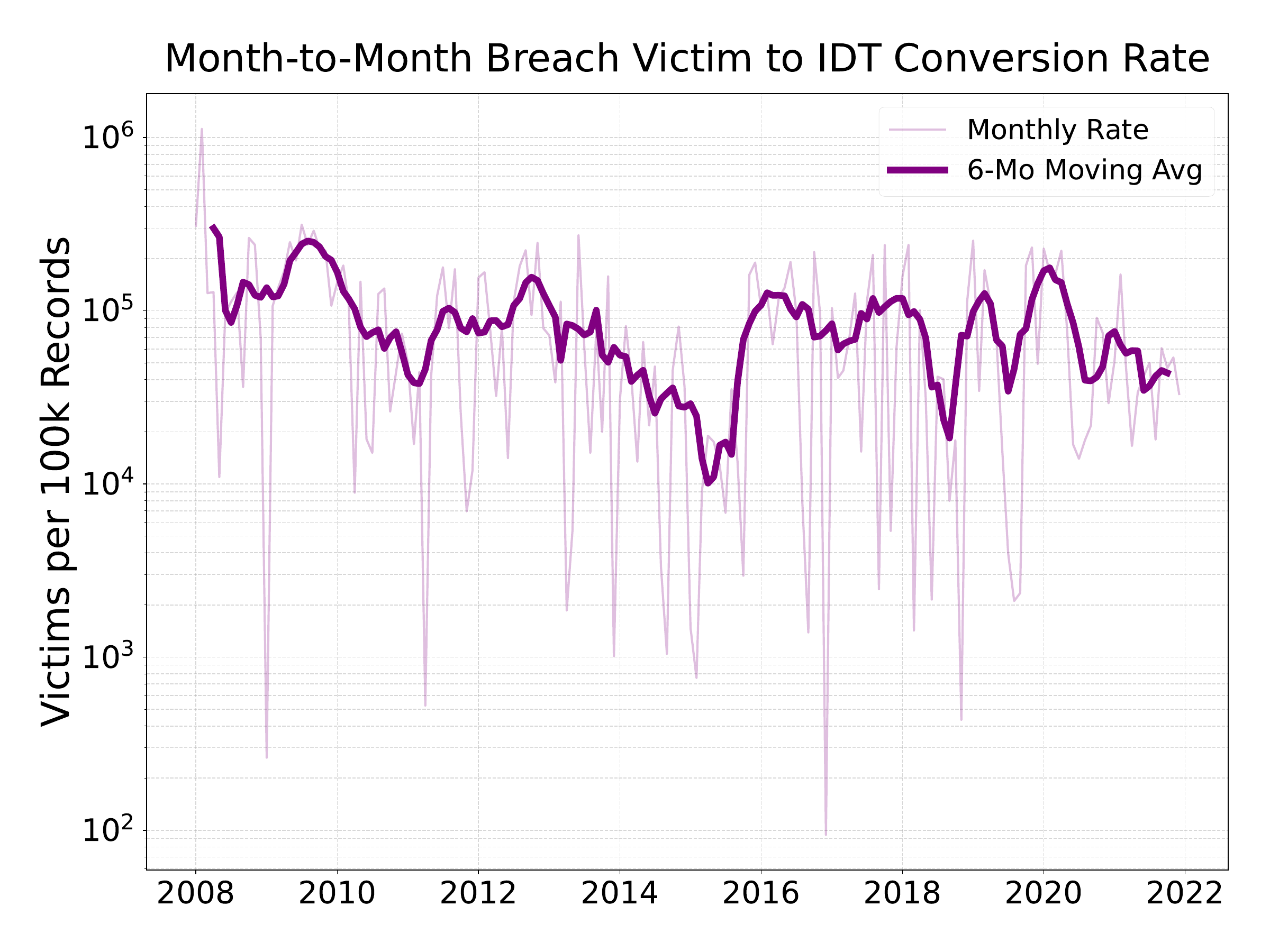}
        \caption{Saturation model conversion rates from data being breached to becoming an IDT victim.}
        \label{fig:saturation_conversion_rate}
    \end{minipage}
    \hfill
    \begin{minipage}[t]{0.32\textwidth}
        \centering
        \includegraphics[width=\textwidth]{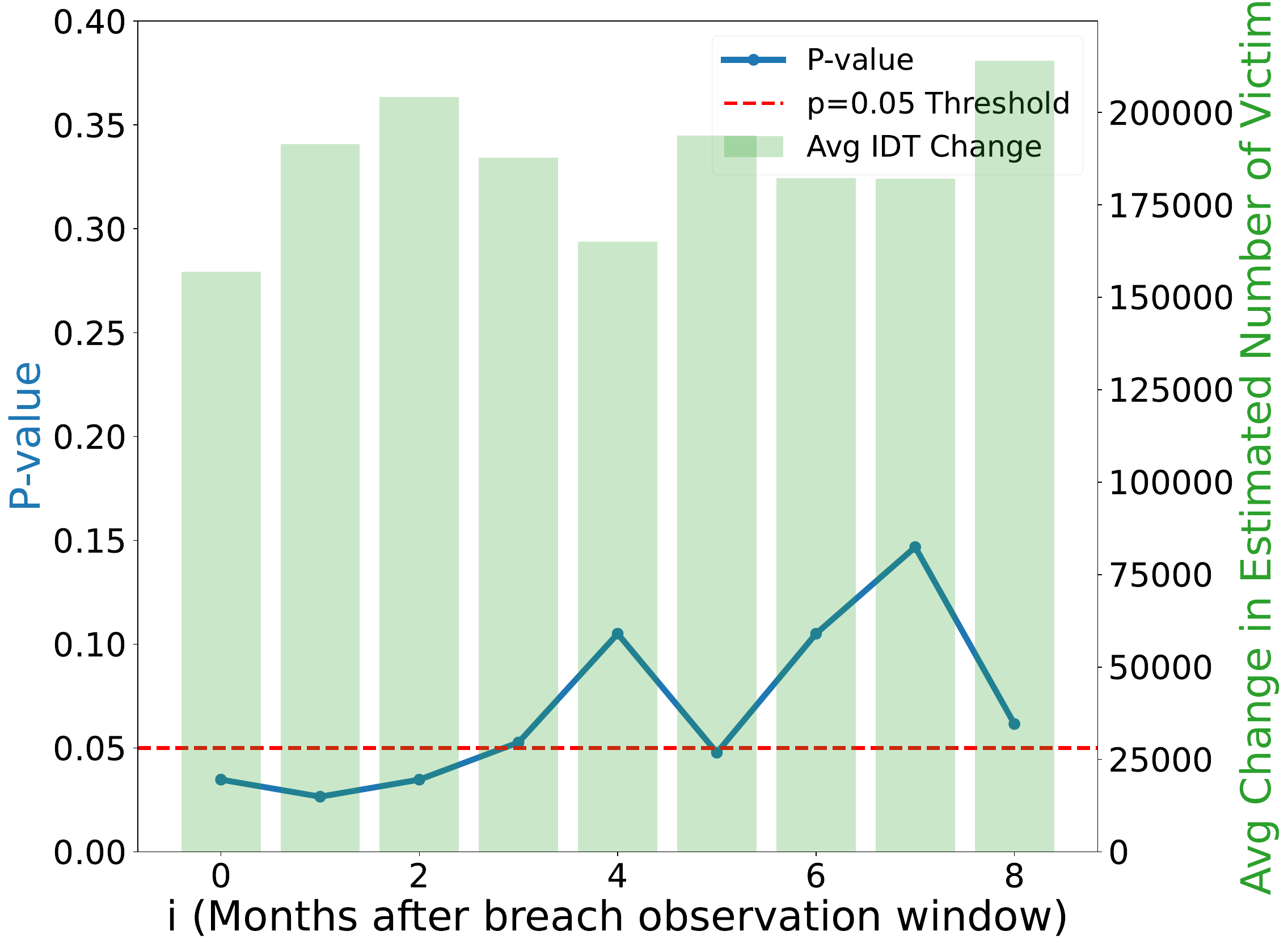}
        \caption{Wilcoxon signed-rank test results for the saturation model PRC data breach chronology data.}
        \label{fig:wilcoxon_sat}
    \end{minipage}
\end{figure}

\section{Harmonization Codes}
\label{app: dataset_harmonization}

The following new variables were created according to the mappings below in order to harmonize similar variables from different years of datasets.

\begin{table}[ht]
\centering
\caption{Harmonization Map for EXISTING\_BANK\_ACCT\_MISUSE\_12MO}
\label{tab:bank_misuse_codes}
\begin{tabular}{llll}
\toprule
\textbf{Year} & \textbf{Variable Name} & \textbf{Yes Code} & \textbf{No Code} \\
\midrule
2008 & \texttt{VS013} & 1  & 2  \\
2012 & \texttt{VS012} & 01 & 02 \\
2014 & \texttt{VS012} & 01 & 02 \\
2016 & \texttt{VS012} & 1  & 2  \\
2018 & \texttt{VS012} & 1  & 2  \\
2021 & \texttt{VS012} & 1  & 2  \\
\bottomrule
\end{tabular}
\end{table}

\begin{table}[ht]
\centering
\caption{Harmonization Map for EXISTING\_CC\_MISUSE\_12MO}
\label{tab:cc_misuse_codes}
\begin{tabular}{llll}
\toprule
\textbf{Year} & \textbf{Variable Name} & \textbf{Yes Code} & \textbf{No Code} \\
\midrule
2008 & \texttt{VS011} & 1  & 2  \\
2012 & \texttt{VS016} & 01 & 02 \\
2014 & \texttt{VS017} & 01 & 02 \\
2016 & \texttt{VS017} & 1  & 2  \\
2018 & \texttt{VS017} & 1  & 2  \\
2021 & \texttt{VS017} & 1  & 2  \\
\bottomrule
\end{tabular}
\end{table}

\begin{table}[ht]
\centering
\caption{Harmonization Map for EXISTING\_OTHER\_ACCT\_MISUSE\_12MO}
\label{tab:other_misuse_codes}
\begin{tabular}{llll}
\toprule
\textbf{Year} & \textbf{Variable Name} & \textbf{Yes Code} & \textbf{No Code} \\
\midrule
2008 & \texttt{VS015}  & 1 & 2 \\
2012 & \texttt{VS018}  & 1 & 2 \\
2014 & \texttt{VS019}  & 1 & 2 \\
2016 & \texttt{VS019}  & 1 & 2 \\
2018 & \texttt{VS019}  & 1 & 2 \\
\midrule 
\multirow{2}{*}{2021} & \texttt{VS019}  & 1 & 2 \\
     & \texttt{VS017E} & 1 & 2 \\
\bottomrule
\end{tabular}
\end{table}

\begin{table}[ht]
\centering
\caption{Harmonization Map for NEW\_ACCT\_OPENED\_12MO}
\label{tab:new_acct_codes}
\begin{tabular}{llll}
\toprule
\textbf{Year} & \textbf{Variable Name} & \textbf{Yes Code} & \textbf{No Code} \\
\midrule
2008 & \texttt{VS017} & 1  & 2  \\
2012 & \texttt{VS041} & 01 & 02 \\
2014 & \texttt{VS041} & 01 & 02 \\
2016 & \texttt{VS041} & 1  & 2  \\
2018 & \texttt{VS041} & 1  & 2  \\
2021 & \texttt{VS041} & 1  & 2  \\
\bottomrule
\end{tabular}
\end{table}

\begin{table}[ht]
\centering
\caption{Harmonization Map for OTHER\_FRAUDULENT\_PURPOSE\_12MO}
\label{tab:other_fraud_codes}
\begin{tabular}{llll}
\toprule
\textbf{Year} & \textbf{Variable Name} & \textbf{Yes Code} & \textbf{No Code} \\
\midrule
2008 & \texttt{VS019} & 1  & 2  \\
2012 & \texttt{VS067} & 01 & 02 \\
2014 & \texttt{VS063} & 01 & 02 \\
2016 & \texttt{VS063} & 1  & 2  \\
2018 & \texttt{VS063} & 1  & 2  \\
2021 & \texttt{VS063} & 1  & 2  \\
\bottomrule
\end{tabular}
\end{table}

\begin{longtable}{c c p{7.5cm}}
\caption{Harmonization Map for How Misuse Was Discovered.} \label{tab:how_discovered_final_long_vars} \\
\toprule
\textbf{Year} & \textbf{Variable} & \textbf{Original Codes} \\
\midrule
\endfirsthead

\toprule

\midrule
\endhead

\bottomrule
\endfoot

\multicolumn{3}{l}{\textbf{1: Noticed Problem with Account or Suspicious Computer Activity}} \\ \addlinespace
2008 & \texttt{VS088} & 2, 3, 5 \\
2012 & \texttt{VS095} & 002, 003, 005, 019, 020, 021 \\
2014 & \texttt{VS090} & 002, 003, 005, 009, 019, 021 \\
2016 & \texttt{VS090} & 2, 3, 5, 21, 22, 18 \\
2018 & \texttt{VS090} & 2, 3, 5, 16, 17, 18 \\
2021 & \texttt{VS090} & 2, 3, 5, 16, 17, 18 \\
\midrule

\multicolumn{3}{l}{\textbf{2: Notified by Financial Institution or Monitoring Service}} \\ \addlinespace
2008 & \texttt{VS088} & 9, 10 \\
2012 & \texttt{VS095} & 009, 010, 011 \\
2014 & \texttt{VS090} & 010, 011, 012 \\
2016 & \texttt{VS090} & 10, 11, 12 \\
2018 & \texttt{VS090} & 10, 11, 12 \\
2021 & \texttt{VS090} & 10, 11, 12 \\
\midrule

\multicolumn{3}{l}{\textbf{3: Notified by Other Person or Organization}} \\ \addlinespace
2008 & \texttt{VS088} & 12, 13, 15 \\
2012 & \texttt{VS095} & 012, 013, 015, 016 \\
2014 & \texttt{VS090} & 013, 014, 016, 017, 022 \\
2016 & \texttt{VS090} & 13, 14, 17, 19 \\
2018 & \texttt{VS090} & 13, 14, 19 \\
2021 & \texttt{VS090} & 13, 14, 19, 20, 21 \\
\midrule

\multicolumn{3}{l}{\textbf{4: Received Bill, Merchandise, or Card Not Ordered/Owed}} \\ \addlinespace
2008 & \texttt{VS088} & 4, 8 \\
2012 & \texttt{VS095} & 004, 008, 018 \\
2014 & \texttt{VS090} & 004, 008, 018 \\
2016 & \texttt{VS090} & 4, 8, 20 \\
2018 & \texttt{VS090} & 4, 8 \\
2021 & \texttt{VS090} & 4, 8 \\
\midrule

\multicolumn{3}{l}{\textbf{5: Checked Credit Report or Monitored Account}} \\ \addlinespace
2008 & \texttt{VS088} & 7, 16 \\
2012 & \texttt{VS095} & 007 \\
2014 & \texttt{VS090} & 007 \\
2016 & \texttt{VS090} & 7 \\
2018 & \texttt{VS090} & 7 \\
2021 & \texttt{VS090} & 7 \\
\midrule

\multicolumn{3}{l}{\textbf{7: Other/Unspecified}} \\ \addlinespace
2008 & \texttt{VS088} & 14 \\
2012 & \texttt{VS095} & 001, 006, 014, 017, 020, 022, 023 \\
2014 & \texttt{VS090} & 001, 006, 009, 015, 020, 023 \\
2016 & \texttt{VS090} & 1, 6, 9, 15, 16, 20 \\
2018 & \texttt{VS090} & 1, 6, 9, 15 \\
2021 & \texttt{VS090} & 1, 6, 9, 15 \\

\end{longtable}

\begin{longtable}{c c p{7.5cm}}
\caption{Harmonization Map for Theft Method.} \label{tab:theft_method_vars} \\
\toprule
\textbf{Year} & \textbf{Variable} & \textbf{Original Codes} \\
\midrule
\endfirsthead

\multicolumn{3}{c}%
{{\bfseries \tablename\ \thetable{} -- continued from previous page}} \\
\toprule
\textbf{Year} & \textbf{Variable} & \textbf{Original Codes} \\
\midrule
\endhead

\bottomrule
\endfoot

\multicolumn{3}{l}{\textbf{1: Lost or Stolen Physical Item (Wallet, Mail, etc.)}} \\ \addlinespace
2008 & \texttt{VS110} & 1, 3, 4 \\
2012 & \texttt{VS100} & 001, 002, 003, 004, 005 \\
2014 & \texttt{VS096} & 001, 002, 003, 004, 005, 020 \\
2016 & \texttt{VS096} & 1, 2, 3, 4, 5, 20 \\
2018 & \texttt{VS096} & 1, 2, 3, 4 \\
2021 & \texttt{VS096} & 1, 2, 3 \\
\midrule

\multicolumn{3}{l}{\textbf{2: Stolen During a Transaction (Online or In-Person)}} \\ \addlinespace
2008 & \texttt{VS110} & 5 \\
2012 & \texttt{VS100} & 006, 007, 017, 019, 020 \\
2014 & \texttt{VS096} & 006, 007, 016, 017, 018 \\
2016 & \texttt{VS096} & 6, 7, 16, 17, 18 \\
2018 & \texttt{VS096} & 5, 6 \\
2021 & \texttt{VS096} & 5, 6 \\
\midrule

\multicolumn{3}{l}{\textbf{3: Stolen from a Computer/Device (Hacking)}} \\ \addlinespace
2008 & \texttt{VS110} & 7 \\
2012 & \texttt{VS100} & 009 \\
2014 & \texttt{VS096} & 008 \\
2016 & \texttt{VS096} & 8, 21 \\
2018 & \texttt{VS096} & 7 \\
2021 & \texttt{VS096} & 4 \\
\midrule

\multicolumn{3}{l}{\textbf{4: Deceived by a Scam (e.g., Phishing)}} \\ \addlinespace
2008 & \texttt{VS110} & 8 \\
2012 & \texttt{VS100} & 010 \\
2014 & \texttt{VS096} & 009 \\
2016 & \texttt{VS096} & 9 \\
2018 & \texttt{VS096} & 8 \\
2021 & \texttt{VS096} & 7 \\
\midrule

\multicolumn{3}{l}{\textbf{5: Stolen from a Company/Employer (Data Breach)}} \\ \addlinespace
2008 & \texttt{VS110} & 9, 12, 14 \\
2012 & \texttt{VS100} & 011, 012, 021 \\
2014 & \texttt{VS096} & 010, 011, 019 \\
2016 & \texttt{VS096} & 10, 11, 19 \\
2018 & \texttt{VS096} & 9, 10 \\
2021 & \texttt{VS096} & 8, 9 \\
\midrule

\multicolumn{3}{l}{\textbf{7: Other}} \\ \addlinespace
2008 & \texttt{VS110} & 6, 11 \\
2012 & \texttt{VS100} & 008, 013, 014, 015, 016, 018, 022 \\
2014 & \texttt{VS096} & 012, 013, 014, 015, 021 \\
2016 & \texttt{VS096} & 12, 13, 14, 15, 22 \\
2018 & \texttt{VS096} & 11, 12 \\
2021 & \texttt{VS096} & 10 \\

\end{longtable}

\begin{table}
\centering
\small 
\caption{Recoding Map for Harmonized Household Income (HOUSEHOLD\_INCOME)}
\label{tab:income_recoding}
\begin{tabular}{llcccccc}
\toprule
\textbf{New} & \textbf{Harmonized Category} & \multicolumn{6}{c}{\textbf{Original Code by Year}} \\
\cmidrule(lr){3-8}
\textbf{Code} & & \textbf{2008} & \textbf{2012} & \textbf{2014} & \textbf{2016} & \textbf{2018} & \textbf{2021} \\
\midrule
1  & Less than \$5,000      & 1  & 1  & 01 & 1  & 1  & 1  \\
2  & \$5,000 to \$7,499       & 2  & 2  & 02 & 2  & 2  & 2  \\
3  & \$7,500 to \$9,999       & 3  & 3  & 03 & 3  & 3  & 3  \\
4  & \$10,000 to \$12,499     & 4  & 4  & 04 & 4  & 4  & 4  \\
5  & \$12,500 to \$14,999     & 5  & 5  & 05 & 5  & 5  & 5  \\
6  & \$15,000 to \$17,499     & 6  & 6  & 06 & 6  & 6  & 6  \\
7  & \$17,500 to \$19,999     & 7  & 7  & 07 & 7  & 7  & 7  \\
8  & \$20,000 to \$24,999     & 8  & 8  & 08 & 8  & 8  & 8  \\
9  & \$25,000 to \$29,999     & 9  & 9  & 09 & 9  & 9  & 9  \\
10 & \$30,000 to \$34,999     & 10 & 10 & 10 & 10 & 10 & 10 \\
11 & \$35,000 to \$39,999     & 11 & 11 & 11 & 11 & 11 & 11 \\
12 & \$40,000 to \$49,999     & 12 & 12 & 12 & 12 & 12 & 12 \\
13 & \$50,000 to \$74,999     & 13 & 13 & 13 & 13 & 13 & 13 \\
14 & \$75,000 and over      & 14 & 14 & 14 & 14 & 15-18* & 15-18* \\
\bottomrule
\multicolumn{8}{p{0.9\textwidth}}{\footnotesize *Note: For 2018 and 2021, the original codes 15, 16, 17, and 18 are all recoded into the harmonized category 14, as the income brackets were expanded in those survey years.} \\
\end{tabular}
\end{table}

\begin{table}[ht]
\centering
\caption{Recoding Map for Harmonized Education Level (PERSON\_EDUCATION)}
\label{tab:education_recoding}
\begin{tabular}{cll}
\toprule
\textbf{Harm.} & \textbf{Harmonized Category} & \textbf{Original Codes*} \\
\textbf{Code} & & \\
\midrule
1 & No Schooling/Kindergarten                 & 0 \\
2 & Elementary School (Grades 1-8)            & 1-8 \\
3 & Some High School (Grades 9-12, no diploma) & 9-12, 27 \\
4 & High School Graduate (or equivalent)      & 28 \\
5 & Some College/Associate Degree             & 21-26, 40, 41 \\
6 & Bachelor's Degree                         & 42 \\
7 & Graduate/Professional Degree              & 43, 44, 45 \\
\bottomrule
\multicolumn{3}{p{0.8\textwidth}}{\footnotesize *Note: Original codes are consistent for all survey years (2008, 2012, 2014, 2016, 2018, 2021). The 2014 survey zero-pads single-digit codes (e.g., '1' becomes '01').} \\
\end{tabular}
\end{table}

\begin{table}[ht]
\centering
\caption{Harmonization Map for OUT\_OF\_POCKET\_LOSS\_RECENT\_INCIDENT}.
\label{tab:map_oop_loss}
\begin{tabular}{ll}
\toprule
\textbf{Year} & \textbf{Source Variable} \\
\midrule
2008 & VS252 \\
2012 & VS299 \\
2014 & VS281 \\
2016 & VS281 \\
2018 & VS281 \\
2021 & VS281 \\
\bottomrule
\end{tabular}
\end{table}

\begin{table}[ht]
\centering
\caption{Harmonization Map for CONTACT\_HIRED\_LAWYER}
\label{tab:map_hired_lawyer}
\begin{tabular}{llll}
\toprule
\textbf{Year} & \textbf{Variable Name} & \textbf{Yes Code} & \textbf{No Code} \\
\midrule
2008 & VS202 & 1  & 2  \\
2012 & VS213 & 01 & 02 \\
2014 & VS197 & 01 & 02 \\
2016 & VS197 & 1  & 2  \\
2018 & VS197 & 1  & 2  \\
2021 & VS197 & 1  & 2  \\
\bottomrule
\end{tabular}
\end{table}

\begin{table}[ht]
\centering
\caption{Harmonization Map for HOURS\_SPENT\_RESOLVING\_PROBLEMS}
\label{tab:map_hours_spent}
\begin{tabular}{ll}
\toprule
\textbf{Year} & \textbf{Source Variable} \\
\midrule
2008 & VS257 \\
2012 & VS304 \\
2014 & VS286 \\
2016 & VS286 \\
2018 & VS286 \\
2021 & VS286 \\
\bottomrule
\end{tabular}
\end{table}

\begin{table}[ht]
\centering
\caption{Harmonization Map for HELP\_TYPE\_COUNSELING}
\label{tab:map_counseling}
\begin{tabular}{llll}
\toprule
\textbf{Year} & \textbf{Variable Name} & \textbf{Selected Code} & \textbf{Not Selected Code} \\
\midrule
2008 & VS228 & 1 & 0 \\
2012 & VS256 & 1 & 0 \\
2014 & VS239 & 1 & 0 \\
2016 & VS239 & 1 & 0 \\
2018 & VS239 & 1 & 0 \\
2021 & VS239 & 1 & 0 \\
\bottomrule
\end{tabular}
\end{table}

\begin{table}[ht]
\centering
\caption{Harmonization Map for HELP\_TYPE\_MEDICATION}
\label{tab:map_medication}
\begin{tabular}{llll}
\toprule
\textbf{Year} & \textbf{Variable Name} & \textbf{Selected Code} & \textbf{Not Selected Code} \\
\midrule
2008 & VS229 & 1 & 0 \\
2012 & VS257 & 1 & 0 \\
2014 & VS240 & 1 & 0 \\
2016 & VS240 & 1 & 0 \\
2018 & VS240 & 1 & 0 \\
2021 & VS240 & 1 & 0 \\
\bottomrule
\end{tabular}
\end{table}

\begin{table}[ht]
\centering
\caption{Harmonization Map for SOUGHT\_HELP\_PHYSICAL\_PROBLEMS}
\label{tab:map_phys_problems}
\begin{tabular}{llll}
\toprule
\textbf{Year} & \textbf{Variable Name} & \textbf{Yes Code} & \textbf{No Code} \\
\midrule
2008 & VS242 & 1  & 2  \\
2012 & VS283 & 1  & 2  \\
2014 & VS265 & 1  & 2  \\
2016 & VS265 & 1  & 2  \\
2018 & VS265 & 1  & 2  \\
2021 & VS265 & 1  & 2  \\
\bottomrule
\end{tabular}
\end{table}

\begin{table}[ht]
\centering
\caption{Harmonization Map for HELP\_TYPE\_VISITED\_MEDICAL\_PROFESSIONAL}
\label{tab:map_med_prof_revised}
\begin{tabular}{llll}
\toprule
\textbf{Year} & \textbf{Source Variable(s)} & \textbf{Selected Code} & \textbf{Not Selected Code} \\
\midrule
2008 & VS230, VS231 & 1 & 0 \\
2012 & VS258, VS259 & 1 & 0 \\
2014 & VS241, VS242 & 1 & 0 \\
2016 & VS241, VS242 & 1 & 0 \\
2018 & VS241A & 1 & 0 \\
2021 & VS241A & 1 & 0 \\
\bottomrule
\end{tabular}
\end{table}

\begin{table}[ht]
\centering
\caption{Harmonization Map for SOUGHT\_HELP\_EMOTIONAL\_DISTRESS}
\label{tab:map_emo_distress_revised}
\begin{tabular}{llll}
\toprule
\textbf{Year} & \textbf{Variable Name} & \textbf{Yes Code} & \textbf{No Code} \\
\midrule
2008 & VS227 & 1 & 2 \\
2012 & VS255 & 1 & 2 \\
2014 & VS238 & 1 & 2 \\
2016 & VS238 & 1 & 2 \\
2018 & VS238 & 1 & 2 \\
2021 & VS238 & 1 & 2 \\
\bottomrule
\end{tabular}
\end{table}

\end{document}